\documentclass[11pt,a4paper]{article}
\pdfoutput=1
\usepackage{pub}
\usepackage[T1]{fontenc}
\usepackage[english]{babel}
\usepackage{lmodern}
\usepackage{bbm}
\usepackage{bm}
\usepackage{color}
\usepackage{slashed}
\usepackage{graphicx}
\usepackage{braket}
\usepackage{bbold}
\usepackage{dsfont}
\usepackage{comment}

\usepackage[usenames,dvipsnames]{xcolor}

\DeclareGraphicsExtensions{.pdf,.png,.jpg}
\DeclareMathOperator{\arccot}{arccot}

\usepackage{multirow}
\usepackage{xcolor,pifont}
\newcommand{\xmark}{\text{\ding{55}}}
\newcommand{\cmark}{\text{\ding{51}}}

\allowdisplaybreaks[1]
\def\simg{{\ \lower-1.2pt\vbox{\hbox{\rlap{$>$}\lower6pt\vbox{\hbox{$\sim$}}}}\ }}
\def\siml{{\ \lower-1.2pt\vbox{\hbox{\rlap{$<$}\lower6pt\vbox{\hbox{$\sim$}}}}\ }}

\makeatletter \@addtoreset{equation}{section} \makeatother

\begin{document}

\flushbottom
\begin{titlepage}
	\begin{centering}
		\vfill
{\Large{\bf
 Non-perturbative effects for dark sectors 
\\
with QCD portals}
}
\\
\vspace{0.5 cm}
\vspace{0.8cm}
{\bf Simone Biondini}\footnote{\color{blue} simone.biondini@unibas.ch}$^{a}$, {\bf Talal  Ahmed  Chowdhury}\footnote{\color{blue} talal@du.ac.bd}$^{b,c,d}$, and {\bf Shaikh Saad}\footnote{\color{blue} shaikh.saad@unibas.ch}$^{a}$

\vspace{0.8 cm}
$^{a}${\em Department of Physics, University of Basel,
\\
Klingelbergstrasse 82, CH-4056 Basel, Switzerland}
\\
\vspace{0.15cm}
$^{b}${\em Department of Physics, University of Dhaka,\\ P.O. Box 1000, Dhaka, Bangladesh} 
\\
\vspace{0.15 cm}
$^{c}${\em Department of Physics and Astronomy, University of Kansas,\\ Lawrence, Kansas 66045, USA}
\\
\vspace{0.15 cm}
$^{d}${\em The Abdus Salam International Centre for Theoretical Physics,\\ Strada Costiera 11, I-34014, Trieste, Italy}

\vspace*{0.8cm}
\end{centering}
\vspace*{0.3cm}
\noindent
	
\textbf{Abstract}: In this work, we consider a class of dark matter (DM) models where the DM does not directly interact with the Standard Model (SM) particles at the tree-level. Therefore, the coannihilation mechanism is crucial in achieving the correct DM relic abundance, which in turn requires the coannihilating partner to be close in mass to the actual DM particle. In our systematisation of the models' class, the mediator and the coannihilation partner are assumed to be charged under QCD interactions. This last feature calls for a scrutiny of non-perturbative effects, namely  Sommerfeld factors and bound-state formation, on the annihilations of the colored partner. 
Such non-perturbative effects are illustrated with an example model comprising a scalar leptoquark mediator, a Dirac vector-like fermion coannihilation partner, and a singlet DM fermion. Phenomenological features of this model, namely DM direct and indirect detection prospects, collider implications, and impact on the muon anomalous magnetic moment, are discussed.

\vfill
\newpage
\tableofcontents
\end{titlepage}

\section{Introduction}
\label{sec:introduction}   

The dark matter comprises about eighty-five percent of the matter
in the universe. However, the Standard Model (SM) - the most successful theory in particle physics - fails to provide a dark matter candidate. Therefore, going beyond the SM (BSM) appears to be inevitable. In the context of particle physics, dark matter can be made of one or more new particles (see e.g.~\cite{Bertone:2004pz,Feng:2010gw} for extensive reviews). Dark matter particles must be stable on cosmological time scales and expected to be uncolored, electrically neutral, and weakly interacting.  Symmetries beyond the SM are typically required to stabilize a dark matter candidate. In the standard thermal freeze-out scenario, dark matter particles were in thermal equilibrium in the early universe, and later on they annihilated into particles of the visible sector. Today, in the universe, we observe the relics, namely the leftover dark matter abundance that has survived the annihilations as of now. The relic dark matter energy density is a precisely measured cosmological quantity, and the Planck collaboration provides $\Omega_{\textrm{DM}} h^2=0.120\pm 0.001$~\cite{Planck:2018vyg}.     

There have been a plethora of particle physics models explaining the origin of dark matter. Depending on the model details, the dark matter may or may not couple directly to the visible sector. Due to increasingly stringent experimental constraints \cite{Arcadi:2017kky}, there has been renovated interest in dark matter candidates that are very weakly coupled to the SM sector. This can be realized in various ways, which include feebly interacting massive particles (FIMP) \cite{McDonald:2001vt,Hall:2009bx,Bernal:2017kxu} or gravitational dark matter \cite{Donoghue:1994dn,Choi:1994ax,Holstein:2006bh,Garny:2015sjg,Mambrini:2021zpp,Barman:2021ugy,Fong:2022cmq} (in the latter case the only interaction between the hidden and visible sector is mediated by gravity).

In this work, we consider a class of models where the dark matter does not have any direct interaction with the SM sector. As a result, the dark matter pair annihilation cross-section to the visible sector is typically small. 
In order to achieve the correct dark matter relic abundance, an additional dark partner with a large coannihilation cross-section is then often required. Such a large cross section is easily attainable if the coannihilation partner carries SM charges, and then has sizable couplings to the SM particles.   For a scenario of this type,  three different sectors are needed:  (i) the visible sector, (ii) the dark sector, and (ii) the mediator sector. In our framework, the dark sector consists of a Majorana or a Dirac fermion dark matter, $\chi$, which is a singlet under the SM. In addition, the dark sector contains a coannihilation partner, which is a dark Dirac fermion, $\psi$, that transforms non-trivially under the SM. As for the mediator sector, that couples to both the SM and dark sectors, we assume it is made of a scalar particle, $\phi$. In particular, we focus on colored mediators and coannihilation partners with non-zero hypercharge, even though they can also be charged under the weak isospin.

In this setup, the dark matter elastic scattering with the SM particles take place only via loops, 
hence the non-observation of dark matter signals in the direct detection experiments can be naturally explained while retaining a rich phenomenology at collider facilities. The presence of a colored scalar $\phi$, in particular a leptoquark option, may also have an interplay with the anomalous magnetic moment of the muon. To stabilize the dark matter, we impose a $\mathcal Z_N$ symmetry under which all dark sector particles, namely, $\chi$ and $\psi$ are charged.

Since the mediator, as well as the coannihilation partner, carry color charges, a standard estimation of the dark matter relic abundance in the coannihilation scenario is not accurate.  More specifically, a precise calculation involving non-perturbative effects, namely the Sommerfeld factors \cite{Sommerfeld,Hisano:2006nn,Cirelli:2007xd} and bound-state formation \cite{Feng:2009mn,vonHarling:2014kha}, must be considered for the pair annihilations of the colored coannihilation partners (we refer to them as \emph{non-perturbative} or \emph{near-threshold} effects along the paper). The two effects are the manifestation of multiple soft exchanges of QCD gluons and, therefore, they should be both included in the relevant cross sections. Recent investigations~\cite{Hisano:2006nn,Cirelli:2007xd,Cirelli:2008id,Feng:2009mn,Cirelli:2009uv,Feng:2010zp,deSimone:2014pda,Beneke:2014gja,Beneke:2014hja,Ibarra:2015nca,Ellis:2014ipa,Liew:2016hqo,Mitridate:2017izz,Garny:2021qsr,Harz:2018csl,Biondini:2018ovz,Biondini:2019int,Becker:2022iso} have shown that the mass benchmarks that give DM energy densities
compatible with cosmological observations are rather different from the case with no threshold effects. Accordingly one finds important changes of the model parameter space that is compatible with the observed DM energy density. 

In this work, our particular focus is to precisely extract the dark matter relic abundance by accounting for the nature of the colored coannihilation partners and mediators of our framework. To this aim, we consider a particular exemplary model where the DM is accompanied by a color-carrying coannihilation partner and a scalar leptoquark, which has been formerly considered in the literature \cite{Baker:2015qna} (see also~\cite{Belanger:2021smw,Manzari:2022iyn,Carpenter:2022lhj}). The importance of coannihilations of the colored coannihilation partner has been highlighted in order to obtain the correct DM energy density.  With respect to earlier studies, we include the Sommerfeld factors, as well as the bound-state formation and decay, for the coannihilating partner, namely, the dark vector-like fermion (a Dirac fermion that carries a dark charge and is vector-like under the SM gauge group). The nature of the dark matter fermion dictates the relevant pair annihilations of the dark vector-like fermion. For the Majorana DM option, particle-particle ($\psi \psi$) annihilations are possible in addition to particle-antiparticle ($\psi \bar{\psi}$) annihilations. Vector-like particle-antiparticle pairs combine either in color-singlet or color-octet states, whereas $\psi \psi$ pairs organize in color antitriplets and color sextets. Color-singlet and antitriplet configurations feature an attractive potential and can sustain bound states. We compute the corresponding bound-state formation cross section and bound-state decays in the framework of non-relativistic effective field theories (NREFTs), and include their effect in the numerical extraction of the DM energy density.

The paper is organized as follows. In Sec.~\ref{sec:model_Setup}, we introduce the framework for our class of models. In Sec.~\ref{sec:DM_relic_density}, the dark matter relic abundance is computed by taking into account non-perturbative effects. Dark matter direct and indirect detection prospects, and collider constraints on the model parameters, as well as correlated phenomenologies, are summarized in Sec.~\ref{sec:pheno}. Finally, conclusions and outlook are given in Sec.~\ref{sec:conclusions}, whereas supplementing material is collected in the appendices.

\section{Model Setup}\label{sec:model_Setup}
In this section, we discuss the construction of the models' class and specify the relevant interactions of the dark sector with the visible sector. We first discuss the general framework and then focus on an exemplary model. 
\subsection{General framework}
\label{sec:general_framework}
In this work, we consider a scenario where the dark matter, $\chi$, is a gauge singlet   under the SM  group. The dark matter talks to the SM sector via a colored scalar mediator, $\phi$, and  a colored fermion, $\psi$, which is the coannihilation partner. To cancel the gauge anomalies, we consider the fermion $\psi$ to be vector-like under the SM and refer to it as a dark vector-like fermion (DVLF), since it belongs to the dark sector. To form a gauge as well as Lorentz invariant interaction of the mediator with the dark sector, the BSM scalar and the fermion must carry the same quantum numbers under the SM group.   

In order to deplete the dark matter abundance through pair annihilations into SM particles via tree-level interactions,  the colored particle $\phi$ can be (i) a scalar leptoquark (LQ) or (ii) a scalar di-quark (DQ).  There are five  possible representations of scalar leptoquarks that can couple to SM quark-lepton bilinears~\cite{Buchmuller:1986zs}, namely, electroweak (i-a) singlets $S_1(\overline 3,1,1/3)~ [QL,u_R\ell_R]$, $\widetilde S_1(\overline 3,1,4/3)~ [d_R\ell_R]$, (i-b) doublets $R_2(3,2,7/6)~ [\overline u_RL, \overline Q \ell_R]$, $\widetilde R_2(3,2,1/6)~ [\overline d_R L]$, and (i-c) triplet $S_3(\overline 3,3,1/3)~ [QL]$. In addition to interaction with a lepton and a quark, three of the scalar LQs can also have di-quark interactions: $S_1~ [QQ, u_Rd_R]$, $\widetilde S_1~ [u_Ru_R]$, and $S_3~ [QQ]$. Here, $Q\sim (3,2,1/6)$, $L\sim (1,2,-1/2)$ are the weak-doublet fermions and   $u_R\sim (3,1,2/3)$, $d_R\sim (3,1,-1/3)$,  $\ell_R\sim (1,1,-1)$ are weak-singlet fermions of the SM.  It is customary to assign a baryon number to these LQs such that their di-quark interactions are absent; hence baryon number violating processes cannot take place. As for the (pure) di-quarks that couple to SM fermion bilinears, the possibilities are (ii-a) a color triplet $\bar S_1(\overline 3,1,-2/3)~ [d_Rd_R]$, (ii-b) color sextets $\omega_6(6,1,1/3)~ [QQ, u_Rd_R]$, $\omega_6^\prime(6,3,1/3)~ [QQ]$, $\omega_6^{\prime\prime}(6,1,2/3)~ [d_Rd_R]$, $\omega^{\prime\prime\prime}_6(6,1,4/3)~ [u_Ru_R]$, and finally (ii-c) a color octet  
$S_8(8,2,1/2)$ $[\overline Q u_R, \overline Q d_R]$. Even though the fermion $\psi$ carries a dark charge, it has direct interactions with the SM through gauge bosons. All these tree-level interactions, including the SM gauge interactions, are summarized in table~\ref{tab:int}.

\begin{table}[th!]
\begin{center}
\begin{tabular}{|c|c|c|c|c|}\hline
particle & type & dark charge & interaction with SM& dark interaction  \\\hline\hline
$\chi$&F&$\cmark$&$\xmark$ & \multirow{4}{1cm}{~~$\overline \psi \phi \chi$}\\ \cline{1-4}
$\psi$&F&$\cmark$&$\overline \psi \psi V$& \\ \cline{1-4} 
$\phi$&S& $\xmark$& $\phi q\ell,\; \overline\phi\phi V,\; \overline\phi\phi VV$~ (LQ)& \\ \cline{4-4}
  & & & $\phi q q,\; \overline\phi\phi V,\; \overline\phi\phi VV$~ (DQ)& \\ \cline{1-5} 
\end{tabular}
\end{center}
\caption{Tree-level interactions of the dark matter $\chi$, the DVLF $\psi$ carrying a dark charge, and the scalar $\phi$. Here, $V$ represents a SM gauge boson. }
\label{tab:int}
\end{table}

\begin{figure}[t!]
\begin{center}
  \includegraphics[width=0.9\textwidth]{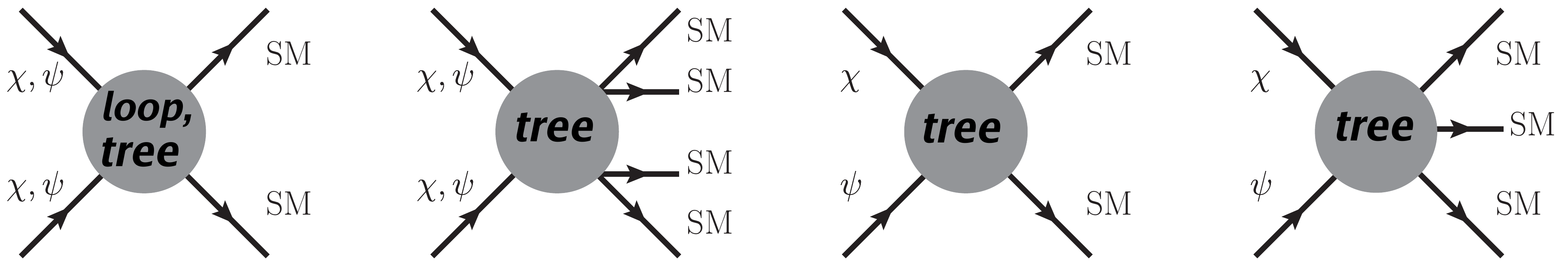}  
\end{center}
\caption{Annihilation and coannihilation channels of the dark sector particles (from left to right). Dark matter annihilation $\chi\chi\to$ SM SM ($\chi\chi\to$ SM SM SM SM) takes place only (already) at the loop-level (tree-level). For the coannihilation partner, both the $\psi\psi\to$ SM SM and $\psi\psi\to$ SM SM SM SM occur at the tree-level. Moreover, coannihilation channels $\chi\psi$ into visible sector particles lead to SM SM and SM SM  SM final states at the tree-level. Processes relevant for dark matter direct detection (for example, from bottom to top in the leftmost diagram for $\chi$) only happen at loop-level.  }
\label{fig:SCH}
\end{figure}

In table~\ref{tab:int}, F and S stand for fermion and scalar  respectively. Moreover, a SM gauge boson is denoted by V.   Note that if the scalar $\phi$ does not couple to SM fermion bilinears, then $\chi\overline \chi$ annihilation to SM sector occurs only at the loop-level. For example, $\chi\overline \chi\to gg$ can occur at one-loop level, where $\phi$ and $\psi$ states propagate inside the loop\footnote{In this work, we do not  include loop-mediated processes for dark matter annihilation.}. However, in a general scenario, dark matter pair annihilation to the visible sector is allowed at the tree-level. Moreover, pair annihilation of the DVLF as well as dark matter-DVLF coannihilation  processes already take place at the tree level.  Annihilation and coannihilation channels of the dark sector particles are schematically presented in figure~\ref{fig:SCH}.

\subsection{An example model}
In this work, we shall explore a specific model realization of the more general framework presented in section ~\ref{sec:general_framework}. In order to assess the relevance of non-perturbative effects, and highlight possible experimental signature and constraints, we  consider $\phi$ to be a LQ that carries hypercharge $ Y_{\phi}=1/3$ and is a singlet under SU(2)$_L$. Although we scrutinize a particular model realization,  we acknowledge that each model merits an independent study on its own. Moreover, we study two scenarios where the dark matter is a (i) Majorana fermion and (ii) Dirac fermion. For the former case, in order to achieve the stability of the dark matter, we assign a $\mathcal{Z}_2$ odd charge to only $\chi$ and $\psi$, whereas the rest of the particles are even. The full quantum numbers of the BSM fields under the SM$\times \mathcal{Z}_2$ are as follows:
\begin{equation}
\chi\sim (1,1,0,-) \, , \quad  \phi\sim (\overline 3,1,1/3,+) \, , \quad \psi\sim (\overline 3,1,1/3,-) \, .
\label{charge_ass_Majorana}
\end{equation}
On the other hand, for a Dirac fermion dark matter scenario, a $\mathcal{Z}_3$ symmetry is imposed 
\begin{equation}
\chi\sim (1,1,0,\omega) \, , \quad  
\phi\sim (\overline 3,1,1/3,\omega^3) \, , \quad 
\psi\sim (\overline 3,1,1/3,\omega^2) \, ,
\label{charge_ass_Dirac}
\end{equation}
with $\omega^3=1$. The dark matter interaction to the DVLF and the mediator, which induces DM annihilations to the visible sector, is realized by the following interaction: 
\begin{align}
\mathcal{L}_\mathrm{Majorana}&\supset y\;\overline \psi \chi \phi + h.c.    =y\;\bigg\{ 
\overline\psi_L \chi_R + \overline\psi_R \left(\chi_R\right)^c 
\bigg\}\phi + h.c. \, ,
\label{int_portal_Majorana}
\end{align}
where we have used  the usual notation $\overline{\chi^c}=\chi^T\mathcal{C}$, where $\mathcal{C}$ is the particle-antiparticle conjugation matrix.  And for the Dirac case, we have,
\begin{align}
\mathcal{L}_\mathrm{Dirac}&\supset y\;\overline \psi \chi \phi + h.c.   =y\;\bigg\{ 
\overline\psi_L \chi_R + \overline\psi_R \chi_L 
\bigg\}\phi + h.c. \, .
\label{int_portal_Dirac}
\end{align}
The coupling $y$ can be made real in both the cases by a field redefinition.

The LQ is neutral under $\mathcal{Z}_N$ and has direct interactions with the SM fermions~\cite{Buchmuller:1986zs},
\begin{align}
\mathcal{L} &= y_{ij}^L\; \overline{Q^c}_i i\sigma_2 \phi L_j +y_{ij}^R\; \overline{u^c_R}_i  \phi \ell_{R j} 
+ \text{h.c.}  \, .
\label{LagS1}
\end{align}
Here, $i,j=e,\mu,\tau$ are the family indices.   We assume baryon number conservation by assigning the LQ and the DVLF the baryonic charge $B=-1/3$ under the global $U(1)_B$.   Consequently, diquark couplings of the LQ are absent, and the theory is safe from proton decay. The theory may also have a global lepton number symmetry $U(1)_L$ as in the SM. Therefore, $\phi$ can carry a negative unit of lepton number; consequently, we assign the same lepton number to $\psi$. This, however, does not lead to any phenomenological implications\footnote{We do not attempt to explain the origin of neutrino mass in this work.}.

The LQ couplings $y_{ij}^L$ and $y_{ij}^R$ to the SM fermions  are a priori free parameters. However, these couplings are constrained from the direct searches at the colliders as well as from flavor violating processes, which will be discussed later in the paper. In the next section~\ref{sec:DM_relic_density}, when discussing the dark matter relic abundance, we turn on only a single coupling from each $y^{L,R}$ and take them to be equal to one. At this point, specifying which non-zero entry is taken, however, is irrelevant. In Section~\ref{sec:pheno}, where we present phenomenological implications of the model, we explicitly specify the  texture of the relevant Yukawa couplings.

\section{Dark matter relic density}
\label{sec:DM_relic_density}
The dark matter cosmological abundance is accurately determined by measuring the CMB
anisotropies and it amounts to $\Omega_{\textrm{DM}} h^2=0.120\pm 0.001$~\cite{Planck:2018vyg}, where $h$ is the reduced Hubble constant. It stands as the main observable that any compelling dark matter model has to comply with. Upon selecting a viable mechanism
to produce dark matter particles in the early universe, one can use the observed relic density as a powerful constraint on the model parameters. 

In this work we consider thermal freeze-out \cite{Gondolo:1990dk,Griest:1990kh}. For such a mechanism to work, dark matter particles have to be kept in equilibrium at high temperatures, and the relevant processes for determining the relic density are dark-matter pair annihilations. When the temperature of
the expanding universe drops below the dark matter mass, the corresponding particle densities become Boltzmann
suppressed and the annihilations cannot keep up with the expansion of the universe.  The chemical freeze-out occurs at temperatures $T \approx m_\chi/25$, therefore, dark matter particles are \emph{non-relativistic}. 

 The presence of additional dark-sector particles during freeze-out may severely affect the relic density: one has to track the (co)annihilations of additional partners when these are close in mass
with the actual dark matter particle \cite{Griest:1990kh,Edsjo:1997bg}. For the model under consideration, the DVLF plays the role of the coannihilating state.\footnote{At variance with the leptoquark mediator $\phi$, the DVLF carries a dark charge as the dark fermion $\chi$. Hence, coannihilations of $\psi$ with $\chi$, as well as $\psi \bar{\psi}$ and $\psi \psi$ pair annihilations, deplete the abundance of the dark sector particles. There are no corresponding processes with leptoquark as coannihilating state.} The thermal freeze-out has been studied for this model in refs.~\cite{Baker:2015qna,Belanger:2021smw}, where it has been highlighted the role of coannihilations in order not to overclose the universe and relative mass splittings as small as $\Delta m  \equiv (m_\psi-m_\chi) /m_\chi \approx 10^{-3}$ are considered (see also \cite{Manzari:2022iyn} for the extreme case $\Delta m =0$). The impact of coannihilating processes depend strongly on (i) the mass splitting between the DM particle ($\chi$) and the coannihilating species ($\psi$); (ii) conversion rates between the dark matter and the coannihilating partner that put them in thermal contact. For the portal coupling $y$, we consider a range $y \in \left[ 0.01, 2\right]$, which well ensures fast conversion rates \cite{Garny:2017rxs,DAgnolo:2017dbv}. 
   A complementary dark-matter production mechanism for the model under study, namely the conversion-driven freeze-out (or co-scattering) \cite{Garny:2017rxs,DAgnolo:2017dbv}, has been addressed in ref.~\cite{Belanger:2021smw}, where much smaller Yukawa couplings $10^{-8} \lesssim y \lesssim 10^{-4}$ are considered. For even smaller portal couplings, the freeze-in mechanism is the viable option \cite{McDonald:2001vt,Hall:2009bx}, and we leave it for future work on the subject. 

In summary, in the coannihilation regime, non-relativistic DVLFs pair annihilation also contribute to the depletion of the dark matter. At variance with the dark matter particle, DVLFs feel QCD strong as well as electroweak interactions. Pair annihilations of slowly moving charged particles get affected by long-range interactions
mediated by soft gauge-boson exchange, that induce near-threshold effects, most notably Sommerfeld and bound-state effects \cite{Sommerfeld,Hisano:2006nn,Feng:2009mn,vonHarling:2014kha}. Due to the observed hierarchy between the corresponding SM gauge couplings, strong interactions are largely dominant, and we focus on them in this work.\footnote{We checked that the additional Sommerfeld and bound-state effects as originated from the photon and the $Z$ boson give few-per-cent corrections to the estimation of the energy density, with respect to the QCD non-perturbative effects.}  
\subsection{Boltzmann equation and cross sections}
The effect of a co-annihilating partner ($\psi$) in thermal equilibrium with the actual dark matter particle ($\chi$) can be captured by a single Boltzmann equation 
\cite{Gondolo:1990dk,Griest:1990kh,Edsjo:1997bg}
\begin{equation}
\frac{dn}{dt} + 3 Hn =-\langle \sigma_{{\rm{eff}}} v_{\textrm{rel}} \rangle (n^2-n^2_{{\rm{eq}}}) \, ,
\label{BE_gen}
\end{equation}
where $H$ is the Hubble rate of the expanding universe and $n$ denotes the total number density of dark sector states $\chi$ and $\psi$. 
By assuming the dark matter being a Majorana fermion, the total equilibrium number density, which accounts for both the particle and antiparticle species of the dark sector is 
\begin{equation}
n_{{\rm{eq}}}= \int_{p}  e^{-E_{p}/T} \left[ g_\chi +  2 g_\psi \, e^{-\Delta M/T} \right] \, , 
\label{n_eq}
\end{equation}
 where $\int_{\bm{p}} \equiv \int d^3 \bm{p}/(2\pi)^3 $, $E_p=m_\chi+p^2/(2 m_\chi)$,   $p \equiv |\bm{p}|$, $g_\chi=2$ and $g_\psi=2 N$ are the particles degree of freedom, i.e.~spin polarizations and color $N=3$, and the effective thermally averaged annihilation cross section reads~\cite{Edsjo:1997bg} 
\begin{equation}
\langle \sigma_{{\rm{eff}}} v_{\textrm{rel}} \rangle = \sum_{i,j} \frac{n^{\hbox{\scriptsize eq}}_i \,  n^{\hbox{\scriptsize eq}}_j}{n_{{\rm{eq}}}^2} \langle \sigma_{ij} v_{\textrm{rel}}  \rangle \,.
\label{co_cross}
\end{equation}

\textbf{$\bm{\chi\chi}$ pair annihilation:}-- 
The cross section of the contributing processes have to be handled with care for several reasons. A first observation is about the relative importance of $\chi \chi$ pair annihilations into SM particles. This process can occur via loops into a two-body final state, $\chi \chi \to 2 \, \textrm{SM}$, or via off-shell decays of leptoquarks into a four-particle final state $\chi \chi \to \phi^* \phi^{\dagger *} \to 4 \, \textrm{SM}$. As noted in \cite{Baker:2015qna}, the first class of processes feature  typical loop-suppression factors, whereas the latter are phase-space suppressed. Only for $\Delta m /m_\chi \gtrsim 0.4$ \cite{Baker:2015qna}, the loop-induced and off-shell decays of the leptoquark is comparable with the $2 \to 2$ coannihilation channels $\chi \psi \to  2 \, \textrm{SM}$ and $\psi \bar{\psi} \to 2 \, \textrm{SM}$. In this work we restrict to smaller mass splittings, namely $\Delta m /m_\chi \lesssim 0.1$, as motivated by the need efficient coannihilation regime~\cite{Baker:2015qna,Belanger:2021smw}, and  we find that the two classess $\chi \chi \to 2 \, \textrm{SM}$ and $\chi \chi \to \phi^* \phi^{\dagger *} \to 4 \, \textrm{SM}$ are negligible with respect to $\chi \psi \to  2 \, \textrm{SM}$ and $\psi \bar{\psi} \to 2 \, \textrm{SM}$ (in agreement with the estimations of ref.~\cite{Baker:2015qna}).  

The phase-space suppression of dark matter annihilations into four-body final state is lifted for $m_\chi > m_\phi$. In this case, the leptoquark can be produced on shell and, by relying on the narrow-width approximation (NWA) \cite{Denner:2005fg,Denner:1999gp,Berdine:2007uv}, the cross section may be factorized for a given final state in the form $\sigma(\chi \chi \to  \phi \phi^{\dagger} \to 4 \, \textrm{SM}) \simeq \sigma(\chi \chi  \to \phi \phi^{\dagger}) \textrm{BR}^2_{\phi \to 2 \, \textrm{SM}}$ (here for the decay of $\phi$ and $\phi^\dagger$ in the same final state). Upon looking at the inclusive annihilation process, namely the sum of all possible four-body SM final states, the relevant cross section reduces to $\sigma(\chi \chi \to  \phi \phi^{\dagger} )$.\footnote{The inclusive cross section comprises all possible finals states, namely all decay modes of the leptoquark. Hence, one obtains the sum squared of the branching ratios that gives indeed unity.} In order to properly use the NWA, there are several conditions to be fulfilled \cite{Berdine:2007uv}. As far as we are concerned, $\Gamma_\phi/m_\phi \simeq 10^{-2} \ll 1$ for the $y_L,y_R$ values adopted in this work (cfr.~eq.~\eqref{leptoquark_width}); the daughter particles are much lighter than the leptoquark; the leptoquark propagator is separable. However, at the opening of the channel, namely $m_\chi \gtrsim m_\phi$,  we are not far from the mass threshold, which is another condition for using safely the NWA to be fulfilled.\footnote{When this condition is not fulfilled, there is cut on the line shape of the Breit Wigener. We find an agreement between the analytical approximation of the cross section and the numerical output from MadGraph \cite{Alwall:2014hca} within few-per cent in this regime.} Moreover, since annihilations happen in a thermal environment, the so-called forbidden region opens up for $m_\chi$ slightly smaller than $m_\phi$, the negative mass gap being compensated by the thermal kinetic energy of the incoming dark matter pair \cite{Griest:1990kh}. We take into account the forbidden region when computing the thermally averaged cross section $\langle \sigma_{\chi \chi} v_{\textrm{rel}} \rangle$. Both in the forbidden region and the allowed region with $(m_\chi-m_\phi)/m_\chi \lesssim 0.1$, the velocity expansion cannot be performed \cite{Griest:1990kh}, and we use the exact expression for the cross section of the processes $\chi \chi \to \phi \phi^\dagger$ and $\chi \bar{\chi} \to \phi \phi^\dagger$ (Majorana and Dirac DM option respectively). Away from the threshold region, we find that $\sigma_{\chi \chi} v_{\textrm{rel}}$ has a leading $p$-wave contribution for the Majorana case, whereas Dirac dark matter annihilation  $\sigma_{\chi \bar{\chi}} v_{\textrm{rel}}$  feature an $s$-wave leading contribution. The exact cross sections for the processes  $\chi \chi \to \phi \phi^\dagger$ and $\chi \bar{\chi} \to \phi \phi^\dagger$, as well as the velocity-expanded ones, are provided in appendix~\ref{sec:app_annihilations_channels}. 

\textbf{$\bm{\chi\psi}$  coannihilation:}-- 
Next, the coannihilation process $\chi \psi \to  q^c \bar{\ell}$ and $\chi \psi \to  Q^c \bar{L} $   proceed via an $s$-channel exchange of a leptoquark. The cross section become resonantly enhanced whenever $m_\chi + m_\psi \approx m_\phi$, which in the coannihilation scenario with small relative mass splittings gives $m_\chi, m_\psi \approx m_\phi/2$. The total decay width of the leptoquark has to be included to properly regulate the annihilation process.\footnote{Here we content ourselves with a zero-temperature analysis of the leptoquark decay width, see ref.~\cite{Laine:2022ner} for a more accurate handling of DM $s$-channel annihilation in a thermal environment.} For order-one Yukawa couplings $y$, $y_L$ and $y_R$,  $\chi \psi \to  q^c \bar{\ell}$ and $\chi \psi \to  Q^c \bar{L} $  give large cross sections in the region $m_\chi \approx m_\phi/2$, that in turn produce prominent effects in the DM energy density (see figures~\ref{fig:Majorana_Mchi_versus_OmegaDM_a} and  \ref{fig:Omega_Mpsi_Mphi_Majorana_a}). There is another coannihilation process, namely $\chi \psi \to \phi \, g \to 3 \, \textrm{SM}$, that proceed both via a $t$-channel exchange of the DVLF as well as an $s$-channel mediated by the leptoquark. The three-body final state (a gluon, a quark and a lepton) is obtained after the decay of the unstable leptoquark. Analogous arguments  as discussed for $\chi \chi$ pair annihilation on the applicability/approximations of the NWA hold.
The leptoquark decay width at leading order reads 
\begin{eqnarray}
\Gamma_\phi=\frac{|y_R|^2 m_\phi}{16 \pi}  + \frac{|y_L|^2 m_\phi}{8 \pi}  + \frac{y^2 m_\phi}{8 \pi} \left( 1-\frac{(m_\chi+m_\psi)^2}{m_\phi^2} \right)^{3/2} \sqrt{1-\frac{(m_\chi-m_\psi)^2}{m_\phi^2}}\, .
\label{leptoquark_width}
\end{eqnarray}
The first two contributions stem for the leptoquark decays into right-handed and left-handed quark and lepton pairs respectively (we treat the SM particles as massless since in this work we take $m_\phi$ of order 1 TeV). The third contribution appears only if $m_\phi > m_\chi +m_\psi$.

\begin{figure}[t!]
    \centering
    \includegraphics[scale=0.53]{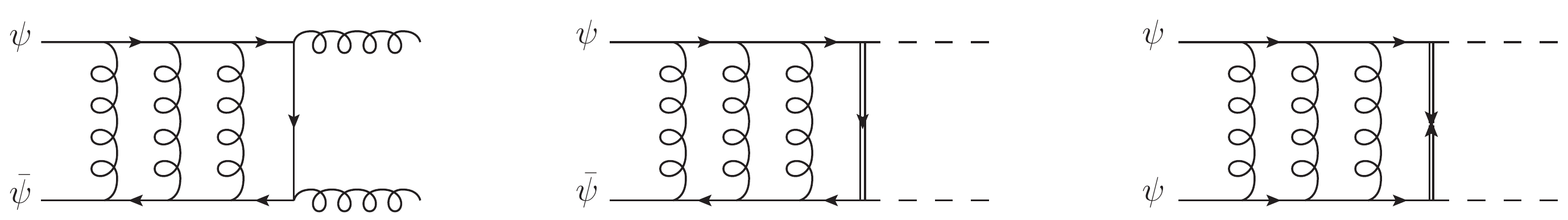}
    \caption{Left: Representative diagram for DVLFs pair annihilation into light SM states; gluons are depicted with curly lines. Middle and right: DVLFs pair annihilations into leptoquarks; leptoquarks are depicted with dashed lines, whereas the solid double arrow stand for the DM fermion $\chi$ (forward-backward arrow for $\chi \chi$ Majorana-type field contractions). Repeated gluon exchange are shown for the incoming DVLFs, both for $\psi \bar{\psi}$ and $\psi \psi$ annihilations. }
    \label{fig:ann_psi_psi_diagrams}
\end{figure}

\textbf{$\bm{\psi\overline\psi}$ and $\bm{\psi\psi}$ pair annihilations:}-- 
Finally, the pair annihilation of non-relativistic DVLFs, both $ \sigma_{\psi \bar{\psi}} v_{\textrm{rel}} $ and $ \sigma_{\psi \psi} v_{\textrm{rel}} $, can be affected by non-perturbative  effects due to repeated gluon exchange, see representative diagrams in figure~\ref{fig:ann_psi_psi_diagrams}. The latter cross section is triggered only by Yukawa interactions and for the Majorana dark matter option via the process $\psi \psi \to \phi \phi$; see rightmost diagram in figure~\ref{fig:ann_psi_psi_diagrams}. For incoming scattering states, long-range interactions induce Sommerfeld factors, which enhance (suppress) the annihilations for an attractive (repulsive) potential experienced by DVLFs pairs. Moreover, there is an additional manifestation of repeated soft gauge-boson exchange: the presence of meta-stable bound states. The bound-state formation process, and the subsequent bound-state decay, triggers an efficient way to deplete further the QCD-charged coannihilating states, and then the overall dark matter abundance \cite{Mitridate:2017izz,Harz:2018csl,Biondini:2018pwp,Biondini:2018ovz,Garny:2021qsr}. The inclusion of bound-states can be implemented in the annihilation cross section of DVLFs in \eqref{co_cross} through the following effective cross section \cite{Mitridate:2017izz,Harz:2018csl}
\begin{eqnarray}
 \langle  \sigma_{\psi \bar{\psi}} \, v_{\textrm{rel}} \rangle  =
   \langle \sigma_{\psi \bar{\psi},\textrm{ann}} v_{\textrm{rel}} \rangle
   + \sum_{n} \langle   \sigma^n_{\psi \bar{\psi},\textrm{bsf}} \, v_{\textrm{rel}} \rangle \, \frac{\Gamma_{\psi \bar{\psi}}^n}{\Gamma_{\psi \bar{\psi}}^n+\Gamma_{\psi \bar{\psi},\textrm{bsd}}^n} \, .
    \label{Cross_section_eff_psi_barpsi}
\end{eqnarray}
 The corresponding effective cross sections for DVLF particle-particle and antiparticle-antiparticle pair annihilation have the same form as eq.~\eqref{Cross_section_eff_psi_barpsi} with the subscripts in the cross sections and widths replaced by $ \psi \psi$ and $\bar{\psi} \bar{\psi}$. In eq.~\eqref{Cross_section_eff_psi_barpsi} the first term stems for the annihilation of scattering (or unbound) states, whereas the second term encodes the reprocessing of an unbound pair into a bound state. Here, the quantities that enter are the thermally averaged bound-state formation cross section $\langle \sigma^n_{\psi \bar{\psi},\textrm{bsf}} \, v_{\textrm{rel}} \rangle$, the decay width of the bound states $\Gamma_{\psi \bar{\psi}}^n$, and the bound-state dissociation $\Gamma_{\psi \bar{\psi},\textrm{bsd}}^n$. The combination of the bound-state decay and dissociation widths $\Gamma_{\psi \bar{\psi}}^n/(\Gamma_{\psi \bar{\psi}}^n+\Gamma_{\psi \bar{\psi},\textrm{bsd}}^n)$  takes into account the ionization of a given bound state in the thermal environment, and dictates how efficiently the bound-state formation contribute to the depletion of colored pairs. Upon the inclusions of bound-to-bound transitions, eq.~\eqref{Cross_section_eff_psi_barpsi} has to be modified \cite{Garny:2021qsr,Binder:2021vfo} and bound-state effects become even more relevant.
As for the DVLF annihilations into leptoquark pairs, which mediate the subsequent decays into SM four-body states, we find that $\psi \bar{\psi} \to 4 \, \textrm{SM}$ and $\psi \psi \to 4 \, \textrm{SM}$ are phase-space suppressed and practically negligible for $m_\psi < m_\phi$ with respect to $\psi \bar{\psi} \to 2\,  \textrm{SM}$ (for the DVLFs the $2 \to 2$ annihilations occur via QCD interactions without the need of the $\phi$ mediator, see e.g.~leftmost diagram in figure~\ref{fig:ann_psi_psi_diagrams}) . At the opening of the on-shell region for the leptoquarks, and in full analogy with $\chi \chi$ annihilations, the suppression is lifted and we approximate the inclusive cross section with  $\sigma(\psi \bar{\psi} \to  \phi \phi^{\dagger} \to 4 \, \textrm{SM}) \simeq \sigma(\psi \bar{\psi}  \to \phi \phi^{\dagger})$ and $\sigma(\psi \psi \to  \phi \phi \to 4 \, \textrm{SM}) \simeq \sigma(\psi \psi  \to \phi \phi)$. 

We have checked our analytical expressions for the  cross sections of $2 \to 2$ annihilation processes $\sigma_{ij}$ that enter eq.~\eqref{co_cross}  with the model implementation in MadGraph~\cite{Alwall:2014hca}, and the corresponding relic density with micrOmegas \cite{Belanger:2006is}. However, in order to go beyond the free annihilations and include non-perturbative effects, an estimation of Sommerfeld and bound-state effects for colored DVLFs is needed. This is the subject of the following sections.

\subsection{Near-threshold effects in NREFTs}
The dark matter is a SM gauge singlet, hence the free cross section accurately account for the corresponding pair annihilation. Conversely, annihilating DVLF pairs, either $\psi \bar{\psi}$,  $\psi \psi$ and $\bar{\psi} \bar{\psi}$ annihilations, are affected by soft gluon exchanges (see figure~\ref{fig:ann_psi_psi_diagrams} for exemplary diagrams). In the following, we assemble existing results, and obtain new ones, in order to 
compute relevant cross sections and decay widths of non-relativistic DVLF pairs in the early universe thermal environment. 
We exploit the hierarchy of energy scales that is typical for non-relativistic particles moving with relative velocity $v_\textrm{rel}$, namely  $m_\psi \gg m_\psi v_\textrm{rel} \gg m_\psi v_\textrm{rel}^2$,
by replacing the fundamental DM theory with a sequence of non-relativistic effective field theories (NREFTs). For Coulombic bound states the relative velocity is fixed by the virial theorem as $v_{\textrm{rel}} \sim \alpha_s$, hence the corresponding hierchy of scales is $m_\psi \gg m_\psi \alpha_s \gg m_\psi \alpha_s^2$. In particular, we shall exploit the framework of NRQCD \cite{Caswell:1985ui,Bodwin:1994jh} and  potential NRQCD (pNRQCD) \cite{Pineda:1997bj,Brambilla:1999xf}, since the DVLF well qualifies as a heavy quark from the QCD perspective.   The original formulation of such EFTs were conceived for heavy quark and antiquark pairs, then leading to color-singlet and color octet states. In our work, it will be relevant to address DVLF particle-particle pairs as well. We shall rely on the corresponding NRQCD and pNRQCD for two heavy colored particles (or antiparticles) as detailed in ref.~\cite{Brambilla:2005yk}. 

pNREFTs are useful for our scope since they stand for the quantum-field theories of non-relativistic interacting pairs, both for scattering or bound states, and allow to systematically describe pair annihilations and pair-to-pair transitions. Since the relevant processes occur in the early universe, we exploit the formulation of pNRQCD at finite temperature~\cite{Brambilla:2008cx,Escobedo:2008sy,Escobedo:2010tu}. There has been a recent effort in transferring and adapting the NREFTs for dark matter freeze-out \cite{Beneke:2014gja,Beneke:2014hja,Kim:2016kxt,Biondini:2018pwp,Biondini:2018ovz,Biondini:2019int,Binder:2020efn,Binder:2018znk,Garny:2021qsr}. A detailed inspection of the interplay with thermal scales in the construction of NREFTs relevant for dark matter annihilations has been recently carried out in ref.~\cite{Biondini:2023zcz}. In this work, we restrict to the bound-state formation process as induced by the radiative emission of a gluon (see Sec.~\ref{sec:bound_state_effects}).  

Moreover, we discuss the applicability of NREFTs when the dark sector particles annihilate into final states with comparable masses (in the model at hand this means DVLF annihilations into leptoquark pairs). The latter situation invalidates the velocity expansion, and hence, some care is needed when one aims to include non-perturbative effects close to the opening of a mass threshold. 
\subsubsection{Sommerfeld factors for pair annihilations}
\label{sec:Sommerfeld_effects}
\begin{figure}
    \centering
    \includegraphics[scale=0.56]{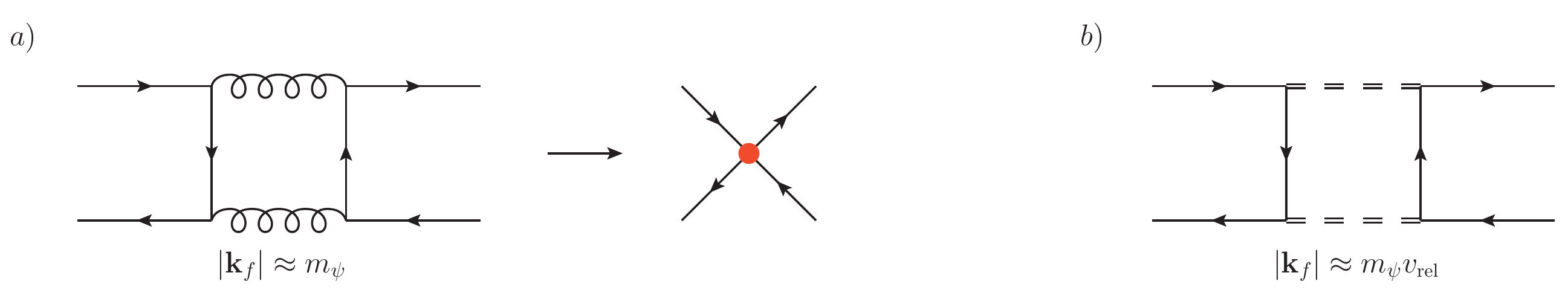}
    \caption{a) A representative diagram  for the matching procedure that leads to four-fermion local operators in NRQCD. The imaginary part of the loop diagrams encode the annihiatlions of DVLFs into gluons with typical momenta $|\bm{k}_f| \approx m_\psi$. b) The box diagram captures the annihilations of DVLF into massive leptoquarks. For small mass splittings, the typical momenta of the final-state leptoquarks is a small scale and comparable with the non-relativistic velocities of the incoming states. }
\label{fig:Local_vs_non_local}
\end{figure}

The annihilation process of fermion-antifermion pairs is encoded in the imaginary part of the matching coefficients of four-fermion operators, that are organised according to spin and color representations~\cite{Bodwin:1994jh}. In pNRQCD, this translates into an imaginary local potential for the pairs, which is inherited from four-fermion operators of NRQCD \cite{Pineda:1997bj,Brambilla:1999xf,Brambilla:2004jw}. Whenever we consider DVLFs directly annihilating into light Standard Model particles, namely quarks and gluons, the large mass gap between initial and final states makes the  wavelength of final-state particles of order $1/m_\psi$ (and the energy scale for such annihilation being of order $m_\psi$). This scale is  much smaller than the corresponding wavelength of incoming non-relativistic  DVLF states, i.e.~$1/m_\psi v_{\textrm{rel}}$. For such a reason heavy-pair annihilations are well described by local interactions, i.e.~the four-fermion effective operators of NRQCD (see figure \ref{fig:Local_vs_non_local}), and the corresponding imaginary local potential in pNRQCD. 

The factorization of hard modes and soft scales is a built-in feature  of NREFTs \cite{Bodwin:1994jh,Pineda:1997bj,Brambilla:1999xf}. 
Soft gluon exchanges, which correspond to energy modes of order $m_\psi \alpha_s$, are encoded in the real part of the potentials of color-singlet and color-octet pairs, which at leading order read\footnote{With some abuse of notation, and as a usual practice in NRQCD and pNRQCD literature, we express potentials, Sommerfeld factors, cross sections and decay widths for a generic SU(N) group, despite we name the relevant representations for the specific case $N=3$.} 
\begin{equation}
    V^{(0)}_{[1]} = -C_{F}\frac{\alpha_s}{r} \,, \qquad\qquad V^{(0)}_{[8]} = \frac{\alpha_s}{2 N r} \, ,
\label{QCD_psi_psibar_pot}
\end{equation}
where $C_F=(N^2-1)/(2N)$.
The fermion-antifermion wavefunctions in pNRQCD, which at leading order in the multipole expansion are the solution of the Schr\"odinger equation
with the potentials in eq.~\eqref{QCD_psi_psibar_pot}, 
accounts by construction for the effect of multiple soft gluon rescattering~ \cite{Pineda:1997bj,Brambilla:1999xf}. Then, by combining the known results for the matching coefficients of  heavy quark-antiquark annihilations \cite{Bodwin:1994jh}, we can obtain the annihilation cross section for $\psi \bar{\psi}$ pairs into Standard Model QCD states, namely gluons and quarks, the latter counted by the number of flavors $n_f$. The main advantage over exploiting the NRQCD framework is a transparent organization of the contributing partial waves, color and spin states to the DVLFs annihilation, which makes manifest the corresponding Sommerfeld factors. In the following, we provide the analytical expressions of the Sommerfeld corrected cross sections at leading order in the velocity expansion. This corresponds to the inclusion of the leading dimension-six operators of NRQCD. 

The color and spin-averaged cross section that accounts for $\psi \bar{\psi}$ annihilation into SM gluons and quarks, $(\sigma_{\psi \bar{\psi}} \, v_{\textrm{rel}})_{gg} + (\sigma_{\psi \bar{\psi}} \, v_{\textrm{rel}})_{q \bar{q}} \equiv (\sigma_{\psi \bar{\psi}} \, v_{\textrm{rel}})_{\hbox{\tiny QCD}}$,  reads
\begin{eqnarray}
     &&(\sigma_{\psi \bar{\psi}} \, v_{\textrm{rel}})_{\hbox{\tiny QCD}} = \frac{C_{\textrm{F}} \pi \alpha_s(\mu_h)^2}{2 N^2 m_\psi^2} \mathcal{S}_{0,[1]}(\zeta) + \frac{C_{\textrm{F}} \pi \alpha_s(\mu_h)^2 }{2 N m_\psi^2}\left( \frac{ (N^2-4)}{2 N} + n_f \right)  \mathcal{S}_{0,[8]}(\zeta) 
     \label{cross_psi_psibar_QCD}\, ,
\end{eqnarray}
where the strong coupling constant that appears in the NRQCD matching coefficients is evaluated at the hard annihilation scale $\mu_h \equiv 2 m_\psi$. 
The Sommerfeld factors $\mathcal{S}_{0,[1]}(\zeta)=|\Psi^{[1]}_{\bm{p}}(\boldsymbol{0})|^2$ and $\mathcal{S}_{0,[8]}(\zeta)=|\Psi^{[8]}_{\bm{p}}(\boldsymbol{0})|^2$ correspond to the squared wave function of the color-singlet and octet pairs evaluated at the origin, because of the local nature of the annihilation process into light states. The Sommerfeld factors encode the soft contribution to pair annihilations. The color-singlet and color-octet Sommerfeld factors,  with orbital angular momentum $\ell=0$, read
\begin{eqnarray}
     \mathcal{S}_{0,[1]}(\zeta) =  \frac{2 \pi C_F\zeta}{1-e^{-2 \pi C_F\zeta}} \, , \quad  \mathcal{S}_{0,[8]}(\zeta) = \frac{\pi \zeta/N}{e^{\pi \zeta/N}-1} \, , 
     \label{Som_1_and_8}
\end{eqnarray}
where $\zeta=\alpha_s(\mu_s)/v_{\textrm{rel}}$. Here the strong coupling constant is evaluated at soft scale, $\mu_{\textrm{s}} \equiv m_\psi \alpha_s$, namely the energy/momentum scale typical of soft-gluon exchanges.\footnote{We evolve the strong coupling constant at one loop with the additional colored states $\phi$ and $\psi$, a scalar and a fermion respectively, as follows $\partial_t g^2_s= \frac{g^4_s}{(4\pi)^2} \left( \frac{4N_G}{3}+\frac{2N_F}{3}+\frac{N_S}{6}-\frac{11N}{3} \right)$, where $t =\ln \bar{\mu}^2$ parameterizes the $\overline{\textrm{MS}}$ renormalisation scale, $N_G=3$ indicates the SM generations, $N_F=1$ and $N_S=1$ stand for the BSM colored fields, and $N=3$ for the specific case of QCD.}
In eq.~\eqref{cross_psi_psibar_QCD}, DVLF pairs in a color-singlet annihilate into gluons only, whereas color-octet pairs can annihilate into both gluons and quarks, as one may see from the appearance of $n_f$.  

Besides direct annihilations into light two-body SM states, there are two additional processes which are driven by the Yukawa-portal and QCD interactions, namely $\psi \bar{\psi} \to \phi \phi^\dagger$ and $\psi \psi \to \phi \phi$ and $\bar{\psi} \bar{\psi} \to \phi^\dagger \phi^\dagger$, that mediate the DVLFs annihilation into a four-body final state via on-shell decays of the leptoquark pairs. For these processes, the velocity expansion breaks down when the DVLF and leptoquark masses are nearly degenerate (this has been highlighted in the context of dark matter freeze-out \cite{Griest:1990kh}). This is because, for small mass splittings $\Delta m_{\psi \phi} \equiv m_\psi-m_\phi$, higher powers of non-relativistic velocity of the incoming DVLFs become of comparable size with $\Delta m_{\psi \phi}$, and it is not sufficient to retain the leading term in the velocity expansion of the cross section.\footnote{In the center of mass of the collisions, the final state particle momentum can be written as follows $|\bm{k}_\phi| = \sqrt{s-4 m_{\phi}^2}/2 = \sqrt{m_\psi^2/(1-v^2_{\textrm{rel}}/4)-m^2_\phi}$. Upon expressing the denominator with the relative velocity as a geometric series and using the definition of the mass splitting $\Delta m_{\psi \phi}= m_\psi -m_\phi$, one obtains for the final-state momentum $|\bm{k}_\phi|= m_\psi \sqrt{\sum_{n=1}^{\infty}(\frac{v^2}{4})^n +  \frac{\Delta m_{\psi \phi}}{m_\psi}\left( 2 - \frac{\Delta m_{\psi \phi}}{m_\psi}\right)}$, which is smaller than the hard energy/momentum scale $m_\psi$.}  In order to recast this situation in the language of effective field theories, let us take the corresponding annihilation diagram into leptoquark pairs, see figure~\ref{fig:Local_vs_non_local} (right, diagram $b$).  The typical momentum of the massive leptoquarks is parametrically of order  $ |\bm{k}_\phi| \approx m_\psi v_{\textrm{rel}} , m_\psi  \sqrt{\Delta m_{\psi \phi}/m_\psi}$, which qualifies as a small energy scale with respect to $m_\psi$. Hence,  the annihilations of DVLF into leptoquarks with slightly smaller masses cannot be described by local annihilations, or equivalently, by local four-fermion operators of NRQCD. The incoming DVLF particles do not have to come very close to annihilate, since the wavelength of the final-state particles is comparable with that of the incoming non-relativistic states. As a result, Sommerfeld and bound-state effects on pair annihilations in this regime, are expected to be less relevant. Lacking of a quantitative assessment for this situation, we do not include any of them at the opening of the mass threshold.\footnote{A qualitative estimate of the Sommerfeld and bound-state effects can be inferred by taking the Coulombic wave functions evaluated at a typical soft scale of order $m_\psi \alpha_s$ or $m_\psi v_{\textrm{rel}}$, rather than at the origin. S-wave bound-state decay widths would be then suppressed by a factor $\sim 1/e^2$. As for the scattering state wave function, the modulus squared of the corresponding hypergeometric function would enter. We checked that, for the attractive channels $|\Psi_p^{[1]}(\bm{r})|^2$ and $|\Psi_p^{[3]}(\bm{r})|^2$, a reduction of an  order of magnitude is found with respect to their value at the origin when one insert $r = 1/ m_\psi \alpha_s$ instead.} 

In practice, we assess the convergence of velocity-expanded cross sections to the exact cross sections as a function of the mass ratio $m_\phi/m_\psi$. We have checked that for $m_\phi/m_\psi \leq 0.8$ the velocity expansion can be used,\footnote{Our finding compares well with the original statements in ref.~\cite{Griest:1990kh}, which considered the velocity expansion to be valid for mass rations of 0.85-0.9. Here we adopt a slightly more conservative condition.} and we include the Sommerfeld effect accordingly. This is also the range that makes the NWA performing rather good because we are sufficiently away from the leptoquark pair threshold, hence, the cross section of the process $\psi \bar{\psi} \to \phi \phi^\dagger$ describes well the annihilations into all possible four-body SM states. 
At leading order in the velocity expansion, we find that $\psi \bar{\psi} \to \phi \phi^\dagger$ contribute to the dimension-six spin-triplet color-singlet and spin-triplet color-octet operators of NRQCD. The corresponding spin- and color-averaged cross section reads
\begin{eqnarray}
    &&(\sigma_{\psi \bar{\psi}} \, v_{\textrm{rel}})_{\phi \phi^\dagger} = \frac{y^4 }{ 16 \, \pi N^2  } \frac{m_\psi^2}{(m_\psi^2+m_\chi^2-m_\phi^2)^2} \left( 1-\frac{m_\phi^2}{m_\psi^2} \right)^{\frac{3}{2}} S_{0,[1]}(\zeta) 
     \nonumber
     \\
     &&\hspace{0.3 cm}+  \frac{C_{\textrm{F}}}{N} \left[  \frac{\pi \alpha_s(\mu_h)^2}{4 m_\psi^2}   -  \frac{y^2 \alpha_s(\mu_h)}{8 (m_\psi^2+m_\chi^2-m_\phi^2)} + \frac{y^4 m_\psi^2}{8 \pi (m_\psi^2+m_\chi^2-m_\phi^2)^2}\right]  \left( 1-\frac{m_\phi^2}{m_\psi^2} \right)^{\frac{3}{2}} S_{0,[8]}(\zeta)\, .
     \nonumber
     \\
     \label{psi_psibar_leptoquarks}
\end{eqnarray}

For Majorana fermion dark matter there is additional annihilation channel for DVLF pairs, namely $\psi \psi \to \phi\phi$ and the complex conjugate process. In our assignation of the quantum numbers, $\bar{\psi}$ has the same SU(3)-color charge of a SM quark, cfr~eq.~\eqref{charge_ass_Majorana}. Hence, annihilating DVLF antiparticle-antiparticle pairs organize either in a color antitriplet or color sextet state, $\bm{3} \otimes \bm{3} = \bar{\bm{3}} \oplus \bm{6}$, whereas particle-particle pairs into the corresponding conjugate representations. In the following we simply denote the representations of the $\psi \psi$ and $\bar{\psi} \bar{\psi}$ pairs with $[3]$ and $[6]$, since a color antitriplet (sextet) has the same symmetry property of a color triplet (antisextet). The corresponding NRQCD and pNRQCD for identical fermions can be found in ref.~\cite{Brambilla:2005yk}. A color triplet pair feels an attractive potential, whereas the color sextet a repulsive one. The leading order potentials read
\begin{eqnarray}
    V_{[3]}^{(0)} =- \frac{N+1}{2 N} \frac{\alpha_s}{r} \, , \quad  V_{[6]}^{(0)} = \frac{N-1}{2 N} \frac{\alpha_s}{r}
\end{eqnarray}
 The spin- and colored averaged cross section is\footnote{The color average amounts to $N(N-1)/2+N(N+1)/2=N^2$, where the dimension of the antisymmetric $[3]$ and symmetric $[6]$ color representations are summed. It gives the same result as for the sum of color singlet and color octet representations.}
\begin{eqnarray}
   && \sigma_{\psi \psi} \, v_{\textrm{rel}}=\sigma_{\bar{\psi} \bar{\psi}} \, v_{\textrm{rel}}= \frac{y^4 (N-1)}{ 16 \, \pi \, N  }  \frac{m_\psi^2}{(m_\psi^2+m_\chi^2-m_\phi^2)^2} \left( 1-\frac{m_\phi^2}{m_\psi^2} \right)^{3/2} S_{0,[3]}(\zeta)  \, ,
   \label{psi_psi_leptoquarks}
\label{QCD_psi_psi_exp}
\end{eqnarray}
and the corresponding attractive Coulombic Sommerfeld factor reads 
\begin{equation}
     S_{0,[3]}(\zeta) =   \frac{2 \pi C_a\zeta}{1-e^{-2 \pi C_a\zeta}} \, ,
\end{equation}
with $C_a \equiv (N+1)/2N$. We notice that, at variance with the annihilation processes $\psi \bar{\psi} \to gg$, $\psi \bar{\psi} \to q \bar{q}$ and $\psi \bar{\psi} \to \phi \phi^\dagger$, only the attractive color-antitriplet  channel contribute at leading order in the velocity expansion. In agreement with general arguments on the symmetry of identical particle annihilations as given e.g.~in ref.~\cite{ElHedri:2016onc}, we find that the velocity-independent cross section \eqref{psi_psi_leptoquarks} corresponds to DVLF pair in a spin triplet. 
\begin{figure}[t!]
    \centering
    \includegraphics[scale=0.7]{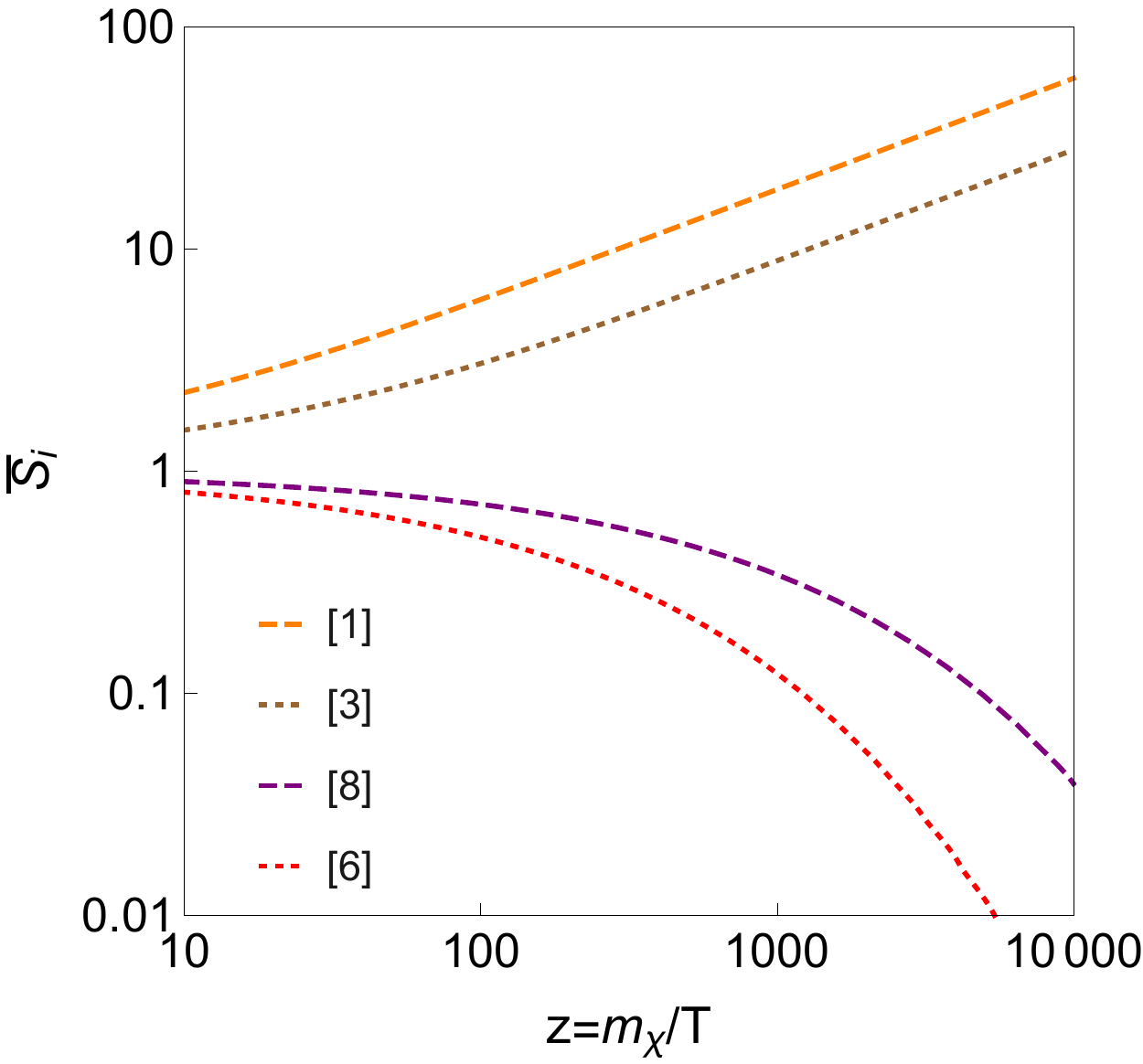}
    \caption{Thermally averaged Sommerfeld factors as a function of the time variable $z=m_\chi/T$ for different color configurations of DVLF pairs. The relative mass splitting is $\Delta m /m_\chi=10^{-2}$.}
    \label{fig:SOM_summary}
\end{figure}

In figure~\ref{fig:SOM_summary}, we show the thermally average Sommerfeld factors for the different color representations. The thermal average is performed in the standard way, see e.g.~\cite{vonHarling:2014kha}, that amounts at a taking Maxwell Botlzmann distribution for the incoming DVLFs. The antitriplet and singlet Sommerfeld factors enhance the corresponding contributions in the free cross section, whereas the sextet and octet suppress them. The mass splitting is fixed to $\Delta m /m_\chi=10^{-2}$, though we find that there is no appreciable  difference for $\Delta m /m_\chi \in [10^{-3},10^{-1}]$. 
\subsubsection{Bound-state formation, dissociation and decays}
\label{sec:bound_state_effects}
Bound-state formation is yet another manifestation of repeated soft-gluon exchanges: in the spectrum of a two-particle system there is an above-threshold continuum of states along with bound states below threshold. In this section, we address the bound-state formation, bound-state decays and bound-state dissociation processes. The latter is a genuine thermal process that happens in a thermal environment and describe the thermal break up of a bound state when hit by a sufficiently energetic thermal gluon. Its interplay with the bound-state decay dictates how efficiently DVLFs are depleted in the form of bound states. All these quantities, which enter the effective cross section \eqref{Cross_section_eff_psi_barpsi}, are needed to estimate the DVLF pair annihilations and, ultimately, their impact on the DM energy density. 
As we have done for the above-threshold states, we rely on the pNRQCD framework to obtain the relevant observables for the bound-state dynamics. A comment is in order. Bound-state decays are the counterpart of the local four-fermion operators projected onto bound states rather than scattering states (see detailed discussions in ref.~\cite{Biondini:2023zcz}). Hence, the same arguments about the applicability of NRQCD and pNRQCD that involve the decay of heavy DVLF pairs into light SM states and/or scalar leptoquarks applies also here.

\textbf{Bound-state formation:}--
For the model at hand, there exist two classes of bound-state formation processes, and corresponding decays, that depends on the DVLF pairs. One find color-singlet bound states, which originate from the combination of a DVLF particle and antiparticle. Moreover, by assembling two vector-like particles color-triplet bound states that appear together with a continuum spectrum of unbound pairs in a color sextet configuration. As for $\psi \psi$ annihilations in a scattering state, see eq.~\eqref{QCD_psi_psi_exp}, the counterpart for the negative-energy part of the spectrum is the decay of a spin-triplet color-triplet bound state (cfr.~eq.~\eqref{decay_singlet}). As done in section, we do not distinguish explicitly between the color representation of $\bar{\psi} \bar{\psi}$ pairs and their conjugate color pairs $\psi \psi$, and simply refer to them as color-triplet bound states and color-sextet scattering (or unbound) states. 

The first ingredient for the estimation of bound-state effects on the DM energy density is the determination of the cross section for the two following processes
\begin{eqnarray}
    (\psi \bar{\psi})^p_{[8]} \to \mathcal{B}^n_{[1]} + g \, , \quad  (\psi \psi)_{[6]}^n \to \mathcal{B}^n_{[3]} + g \, ,
    \label{bsf_processes_eq}
\end{eqnarray}
where a color-singlet bound state $\mathcal{B}^n_{[1]}$ is formed from a color-octet scattering state $(\psi \bar{\psi})^p_{[8]}$ via the emission of
an ultrasoft gluon.\footnote{Bound states cannot form by the gluon emission from color-singlet or color-antitriplet scattering states because of SU(3) charge conservation, namely the two processes $(\psi \bar{\psi})^n_{[1]} \to \mathcal{B}^n_{[1]} + g$ and $(\psi \bar{\psi})^n_{[3]} \to \mathcal{B}^n_{[3]} + g$  do not occur.} The same hold for the second process that involves unbound pairs in a color-sextet $ (\psi \psi)^p_{[6]}$ and a bound state in a color triplet $\mathcal{B}^n_{[3]}$. The subscripts indicate the color representation of the pairs, $n$ stands for the collective discrete quantum numbers of a given bound state ($n \ell m$), and the unbound scattering state is labeled with the momentum of the relative motion $p=m_\psi v_{\textrm{rel}}/2$.   

\begin{figure}
    \centering
    \includegraphics[scale=0.55]{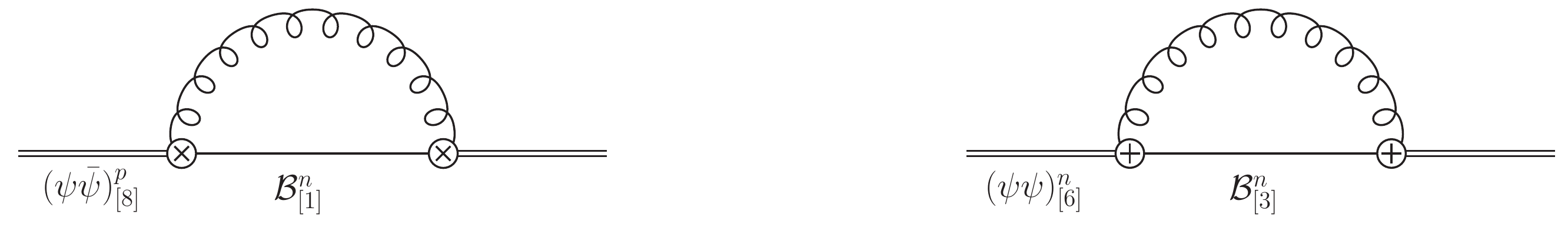}
    \caption{Self-energy diagram in pNRQCD with an initial scattering state in a color-octet (sextet) and an intermediate color-singlet (color-antitriplet) 
bound state. The imaginary part of the one-loop diagram on the left is responsible for the bound-state formation
process $(\psi \bar{\psi})^p_{[8]} \to \mathcal{B}^n_{[1]} + g $, whereas the imaginary part of the right diagrams account for  $ (\psi \psi)^p_{[6]} \to \mathcal{B}^n_{[3]} + g $.}
    \label{fig:pnEFT_DM_self_8_6}
\end{figure}

The bound-state formation cross section can be computed at leading order from the imaginary part of the one-loop self energy in pNRQCD. This has been recently discussed and detailed in refs.~\cite{Binder:2020efn,Biondini:2023zcz} for abelian and non-abelian dark matter models (see ref.~\cite{Biondini:2021ycj} for the case of soft scalar exchange and the corresponding pNREFT). We show the diagrams for the processes in eq.~\eqref{bsf_processes_eq} in figure~\ref{fig:pnEFT_DM_self_8_6}. In pNRQCD transitions among pairs are induced, at leading order, by \emph{chromoelectric-dipole} vertices.  
The bound-state formation process $(\psi \bar{\psi})^p_{[8]} \to \mathcal{B}^n_{[1]} + g$ has been computed formerly in the literature \cite{Mitridate:2017izz,Harz:2018csl,Garny:2021qsr,Biondini:2023zcz}. We instead have to derive the corresponding process for transition from a color-sextet unbound state to a color-triplet bound state. We shall follow the procedure outlined in ref.~\cite{Biondini:2023zcz}.

Because we are interested in the bound-state formation process happening in the early universe, the gluon can be thermal, and, therefore, the computation needs to be performed in the thermal field theory version of pNRQCD.
As long as the temperature scale is not larger than the inverse Bohr radius, one can rely on the in-vacuum derivation pNRQCD (see \cite{Brambilla:2008cx,Brambilla:2011sg} for heavy-quarkonium and for dark matter \cite{Biondini:2023zcz}). In the so-obtained EFT, the dynamical scales are the ultra-soft scale, $m_\psi \alpha_s^2$, and the temperature scale. The main relevant aspect is that transitions among pairs are still described by the in-vacuum electric dipole transitions.\footnote{The multipole expansion holds
for thermal gluons as long as the typical distance of the fermion-antifermion or fermion-fermion pairs is
smaller than $1/T$. At large temperatures, $T \gtrsim m_\psi \alpha_s$, the multipole expansion breaks down. In our numerical study, we solve the Boltzmann equation \eqref{BE_gen} starting from $m_\chi/T=10$ down to smaller temperatures. For $\alpha_s \approx 0.1$, the multiple expansion holds to a large extent for the whole temperature window, including  the chemical freeze-out occurring for $m_\chi/T \sim 1/25$.} 

In order to compute the bound-state formation process $(\psi \psi)_{[6]}^p \to \mathcal{B}^n_{[3]} + g$, we use pNRQCD for two heavy quarks \cite{Brambilla:2005yk}, which well applies to the vector-like quarks of our model. The non-relativistic DVLF fields define a pair in the color space as follows 
\begin{equation}
    \psi_{\textrm{NR},i}(t,\bm{x}_1)  \psi_{\textrm{NR},j}(t,\bm{x}_2) \sim \sum_{\ell=1}^3 {\rm{T}}^\ell(\bm{r},\bm{R},t) \bm{{\rm{T}}}^\ell_{ij} +   \sum_{\sigma=1}^6 \Sigma^\sigma(\bm{r},\bm{R},t) \bm{{\rm{\Sigma}}}^\sigma_{ij} \, ,
\end{equation} 
where ${\rm{T}}^\ell(\bm{r},\bm{R},t)$ and $\Sigma^\sigma(\bm{r},\bm{R},t)$ are the bi-local fields of pNRQCD that accounts for the wave-function of the corresponding color configurations, $\bm{r} \equiv \bm{x}_1-\bm{x}_2$ is the distance between a fermion located at $\bm{x}_1$ and an antifermion located at $\bm{x}_2$ and $\bm{R}\equiv(\bm{x}_1+\bm{x}_2)/2$ is the center of mass coordinate. Then $\ell=1,2,3$, $\sigma=1,...,6$ and $i,j=1,2,3$ and the tensors $\bm{{\rm{T}}}^\ell_{ij}$ and $\bm{{\rm{\Sigma}}}^\sigma_{ij}$ can be found in ref.~\cite{Brambilla:2005yk}. The relevant electric-dipole interactions read~\cite{Brambilla:2005yk}
\begin{eqnarray}
\mathcal{L}^{\psi \psi}_{\hbox{\tiny pNRQCD},\textrm{dipole}}& = & \int d^3 r \sum_{a=1}^8 \sum_{\ell=1}^3 \sum_{\sigma=1}^6 \left[  \left( \bm{{\rm{\Sigma}}}^\sigma_{ij} T^a_{jk} \bm{{\rm{T}}}^\ell_{ki}  \right) \Sigma^{\sigma \dagger} \bm{r} \cdot g_s \bm{E}^a \, {\rm{T}}^\ell  \right.
\nonumber
\\
&& \hspace{3.5 cm} \left. -  \left(\bm{{\rm{T}}}^\ell_{ij}  T^a_{jk}   \bm{{\rm{\Sigma}}}^\sigma_{ki} \right) {\rm{T}}^{\ell \dagger} \bm{r} \cdot g_s \bm{E}^a \, \Sigma^\sigma \right] \, ,
\label{pNRQCD_FF}
\end{eqnarray}
where $T^a$ are the SU(3) generators.
Having clarified the relevant vertices, we move to the evaluation of the sextet self-energy in figure~\ref{fig:pnEFT_DM_self_8_6}. 
We use the real-time Schwinger--Keldysh formalism~\cite{Bellac:2011kqa,Laine:2016hma}.
The real-time formalism leads to a doubling of the degrees of freedom called of type 1 and 2. The type 1 fields are the physical ones, namely those that appear in the initial and final states. Propagators are represented by $2\times 2$ matrices, as they may involve fields of both types. As for heavy non-relativistic particles at finite temperature, it has been shown in~\cite{Brambilla:2008cx} that the 12 component of a heavy-field propagator vanishes in the heavy-mass limit, as a result, 
the physical heavy fields do not propagate into type 2 fields.
Hence, the type 2 fermion-antifermion fields decouple and may be ignored in the heavy-mass limit, which reduces the relevant self-energies that we need to compute. In practice, it suffices to obtain the 11 component of the self-energy diagrams given in figure~\ref{fig:pnEFT_DM_self_8_6}. As for the thermal gluon propagator, we shall adopt its form in Coulomb gauge \cite{Brambilla:2008cx,Biondini:2023zcz}.

The self-energy for the color-sextet field $\Sigma^\sigma(\bm{r},\bm{R},t)$ in the right panel of figure~\ref{fig:pnEFT_DM_self_8_6}, reads\footnote{The self-energy $\Sigma^{11}$ should not be confused with the bilocal sextet field $\Sigma^\sigma$.} 
\begin{align}
\Sigma^{11}(p_0) = -i\frac{g_s^{2}}{2} \frac{N(N+1)}{2}\frac{d-2}{d-1} ~& \mu^{4-d}r^{i}  \int \frac{\textrm{d}^{d}k}{(2\pi)^{d}} \frac{i}{p^{0}-k^{0}-H + i\epsilon}\nonumber\\
&\times k_{0}^{2}\left[\frac{i}{k_{0}^{2} -|\bm{k}|^{2}+i\epsilon} + 2\pi \delta(k_{0}^{2} - |\bm{k}|^{2})n_{B}(|k_{0}|) \right]r^{i} \,,
\label{self_sextet_antitriplet_11}
\end{align}
where $p_0$ is the energy of the incoming pair and $n_B(x)=1/(e^{x}-1)$ the Bose-Einstein distribution.
In eq.~\eqref{self_sextet_antitriplet_11} one can distinguish the in-vacuum and thermal contributions originating from the gluon propagator. The next steps are to extract the imaginary part of the self-energy in eq.~\eqref{self_sextet_antitriplet_11}, project the self-energy onto external scattering states and use the optical theorem in order to obtain the corresponding cross section  
\begin{eqnarray}
 (\sigma_{(\psi \psi)^p_{[6]} \to \mathcal{B}^n_{[3]} + g} \, v_{\hbox{\scriptsize rel}})(\bm{p})
= -2 \langle \,\bm{p}|  {\rm{Im}}[\Sigma^{11}(E_p)] | \bm{p} \, \rangle        \, .      \end{eqnarray}
Finally, we project onto intermediate color-antitriplet bound states, introduce a short-hand notation for the process $(\psi \psi)^p_{[6]} \to \mathcal{B}^n_{[3]} + g$ with simpler subscript and superscript for the cross section, and perform the color average of the cross section. Our result reads 
\begin{align}
(\sigma^{6\to 3}_{\hbox{\scriptsize bsf}} \, v_{\hbox{\scriptsize rel}})(\bm{p})
  =\frac{N(N+1)}{3N^2} \, \alpha_s(\mu_\textrm{us}) \sum_{n} \left[ 1 + n_{\text{B}}(\Delta \tilde{E}_{n}^{p}) \right] |\langle n_{[3]} | \bm{r} | \bm{p}_{[6]}\,\rangle|^2  (\Delta \tilde{E}_{n}^{p})^3 \, ,
\label{bsfn_6_3}  
\end{align}
where we make explicit that one power of the strong coupling constant is evaluated at the ultrasoft scales $\mu_{\textrm{us}} \equiv m_\psi \alpha_s^2,T$, as dictated by the ultrasoft interaction in eq.~\eqref{pNRQCD_FF} .  Then, the energy difference between the incoming scattering state and outgoing bound state is at leading order
\begin{equation}
\Delta \tilde{E}_{n}^{p} = E_p-E_{n_{[3]}} = \frac{M_\psi}{4}v_{\textrm{rel}}^2\left(1+\frac{C_a^2  \alpha_s^2}{ n^2v_{\textrm{rel}}^2}\right) \, ,
\label{energy_photon_bsf}
\end{equation}
where we used $ C_a =(N+1)/(2N)$ in order to write the Coulombic energy levels in a compact way. 

As a relevant example, which we shall use in the numerical extraction of the DM energy density in section~\ref{sec:DM_energy_numerics}, we specify the general result in eq.~\eqref{bsfn_6_3} to the formation of the lowest-lying $1\textrm{S}$ bound state.
In this case, only scattering states in the partial wave $\ell=1$ contribute, 
and the bound-state formation cross section reads
\begin{eqnarray}
  &&(\sigma^{1\textrm{S}[3]}_{\hbox{\scriptsize bsf}} v_{\hbox{\scriptsize rel}})(\boldsymbol{p}) = \frac{N(N+1)}{3N^2}\,\alpha_s(\mu_{\textrm{us}})
  \left[ 1 + n_{\text{B}}(\Delta \tilde{E}_{1}^{p}) \right] \left|\langle 1\textrm{S}_{[3]} | \boldsymbol{r} | \boldsymbol{p}_{[6] } 1 \rangle\right|^2
  (\Delta \tilde{E}_{1}^{p})^3 \nonumber 
  \\
  &&=  \alpha_s(\mu_{\textrm{us}}) \frac{2^{8} \pi^2 \alpha_s^6}{3 \, m_\psi^2 \, v_{\hbox{\scriptsize rel}}^5}\frac{C_a^4(C_a+1)}{N}\left( C_a-\frac{1}{N}\right)
      \nonumber
      \\
      &&\hspace{3.5 cm}\times \frac{1+\left(\frac{\alpha_s (N-1)}{2Nv_{\textrm{rel}}}\right)^2}{\left[ 1+ \left(\frac{C_a\alpha_s}{v_{\textrm{rel}}}\right)^2\right]^3}  \frac{e^{ \frac{2(N-1)}{N} \frac{\alpha_s}{v_{\textrm{rel}}} \hbox{\scriptsize arccot} \frac{C_a\alpha_s}{v_{\textrm{rel}}} }}{e^{\frac{\pi(N-1)}{N} \frac{\alpha_s}{v_{\textrm{rel}}}}-1} \left[ 1 + n_{\text{B}}(\Tilde{\Delta} E_{1}^{p}) \right] .
  \label{bsf_sextet_to_bound_1S}
\end{eqnarray}
In the dipole matrix element, the natural renormalization scale of the coupling  is the soft scale\footnote{We do not distinguish the soft scale between antitriplet bound states and unbound sextets.}. However, in order to avoid clutter, we dropped the corresponding scale dependence for the strong coupling in the bound-state formation expression \eqref{bsf_sextet_to_bound_1S}. Details on the Coulombic wave functions and the general expression for the electric-dipole matrix elements for sextet-to-antitriplet transitions are given in appendix~\ref{sec:app_electri_dipole}. 

In order to compare with the bound-state formation process $(\psi \bar{\psi})^p_{[8]} \to \mathcal{B}^n_{[1]} + g$, that is also needed for the determination of the DM energy density in section~\ref{sec:DM_energy_numerics}, we provide the cross section for the formation of the color-singlet ground state, which reads \cite{Mitridate:2017izz,Harz:2018csl} (see refs.~\cite{Garny:2021qsr,Biondini:2023zcz} for a derivation in pNRQCD)
\begin{equation}
\begin{aligned}
  &(\sigma^{1\textrm{S}[1]}_{\hbox{\scriptsize bsf}} v_{\hbox{\scriptsize rel}})(\boldsymbol{p}) = \frac{4C_F}{3N^2}\,\alpha_s(\mu_{\textrm{us}})
  \left[ 1 + n_{\text{B}}(\Delta E_{1}^{p}) \right] \left|\langle 1\textrm{S}_{[\bm{1}]} | \boldsymbol{r} | \boldsymbol{p}_{[\bm{8}]} 1 \rangle\right|^2
  (\Delta E_{1}^{p})^3 \\
  &= \alpha_s(\mu_{\textrm{us}})\frac{2^7 \pi^2 \alpha_s^6}{3 \, m_\psi^2 \, v_{\hbox{\scriptsize rel}}^5}\frac{C_F^4}{N^3}\left(2C_F+N\right)^2
  \frac{1+\left(\frac{\alpha_s}{2Nv_{\textrm{rel}}}\right)^2}{\left[ 1+ \left(\frac{C_F\alpha_s}{v_{\textrm{rel}}}\right)^2\right]^3}
      \frac{e^{\frac{2\alpha_s}{Nv_{\textrm{rel}}} \hbox{\scriptsize arccot} \frac{C_F\alpha_s}{v_{\textrm{rel}}} }}{e^{\frac{\pi}{N} \frac{\alpha_s}{v_{\textrm{rel}}}}-1}
      \, \left[ 1 + n_{\text{B}}(\Delta E_{1}^{p}) \right] .
  \label{bsf_octet_to_bound_1S}
  \end{aligned}
\end{equation}
The corresponding energy difference between the incoming color-octet scattering state and the color-singlet bound state is 
\begin{equation}
  \Delta E_{1}^{p}
  =\frac{Mv_{\textrm{rel}}^2}{4} \left( 1+ \frac{C_F^2\alpha_s^2}{v_{\textrm{rel}}^2}\right)\,.
    \label{delta_energy_octet_singlet}
\end{equation} 
\begin{figure}[t!]
    \centering
    \includegraphics[scale=0.555]{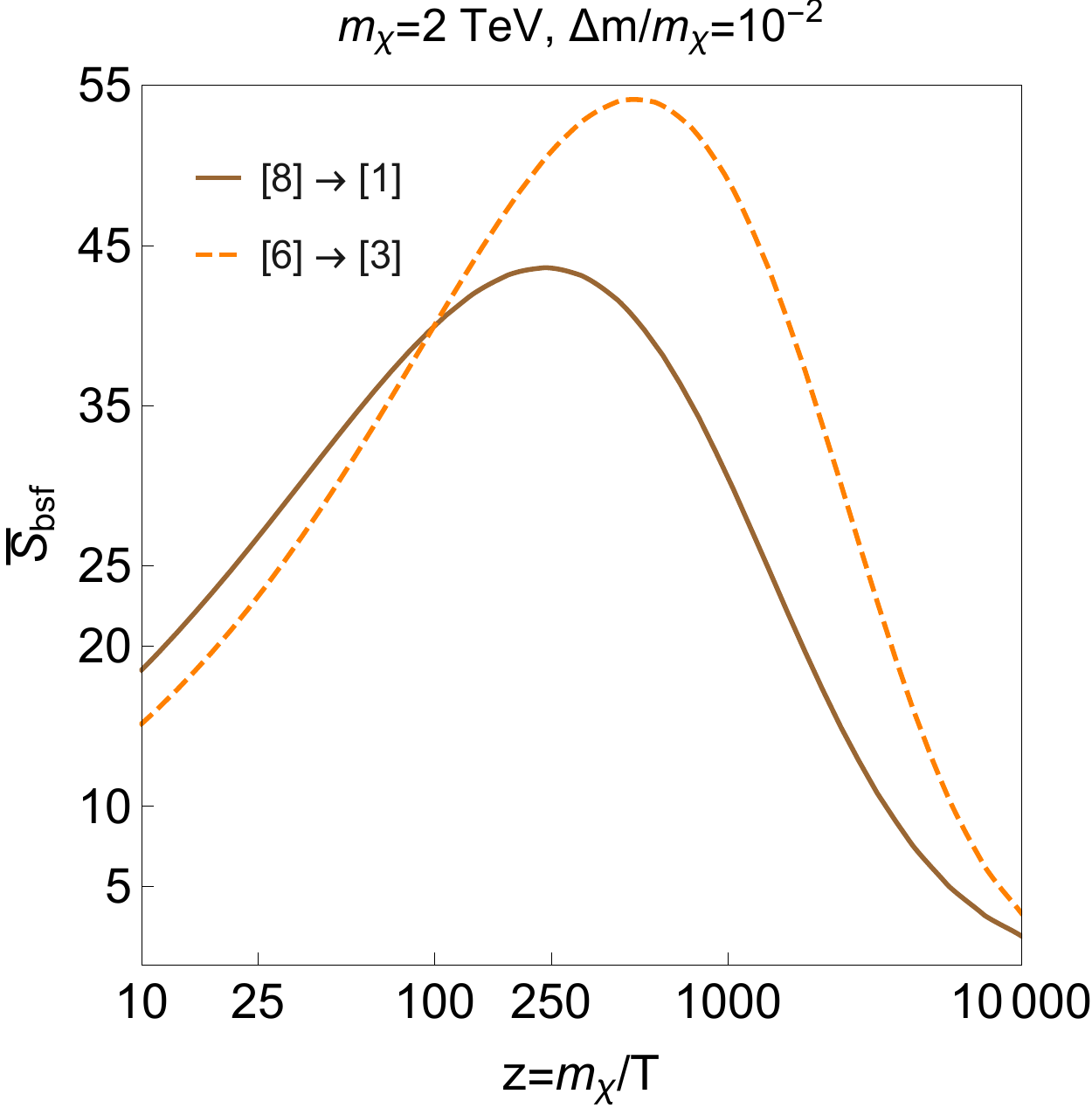}
    \hspace{0.1 cm}
     \includegraphics[scale=0.59]{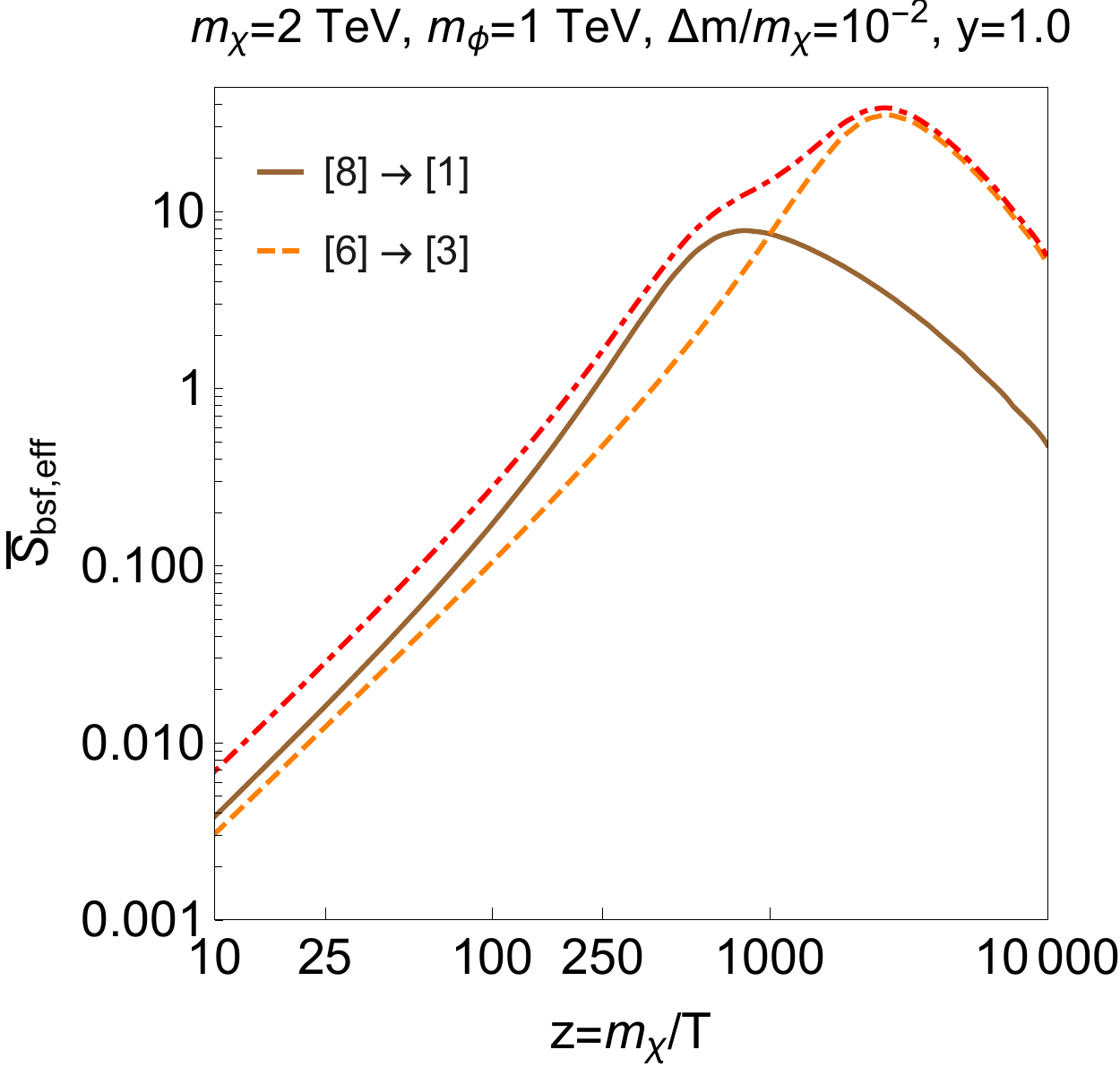}
    \caption{(Left) Thermally averaged bound-state formation cross sections for the octet-to-singlet in eq.~\eqref{bsf_octet_to_bound_1S} and sextet-to-triplet eq.~\eqref{bsf_sextet_to_bound_1S} divided by $\pi \alpha_s(\mu_h)^2 / M_\psi^2$. The relative mass splitting is $\Delta M / M_\chi=10^{-2}$. (Right) Bound-state formation cross section times the widths combination as in the second term of the right-hand side of eq.~\eqref{Cross_section_eff_psi_barpsi}, also divided by $\pi \alpha_s(\mu_h)^2 / M_\psi^2$. Bound-state formation, dissociation and decays of the ground state is considered.}
    \label{fig:bsf_and_bsd}
\end{figure}
In figure~\ref{fig:bsf_and_bsd} we show the thermally averaged bound-state formation cross section for the color-singlet and color-triplet 1S bound states divided by $\pi \alpha_s(\mu_h)^2 /m_\psi^2$.\footnote{The normalization factor is just needed to plot a dimensionless quantity, $\bar{S}_{\textrm{bsf}} \equiv (\sigma^{1\textrm{S}[R]}_{\hbox{\scriptsize bsf}} v_{\hbox{\scriptsize rel}})/(\pi \alpha_s(\mu_h)^2 /m_\psi^2)$. One may consider different renormalization scale for the strong coupling in the normalization factor, e.g.~$\pi \alpha_s(\mu_s)^2 /m_\psi^2$, which would make the curves lower.} The thermal average is performed in the standard way, see e.g. \cite{vonHarling:2014kha,Biondini:2023zcz}, with Maxwell-Boltzmann distribution of the incoming DVLF pair. On the one hand, one may see how the two different bound-state formation processes are comparable at typical freeze-out temperatures, with the cross section for  $(\psi \bar{\psi})^p_{[8]} \to \mathcal{B}^n_{[1]} + g$ being marginally larger than $(\psi \psi)_{[6]}^n \to \mathcal{B}^n_{[3]} + g$. On the other hand, at smaller temperatures, the bound-state formation for the color-triplet 1S state is larger and peaks at later time with respect to the color-singlet ground state. This latter aspect is due to a smaller absolute value of the binding energy for the triplet, $|E_{1_{[1]}}|/|E_{1_{[3]}}|=4$ (as for the latter aspect, a qualitative similar behaviour is found when comparing the ground state with excited states \cite{Binder:2021vfo,Garny:2021qsr,Biondini:2023zcz}).  

When inserting the bound-state formation cross sections \eqref{bsf_octet_to_bound_1S}  and \eqref{bsf_sextet_to_bound_1S} into eq.~\eqref{Cross_section_eff_psi_barpsi} and into the corresponding expression for $(\psi \psi)/(\bar{\psi} \bar{\psi})$ pairs, one performs the spin average and the factors $1/4$ and $3/4$ appears respectively.

\textbf{Dissociation and decays:}--
Once bound state form, it can either decay or get dissociated by thermal particles of the early universe plasma. In this work, we consider the gluodissociation process \cite{Kharzeev:1994pz,Xu:1995eb,Brambilla:2011sg}, namely $\mathcal{B}^n_{[3]} + g \to (\psi \psi)_{[6]}^n$. Whenever thermal gluons in the early universe plasma have sufficient energy, they can break the bound state and turn it into an above-threshold scattering state. The corresponding rate is a thermal width, or dissociation width, of a bound state.  The efficiency of the conversion of bound states into its decay product depend on the interplay of the dissociation and decay width, as displayed in the effective cross section in eq.~\eqref{Cross_section_eff_psi_barpsi}. The bound-state dissociation can be obtained in two ways. One powerful argument is that, whenever the ionization equilibrium is maintained, the bound-state dissociation and bound-state formation cross section are related via the Milne relation \cite{vonHarling:2014kha,Mitridate:2017izz} (one can find a recent derivation for it in ref.~\cite{Harz:2018csl}). More specifically, the Milne relation links the bound-state formation with the ionization cross section. The latter is used to obtain the  bound-state dissociation width through a convolution integral with the incoming thermal gauge boson momentum, that may break the bound state if sufficiently energetic. We write the dissociation width for a generic color configuration $[R]=[1],[3]$,  as follows
\begin{eqnarray}
    \Gamma^n_{\textrm{bsd},[R]} = g_g \int_{|\bm{k}|>|E^n_{[R]}|} \frac{d^3 k}{( 2\pi)^3} n_B(|\bm{k}|) \sigma^n_{\textrm{ion},[R]} (|\bm{k}|) \, ,
\end{eqnarray}
where $|\bm{k}|$ is the energy of the gluon, $[R]$ stands for the representation of the bound state either in a color singlet or a color antitriplet, $g_g$ are the gluon degrees of freedom and $\sigma^n_{\textrm{ion},[R]} (|\bm{k}|)$ is the ionization cross section, which is related to the bound state formation cross section via the Milne relation 
\begin{eqnarray}
     \sigma^n_{\textrm{ion},[R]} (|\bm{k}|) = \frac{g_\psi^2}{g_g \, g_{\mathcal{B}_{[R]}}} \frac{M_\psi^2 v_{\textrm{rel}}}{4 |\bm{k}|^2} \sigma_{\textrm{bsf}}^{[R'] \to [R]} \, . 
\end{eqnarray}
Here $g_{\mathcal{B}_{[R]}}$ stands for the degrees of a bound state, whereas $[R']$ for the unbound scattering states in a color-octet or color-sextet represenation. 

Alternatively, and as a non-trivial check, one can derive the bounds-state dissociation from the imaginary part of the bound-state in pNRQCD. This has been shown in refs.~\cite{Biondini:2021ycj,Biondini:2023zcz} for vector as well as for scalar force mediators in the context of dark matter, and earlier in refs.~\cite{Brambilla:2008cx,Brambilla:2011sg} for heavy quarkonium phenomenology.

We do not include the additional dissociation mechanism as induced by $2 \to 2$ scatterings with the in-medium constituents. This is known as inelastic parton scattering \cite{Grandchamp:2001pf,Grandchamp:2002wp,Laine:2006ns,Brambilla:2013dpa} in heavy quarkonium literature. The counterpart for cosmological applications to dark matter freeze-out was considered in refs.~\cite{Biondini:2017ufr,Biondini:2018pwp,Binder:2018znk} in the screening regime, namely for temperature larger than the typical Bohr radius $T \gg m_\psi \alpha_s$. Here, a non-trivial interplay with another thermal scale, a Debye mass for the gluons, is established and the extraction of the relevant cross sections and widths of the pairs becomes rather challenging for a broad temperature range. The bound-state formation process has been computed at fixed order, without a resummation of collective plasma effects that generates a Debye mass for the gluons, in ref.~\cite{Binder:2021otw} (the corresponding cross section for an abelian dark matter model is given in ref.~\cite{Binder:2019erp}). A careful investigation of the Debye mass scale within the
framework of non-relativistic effective field theories, in particular its role in the bound-state formation via gauge boson emission, is still ongoing and we do not account for it in our work (however see \cite{Garny:2021qsr} for an exemplary implementation of these effects in a dark matter model with colored coannihilators). Hence, the DM energy density as derived in section~\ref{sec:DM_energy_numerics} has to be understood as upper bound. 

The last ingredient is the decay width of the bound states. The color and spin-averaged bound state decay widths for DVLFs pairs, at leading order in the coupling and in the non-relativistic expansion, read as follows
\begin{eqnarray}
&& \Gamma^n_{[1],\textrm{ann}} = \frac{C_F^3 m_\psi  \alpha_s(\mu_{\textrm{s}})^3}{4 n^3} \left[ C_F\alpha_s(\mu_h) +  \frac{y^4  \, m_\psi^4 }{24 \pi^2  \,  (m_\psi^2+m_\chi^2-m_\phi^2)^2}  \left( 1-\frac{m_\phi^2}{m_\psi^2} \right)^{\frac{3}{2}} \right]  \, , 
\label{decay_singlet}
\\
&& \Gamma^n_{[3],\textrm{ann}} =  C_a^3\frac{y^4 \alpha_s(\mu_{\textrm{s}})^3 \, m_\psi^5 }{24 \pi^2 (m_\psi^2+m_\chi^2-m_\phi^2)^2}  \left( 1-\frac{m_\phi^2}{m_\psi^2} \right)^{\frac{3}{2}} \, .
\label{decay_antitriplet}
\end{eqnarray}
$\Gamma^n_{[1],\textrm{ann}}$ stands for decay width of $n$S color-singlet bound states, which receive contributions from decays into gluon and leptoquark pairs, whereas  $\Gamma^n_{[3],\textrm{ann}}$ is the color-triplet $n$S decay width, which encompasses only decay into leptoquark pairs. We have explicitly indicated the scale for $\alpha_s$ at the hard scale, $\mu_h =2 m_\psi$, and the soft scale $\mu_s$, which originates from the wave function. 

Finally, in the right panel of figure~\ref{fig:bsf_and_bsd}, we show the effective bound state formation cross section, namely the second term in the right-hand side of eq.~\eqref{Cross_section_eff_psi_barpsi}, once again normalized by $\pi \alpha_s(\mu_h)^2 /m_\psi^2$ in order to display a dimensionless quantity. Here, the dissociation widths as well as the decay widths enter. Further parameters are specified in the plot label. One can see how the bound-state formation of a color-triplet 1S state (orange-dashed curve) gives a sizeable contribution to the total bound-state formation (red dash-dotted line) for $y=1.0$. We have checked that the $(\psi \bar{\psi})^p_{[8]} \to \mathcal{B}^n_{[1]} + g $ is largely dominant for $y \lesssim 0.5$.
\subsection{Numerical results for the DM energy density}
\label{sec:DM_energy_numerics}
In this section we solve the effective Boltzmann equation \eqref{BE_gen} with  the relevant cross sections and decay widths that have been discussed in sections \ref{sec:Sommerfeld_effects} and \ref{sec:bound_state_effects}. The main scope is to address the impact of near-threshold effects on the DM energy density in the coannihilation regime, namely for small mass splittings between the dark fermion $\chi$ and the DVLF. We recall at this stage that the Sommerfeld and bound-state effects play a role for the DVLF pair annihilations, namely when the incoming states are $\psi \bar{\psi}$, $\psi \psi$ and $\bar{\psi} \bar{\psi}$.
\begin{figure}[t!]
    \centering
    \includegraphics[scale=0.57]{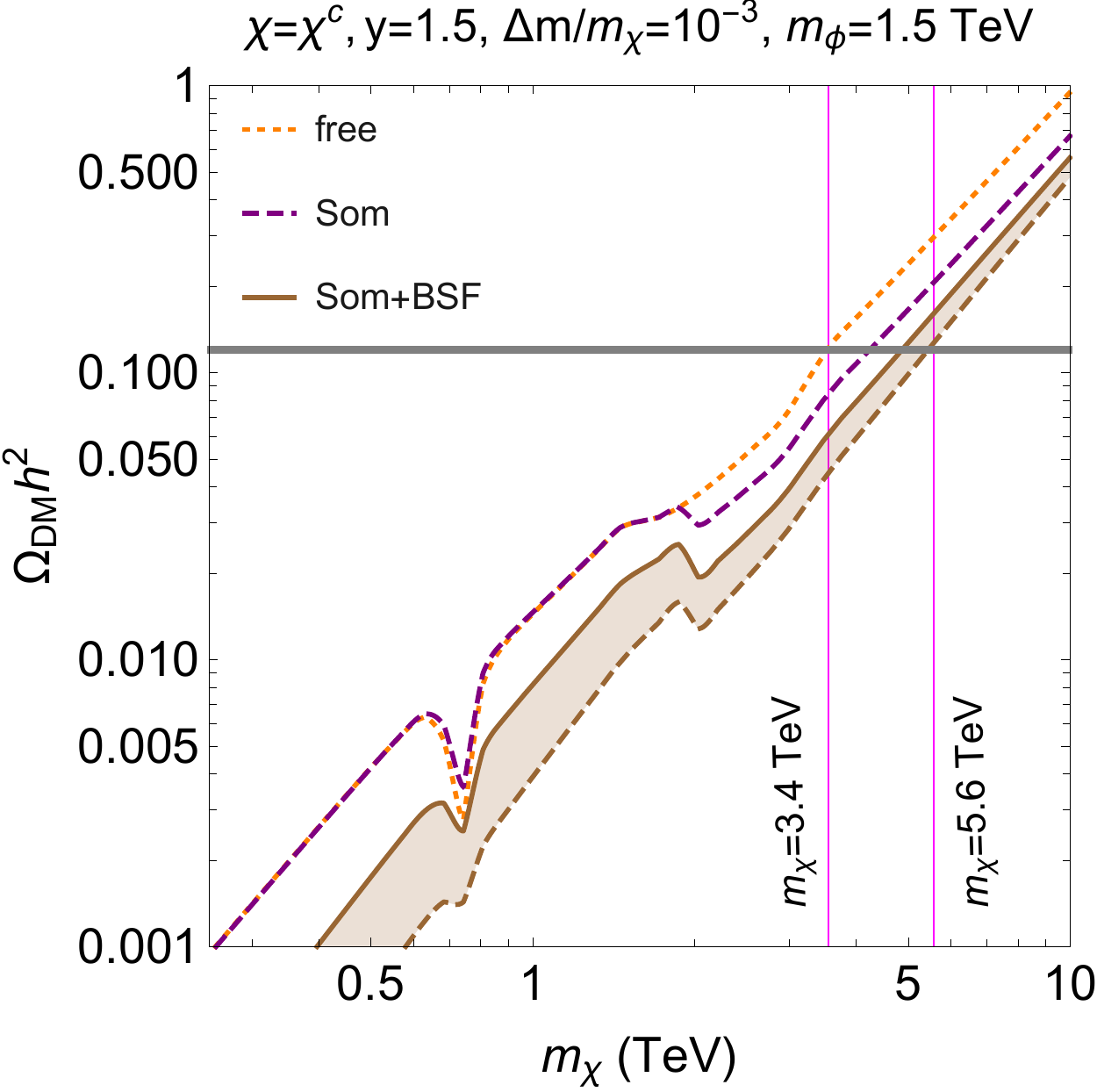}
    \hspace{0.1 cm}
    \includegraphics[scale=0.57]{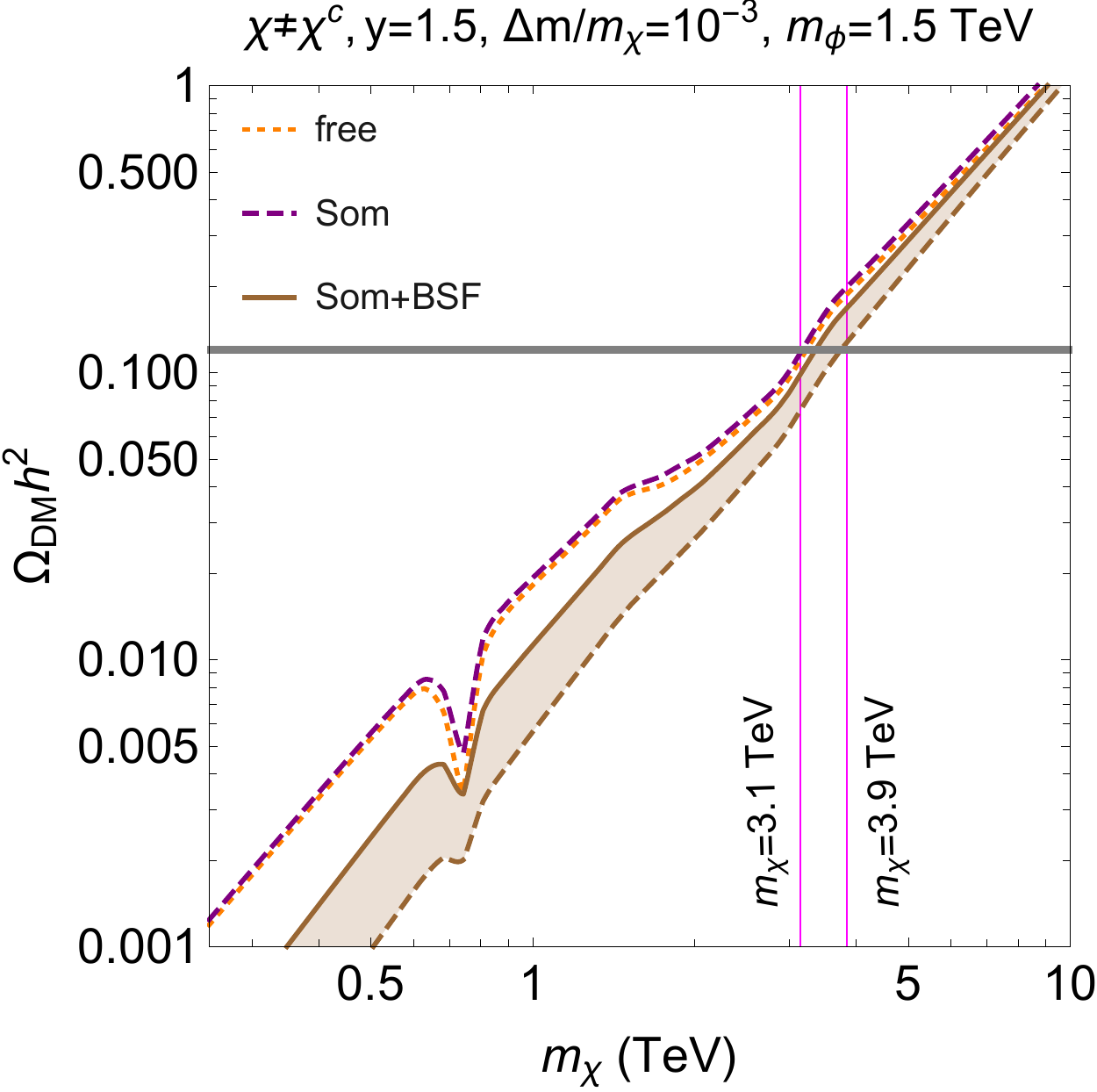}
    \caption{Dark matter energy density $\Omega_{\textrm{DM}}h^2$ as a function of the dark matter mass $m_\chi$. The model parameters and the Majorana/Dirac options are indicated at the top of each plot. Magenta vertical lines serve as orientation to display the largest dark mater mass compatible with the observed energy density as obtained with a free cross section and with near-threshold effects (both Sommerfeld and bound states) respectively. }
    \label{fig:Majorana_Mchi_versus_OmegaDM_a}
\end{figure}

In figure~\ref{fig:Majorana_Mchi_versus_OmegaDM_a} the dark matter energy density is given as a function of the dark matter mass $m_\chi$ for the two options, Majorana and Dirac dark matter, respectively in the left and right panel. We take the leptoquark mass to be $m_\phi=1.5$ TeV, the relative mass splitting is $\Delta m /m_\chi=10^{-3}$ and the portal coupling is $y=1.5$.  In this section, we fix the leptoquark-to-SM couplings $y_L=y_R=1$ (we elaborate more on varying these couplings in section~\ref{sec:muon_gminus2_pheno}). The orange-dotted, purple-dashed and brown lines correspond to the energy density as extracted with free annihilation cross section, the Sommerfeld-only corrected cross section and with the inclusion of both Sommerfeld effects and bound-state formation respectively. More specifically, the solid-brown curve accounts for the lowest lying 1S bound state, whereas the brown-dashed line comprises the effect of the 2S state as well, in the no-transition limit. In the Majorana case, both bound states in a color singlet and color antitriplet contribute.  

In addition to the dip at $m_\chi \approx m_\phi/2$, which captures the resonant enhancement of the coannihilation process $\chi \psi \to q^c \bar{\ell}$, one finds a further dark matter mass range where the energy density is locally decreased. This is due to the opening of additional annihilation channels for the dark fermion $\chi$ and DVLF into leptoquark pairs. In the Majorana dark matter scenario, the Sommerfeld effect as well as formation and decays of bound states make this feature more prominent. As a general trend, below the leptoquark mass threshold, the overall Sommerfeld corrections have practically no impact (a small effect can be seen at the resonant window $m_\chi \approx m_\phi/2$). One can trace this back to a competing enhancement and suppression of the color-singlet and color-octet contributions to the annihilations into light Standard Model QCD states, cfr.~eq.~\eqref{cross_psi_psibar_QCD}, that makes the Sommerfeld-corrected cross section slightly smaller than the free cross section at the freeze-out. However, the situation is different above the leptoquark mass, where DVLF annihilations experience an overall enhancing effect (accordingly the purple-dashed line is below the orange-dotted line because of a larger cross section that results in a smaller DM energy density). For the Majorana option, the enhancement of the cross section is more important because of $\psi \psi \to \phi \phi$ and its conjugate process, whose leading contribution to the annihilation cross section originates from an attractive color-triplet channel, see eq.~\eqref{psi_psi_leptoquarks}.  As for the Dirac case, only $\psi \bar{\psi} \to \phi \phi^\dagger$ can occur, for which competing color-singlet and color-octet effects make the cross section only slightly larger than the free cross section, see eq.~\eqref{psi_psibar_leptoquarks}. 

The bound-state effects have a rather different behaviour with respect to the Sommerfeld-only scenario,  see solid-brown lines in figure~\ref{fig:Majorana_Mchi_versus_OmegaDM_a}.
Bound-state formation is effectively active below the leptoquark mass threshold due to formation of color-singlet bound states and their decays into light SM quarks and gluons. Above the leptoquark mass, the formation and decays of color-antitriplet bound states also contribute in depleting the dark matter, since the corresponding decays becomes kinematically allowed.\footnote{The bound-state formation $(\psi \psi)_{[6]}^n \to \mathcal{B}^n_{[3]} + g$ and dissociation process are active even below the leptoquark mass, as they are independent of $m_\phi$. However, in order for color-triplet bound states to contribute to the DVLF pair annihilations, the decay process $\mathcal{B}^n_{[3]} \to \phi \phi$ has to be possible.} For $m_\psi$ larger than the leptoquark mass, color singlet bound-state can also decay into four-body SM states through unstable $\phi \phi^\dagger$ pairs. Since bound-state effects are efficiently annihilating the DVLF pairs, and hence the DM fermion in the coannihilation regime, the DM energy density is systematically below the free-annihilation scenario for the entire dark-matter mass range. As one may see from the comparison of the left and right plots in figure~\ref{fig:Majorana_Mchi_versus_OmegaDM_a}, bound-state effects are more important for the Majorana case, because of the additional bound-state formation processes for $(\psi \psi)$ and $(\bar{\psi} \bar{\psi})$ pairs and corresponding bound-state decays. 

For the choice of the parameters as given in figure~\ref{fig:Majorana_Mchi_versus_OmegaDM_a}, and accounting for 2S excited state in the non-transition limit, the dark matter mass that is consistent with the observed energy density shift from $m_\chi=3.4$ TeV ($m_\chi=3.1$ TeV) to $m_\chi=5.6$ TeV ($m_\chi=3.9$ TeV) for the Majorana (Dirac) case. Having clarified the impact of the Sommerfeld-only corrected cross section, in the following we present the numerical results by accounting for both manifestation of non-perturbative effects, i.e.~Sommerfeld and bound states. 
Changing the leptoquark mass does not affect the qualitative behavior of the DM energy density curves displayed in figure~\ref{fig:Majorana_Mchi_versus_OmegaDM_a}.

\begin{figure}[t!]
    \centering
    \includegraphics[scale=0.57]{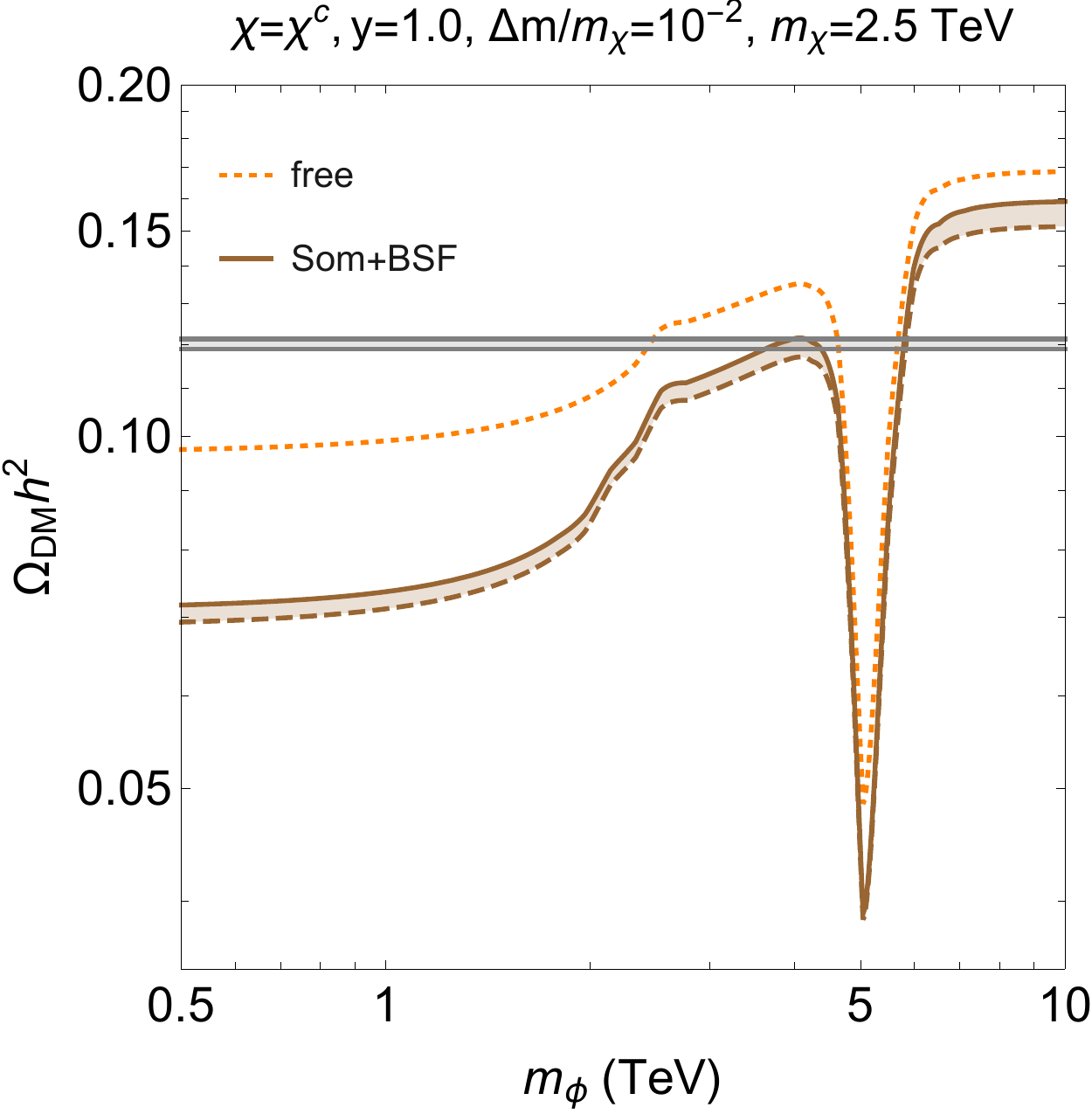}
     \includegraphics[scale=0.57]{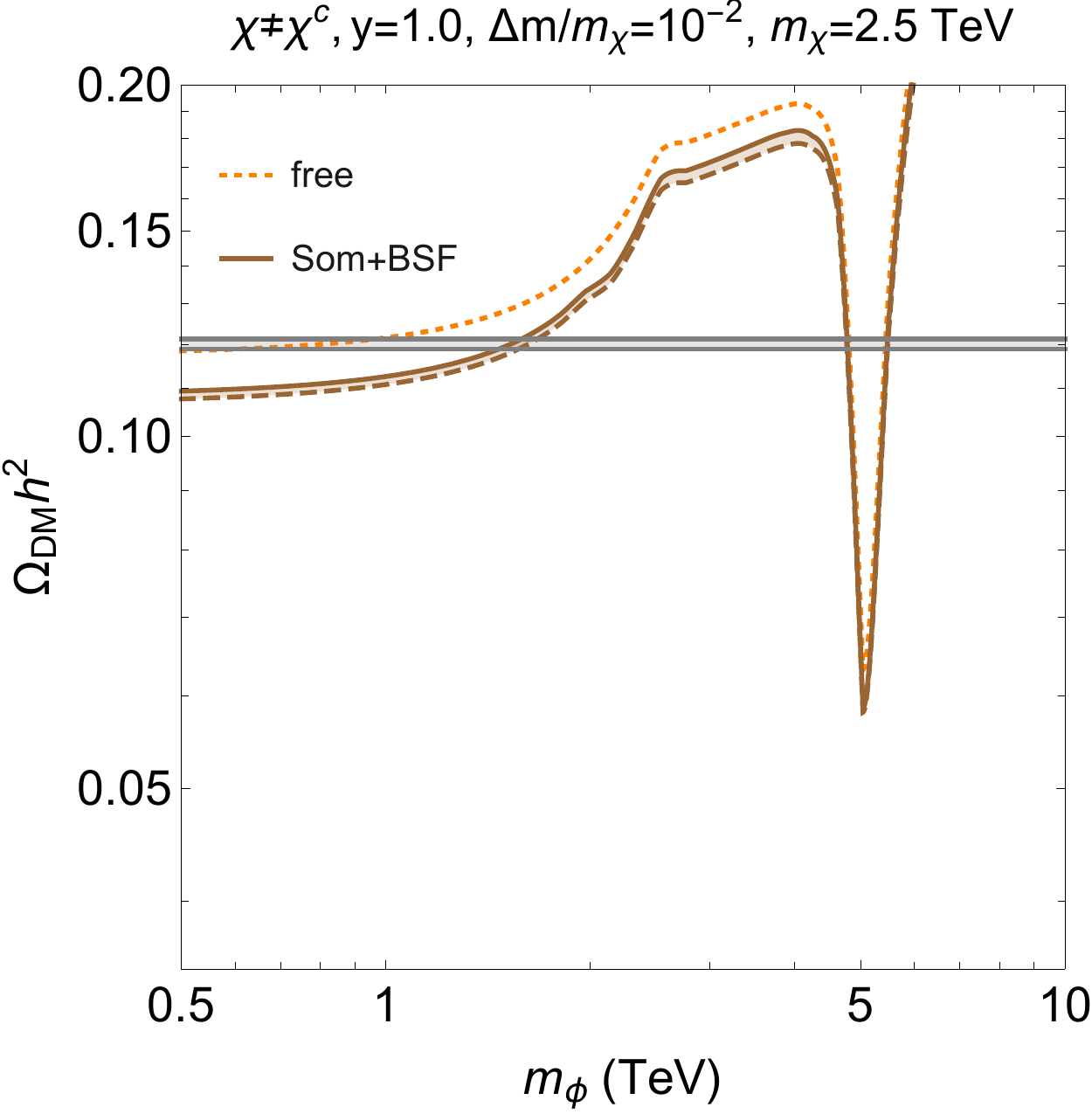}
    \caption{Dark matter energy density $\Omega_{\textrm{DM}}h^2$ as a funtion of the leptoquark mass $m_\phi$. The other model parameters indicated at the top of each plot, as well as the nature of the DM fermion (Majorana left panel, Dirac right panel).}
    \label{fig:Omega_Mphi_plane_a}
\end{figure}
In the next example, we consider the DM energy density as a function of the leptoquark mass. In figure~\ref{fig:Omega_Mphi_plane_a}, left panel, we notice how the non-perturbative effects provide a much wider range for $m_\phi$ that is compatible with $\Omega_{\textrm{DM}}h^2 \leq 0.1200 \pm 0.001$, so that we do not overclose the universe up to $m_\phi=5.6$ TeV. On the contrary, if one estimates the DM energy density without Sommerfeld and bound-state effects, leptoquark masses 2.4 TeV $\leq m_\phi \leq$  4.5 TeV are excluded by the Planck collaboration measurement, and one has to rely on the resonant dip that allows for a viable mass range 4.6 TeV $\leq m_\phi \leq$ 5.6 TeV. For the same choice of parameters, the Dirac case displays  differences with the Majorana dark matter scenario. The overall annihilation cross section is smaller and the curves shift at a higher DM energy density. The less prominent non-perturbative effects make the orange-dotted line and brown curves closer for $m_\chi$ and $m_\psi$ larger than  $m_\phi$. Moreover, the leptoquark mass window compatible with $\Omega_{\textrm{DM}}h^2 \leq 0.1200 \pm 0.001$ is $m_\phi \leq 1.0$ TeV for the free case, whereas it is extended to $m_\phi \leq 1.5$ when Sommerfeld and bound-state effects are included. Then, a second mass region is available at around $m_\phi \approx 2 m_\chi$, as the resonant enhancement is sufficient to reduce the DM energy density below the observed value.  

Finally we aim to explore the parameter space of the model which is compatible with the observed dark matter energy density. The model contains three mass parameters ($m_\chi, m_\psi, m_\phi$) and three new couplings $(y,y_L,y_R)$. We focus on $(m_\psi,m_\phi)$ contours for different values of the relative mass splitting $\Delta m/m_\chi$ and the portal coupling $y$ (we remind the reader we fix the leptoquark-to-SM couplings $y_L=y_R=1$).  Our choice to consider the $(m_\psi,m_\phi)$ mass plane is motivated by the present collider limits that are largely applicable to the colored states of the model, see discussion in section~\ref{sec:collider_constraints}. 
\begin{figure}[t!]
    \centering
 \includegraphics[scale=0.55]{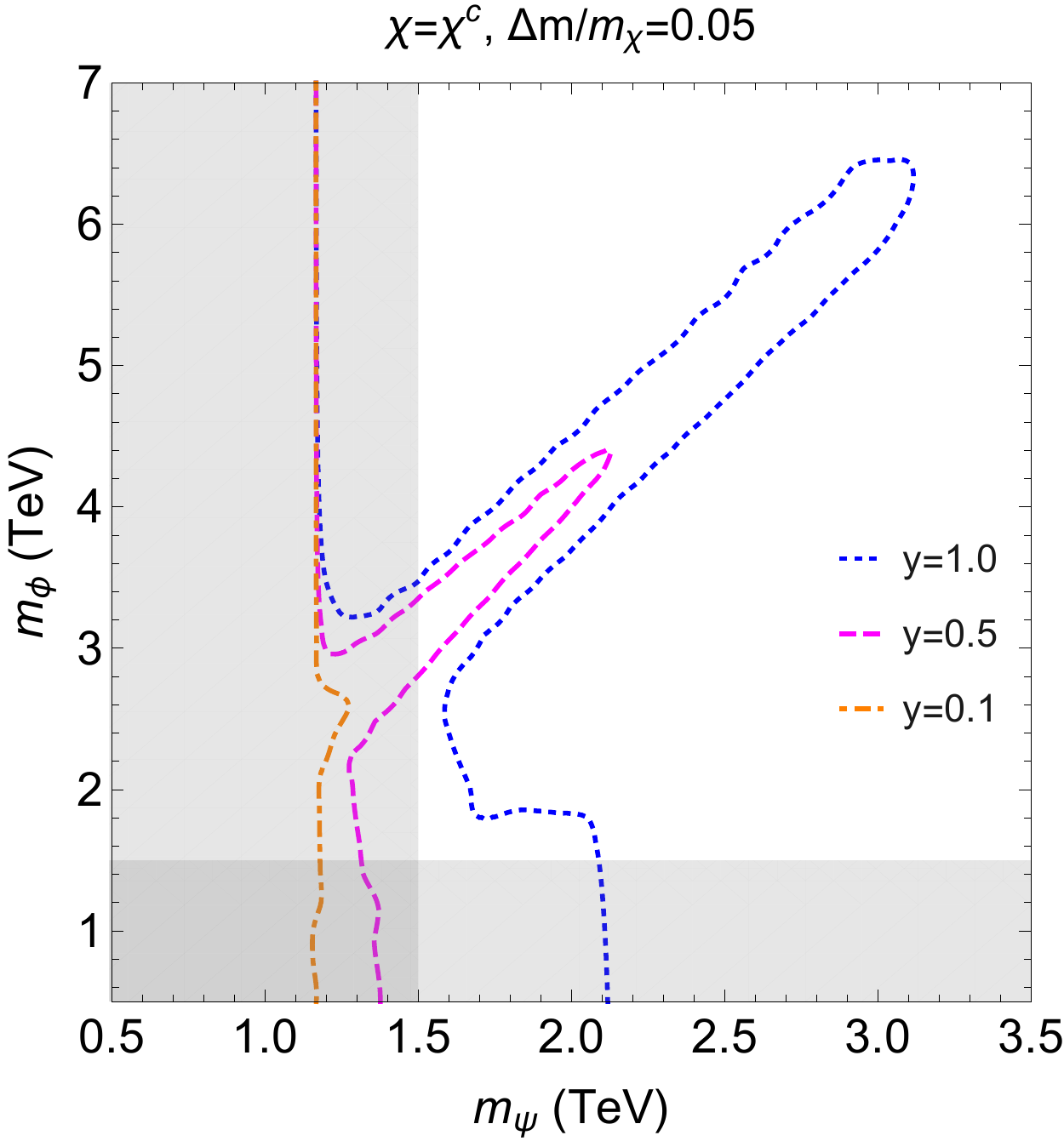}
    \includegraphics[scale=0.55]{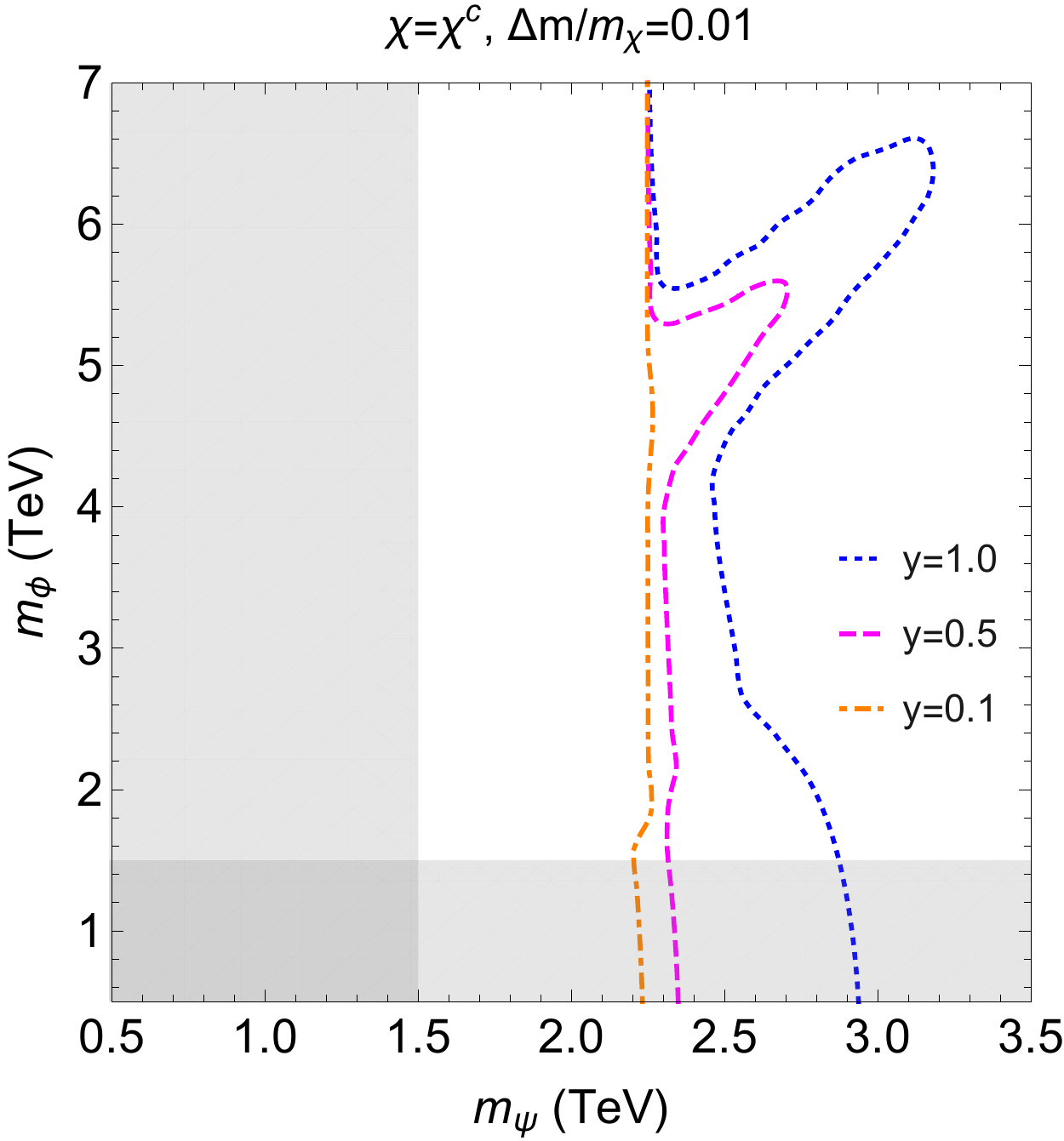}
    \\
    \vspace{0.5 cm}
     \includegraphics[scale=0.55]{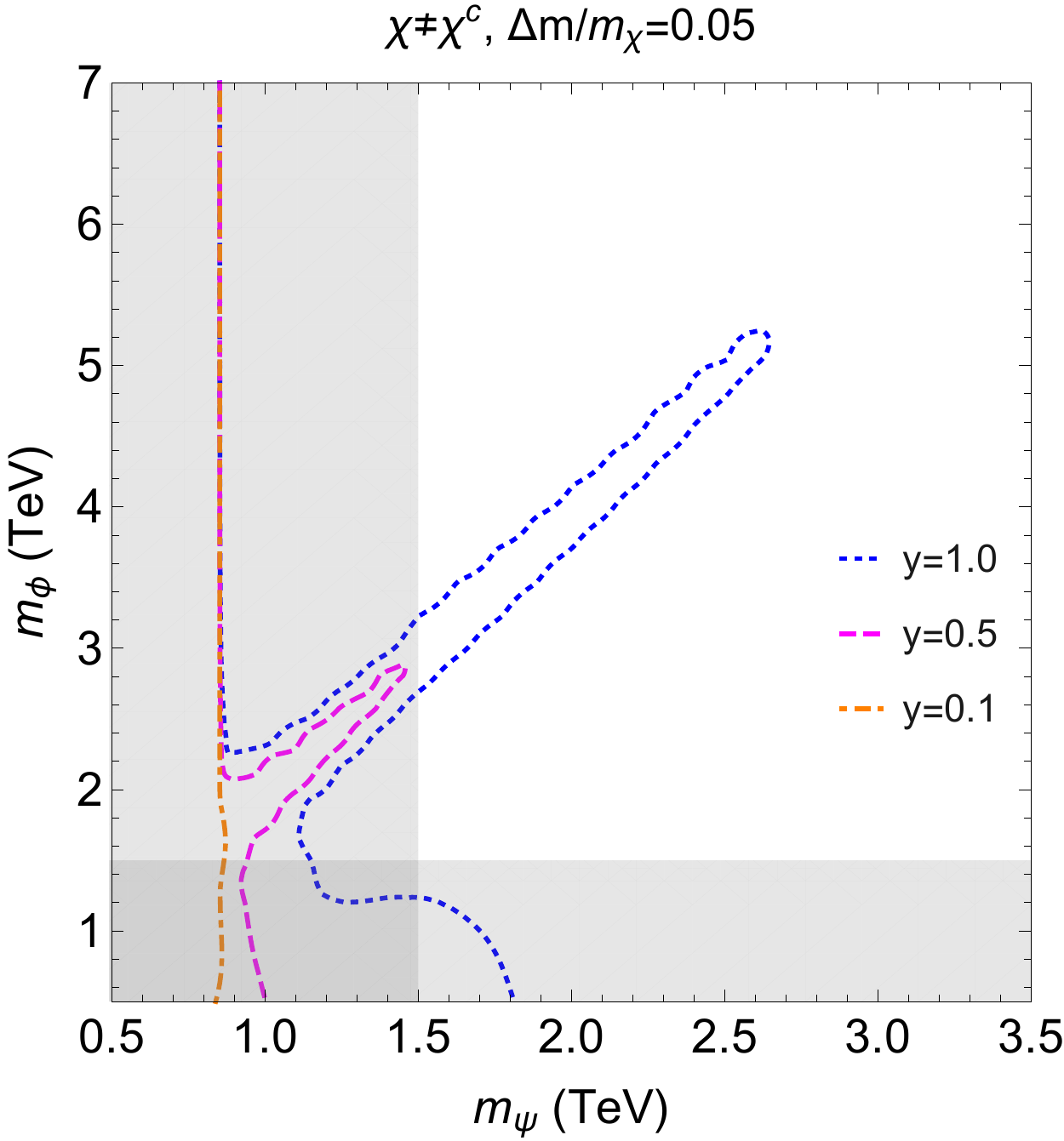}
    \includegraphics[scale=0.55]{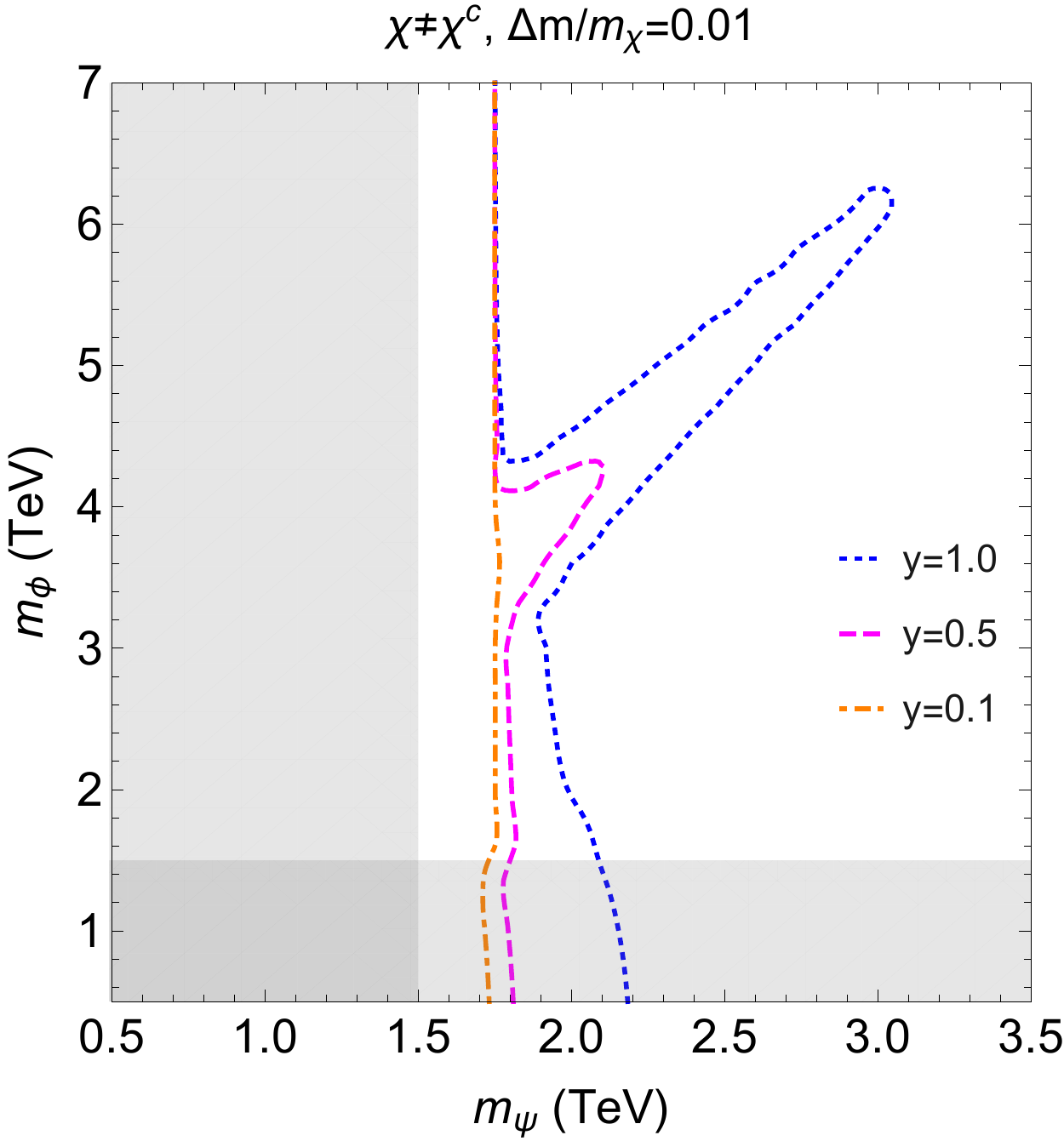}
    \caption{Curves in the parameter space $(m_\psi,m_\phi)$ that reproduces the observed DM energy density for $y=0.1,0.5,1.0$, and for the Majorana and Dirac DM options respectively in the upper and lower panels. Two benchmark values for relative mass splitting are chosen, $\Delta m/m_\chi=0.05$ and  $\Delta m/m_\chi=0.01$; the leptoquark-to-SM couplings are $y_L=y_R=1$. The gray shaded area are the LHC limits  on the masses of the  DVLF  and leptoquark of our model (for simplicity, here we quoted the most stringent bounds, which can be relaxed depending on the details, as explained later in the text).   }
    \label{fig:Omega_Mpsi_Mphi_Majorana_a}
\end{figure}

In figure~\ref{fig:Omega_Mpsi_Mphi_Majorana_a} we provide the curves that reproduce the observed energy density for Majorana (upper row) and Dirac (lower row) dark matter. Let us discuss the first scenario. We fix the relative mass splitting to $\Delta m / m_\chi = 0.05$ in the left panel and to $\Delta m / m_\chi = 0.01$ in the right plot. We select three benchmark values for the portal coupling to be $y=0.1, 0.5, 1.0$. The qualitative difference between the left and right panels of figure \ref{fig:Omega_Mpsi_Mphi_Majorana_a} can be explained as follow. The mass splitting $\Delta m/m_\chi =0.05$ makes the role of $\psi \psi $, $\bar{\psi} \bar{\psi} $  and $\psi \bar{\psi}$ annihilations less important than the case with $\Delta m/m_\chi =0.01$. Accordingly, in order to reproduce the same energy density, higher dark matter mass, and hence $m_\psi$, are required. As a result the curves are shifted towards larger DVLF masses, irrespective of the portal coupling $y$. Another aspect worth explaining is the behaviour with varying $y$. The effect is well visible in both panels of figure \ref{fig:Omega_Mpsi_Mphi_Majorana_a} (upper row). By increasing the value of $y$, it corresponds to a larger cross section of the resonantly-enhanced annihilation $\psi \chi \to q^c \bar{\ell}/Q^c\bar{L}$, with the leptoquark in the $s$-channel. For $m_\chi \approx m_\psi \approx m_\phi/2$, the enhancement demands large dark matter and DVLF masses in order to reproduce the observed energy density. The effects gradually fades away for decreasing $y$. For large $m_\phi$ masses, the contours merge into straight vertical lines independent of $y$, which signals that the energy density is determined by DVLF annihilations into SM quarks and gluons. The gray shaded areas implements the relevant collider exclusion limits, $m_\phi, m_\psi \leq 1.5$ TeV. For $\Delta m/m_\chi =0.05$, a good portion of the cosmologically favoured parameter is probed and ruled out for small $y$'s. The surviving regions are those along the resonant condition, $m_\psi>m_\phi$, and for large Yukawa-portal coupling $y=1$. For the smaller relative mass splitting $\Delta m /m_\chi 0.01$, the main constraint comes from the leptoquark exclusion limit, since the required DVLF masses are $m_\psi \gtrsim 2.2$ TeV and are out of the reach of current collider limits.

The corresponding parameter space for the Dirac dark matter option is displayed in figure~\ref{fig:Omega_Mpsi_Mphi_Majorana_a} (lower row). One find the main same qualitative features as for the Majorana case in figure~\ref{fig:Omega_Mpsi_Mphi_Majorana_a}. However, since the total annihailtion cross is smaller in the Dirac case, the parameter space compatible with the observed DM energy density shifts to smaller $m_\psi \approx m_\chi$ masses. As a result, for the larger relative mass splitting $\Delta m / m_\chi=0.05$, only the case $y=1.0$ remains still viable, and only along the resonant region. For the smaller splitting, the stronger coannihilations also make the DVLF masses out of the present collider limit, though in a less severe way with respect to the Majorana option.

\section{Correlated observables, dark matter and collider phenomenologies}
\label{sec:pheno}
In this section, we discuss the interplay among different observables as well as summarize the collider constraints on the masses and the relevant Yukawa couplings. 

\subsection{Dark matter direct and indirect detection prospects}
The DM fermion $\chi$ can interact with the nucleon constituents only via loop processes, where  the DVLQ and the leptoquark run in the loops. Following ref.~\cite{Mohan:2019zrk}, we calculate the spin-independent cross-section $\sigma_{\mathrm{SI}}$ for the Majorana (Dirac) DM option at one-loop, and we find that the typical cross section are beyond the current, and most likely future, sensitivities. For example,   
for the benchmark point satisfying the correct relic abundance as given in figure~\ref{fig:Majorana_Mchi_versus_OmegaDM_a}, namely $m_{\chi}=5\,(3.5)$ TeV, $\Delta m/m_{\chi}=0.001$, $m_{\phi}=1.5$ TeV, the corresponding cross-section is $\sigma_{\mathrm{SI}}=2.6\,(4.7)\times 10^{-51}\,\mathrm{cm}^{2}$, which  is not only several orders of magnitude below the current experimental limit, $1.45\times 10^{-45}\,(9\times 10^{-46})\,\mathrm{cm}^{2}$ \cite{LZ:2022ufs}, but also below the neutrino coherent scattering floor, $\sigma_{\nu N}\sim 10^{-48}\,\mathrm{cm}^{2}$ \cite{Strigari:2009bq}. Therefore, both Majorana and Dirac dark matter candidates are beyond the reach of direct detection experiments in the foreseeable future. 

An additional potential signal is given by the present-day annihilations of DM particles from astrophysical sources, that can be searched with indirect detection strategies. In our case, Majorana or Dirac DM can leave an imprint via annihilation into leptoquark pairs and their subsequent decays into the SM quarks and leptons at the tree-level, namely $\chi\chi (\chi \bar{\chi})\rightarrow \phi\phi^{*}\rightarrow q\, \ell \,\overline{q}\,\overline{\ell}$, as well as via loop-induced processes, i.e.~$\chi\chi(\chi \bar{\chi})\rightarrow\gamma\gamma,\,g g,\,Z\,Z,\,\gamma\, Z$. Here, we simply give an estimation of the cross section for dark matter being heavier than the leptoquark. In this case, the tree-level DM annihilation into four-body final states via the decay of unstable leptoquarks is not phase-space suppressed, at variance with the off-shell region $m_\phi > m_\chi$, whereas the loop-processes remain suppressed.  For Majorana DM, the corresponding pair annihilation into leptoquarks is p-wave suppressed (cfr.~eq.~\eqref{chi_chi_ann_exp}) and, therefore, the present-day annihilation rate of Majorana DM will be several orders of magnitude smaller than the case of Dirac DM. We then focus on the latter option in the following.

\begin{figure}[t!]
\centerline{\includegraphics[width=0.51\textwidth]{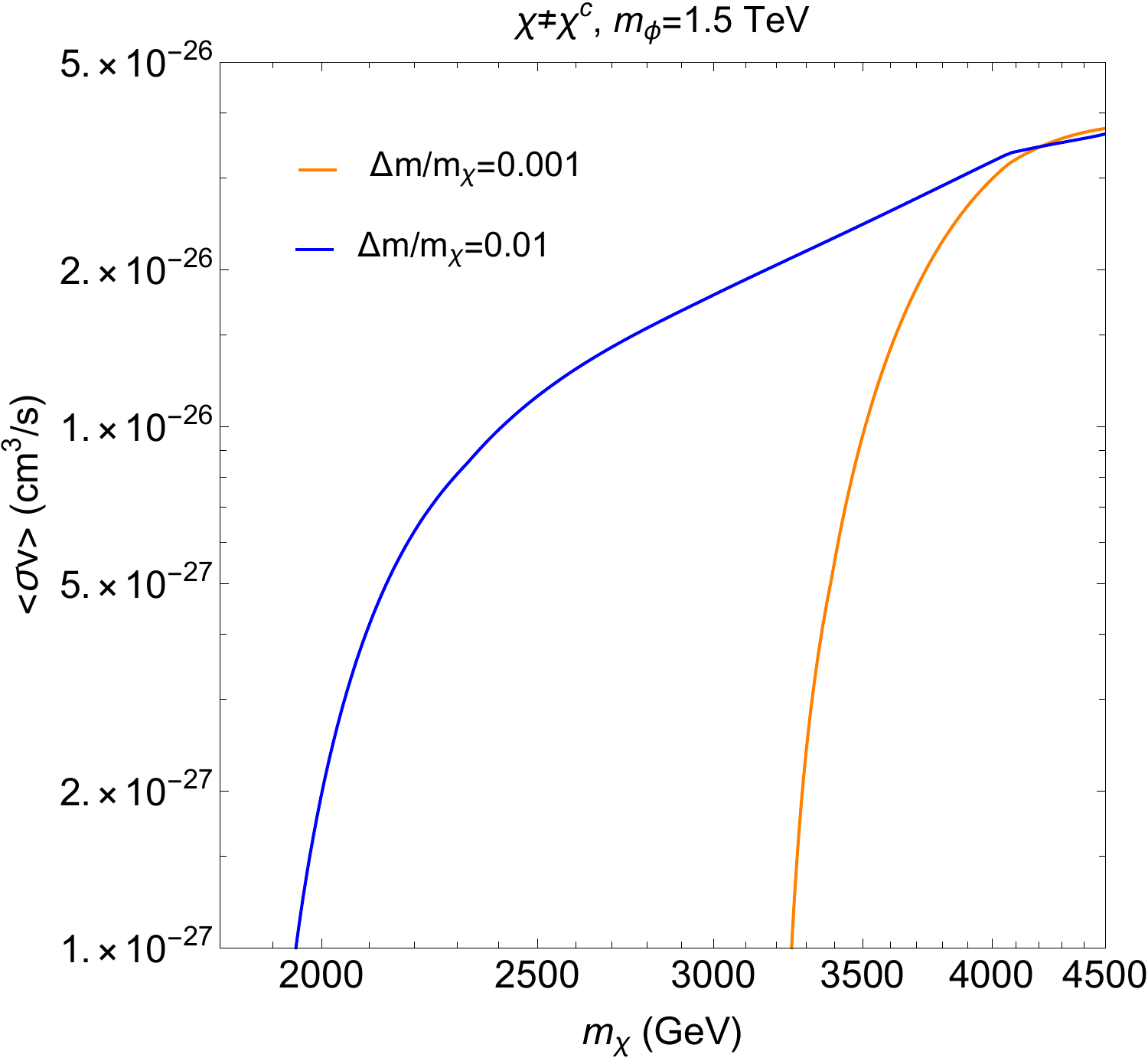}\hspace{1mm}
\includegraphics[width=0.47\textwidth]{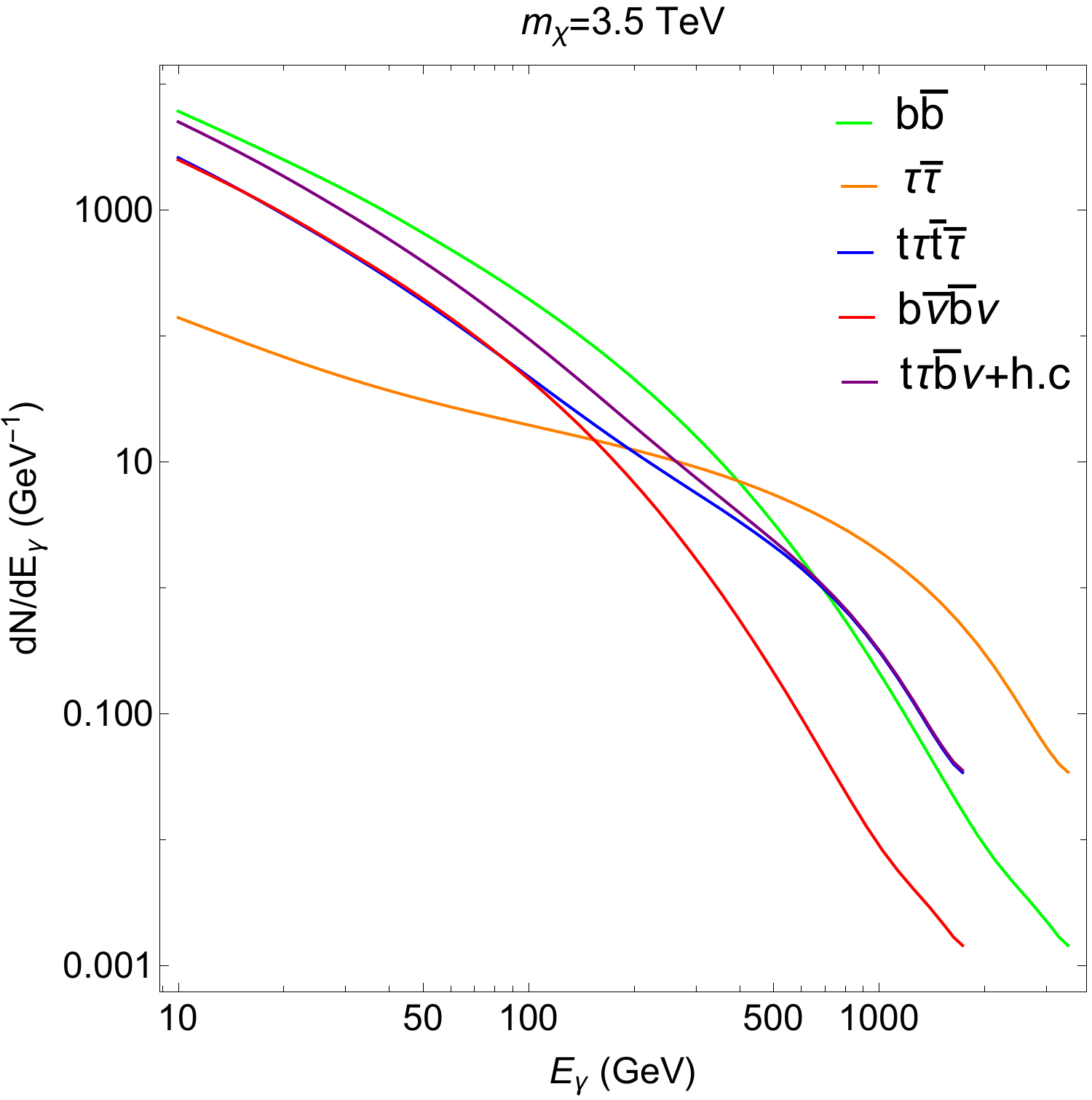}}
\caption{Left panel: The Dirac DM annihilation cross-sections into leptoquarks, $\chi\overline{\chi}\rightarrow\phi\phi^{\dagger}$ with respect to the DM mass, $m_{\chi}$ for the leptoquark mass, $m_{\phi}=1.5$ TeV and two mass-splittings between DM and DVLQ, $\Delta m/m_{\chi}=0.001$ and $0.01$, respectively, where all satisfy the correct relic abundance. Right panel: Photon spectra generated by different two-body and four-body final SM states, calculated using PPPC4DMID \cite{Cirelli:2010xx}.} \label{fig:anncross}
\end{figure}

In figure~\ref{fig:anncross} (left), we present the present-day Dirac DM annihilation cross-section into on-shell leptoquarks for a fixed value $m_{\phi}=1.5$ TeV and for $(m_{\chi},\,y)$ pairs that give the correct relic abundance. Two relative mass splittings are considered. Given that DM fermion masses lie in the TeV range, very high energy (VHE) gamma-rays can be expected from the energetic SM final state particles as produced from DM annihilation in typical DM-rich environments like the Galactic Center (GC) and Dwarf Spheroidal (dSph) galaxies. For DM in the mass range of 1-10 TeV, the current combined limits from VHE gamma-rays searches at 20 dSph galaxies by the Fermi-LAT, HAWC, H.E.S.S., MAGIC, and VERITAS experiments \cite{Hess:2021cdp} on the DM annihilation are in the range $1.2\times 10^{-25}-3.3\times 10^{-24}\,\mathrm{cm}^{3}/s$ for the $b\overline{b}$ final state, and $4.4\times 10^{-25}-4.3\times 10^{-24}\,\mathrm{cm}^{3}/s$ for the $\tau^+\tau^-$ final state, respectively.
Moreover, the future sensitivities of the upcoming Cherenkov Telescope Array (CTA) on the $b\overline{b}$ and $\tau^+\tau^-$ final states for the Galactic Center with Einasto DM profile and an observation time of 525 hr \cite{CTA:2020qlo}, are $(1.4-2)\times 10^{-26}\,\mathrm{cm}^{3}/s$ and $(3-9)\times 10^{-26}\,\mathrm{cm}^{3}/s$, respectively, for 1-10 TeV DM mass. 

Despite the annihilation cross-section curves cover the same order of magnitude of the current combined limits or future sensitivities, we stress that the experimental limits are derived for two-body final states. In our case, DM annihilating to leptoquark pairs produces in turn four-body final states consisting of two quarks and two leptons (accordingly we refrain from superimposing the experimental limits with the cross section curves of the model). As we can see from figure~\ref{fig:anncross} (right), for a banchmark value $m_{\chi}=3.5$ TeV, the end-point energy and photon spectra differ for the four-body final states, $t\,\tau\,\overline{t}\,\overline{\tau}$, $b\overline{\nu}\overline{b}\nu$, and $t\,\tau\,\overline{b}\,\nu\,+\mathrm{h.c}$ compared to the two-body final states, $b\overline{b}$ and $\tau^+\tau^-$. Therefore, a dedicated statistical analysis is required to quantitatively derive the limit from the current experiments or the sensitivity of the future experiments on the Dirac DM annihilation of the model, and assess the indirect detection prospects.

\subsection{Muon anomalous magnetic moment}
\label{sec:muon_gminus2_pheno}
Lepton flavor universality is not a fundamental property of nature. New physics can, in principle, couple more strongly to a specific fermion generation.
In fact, there is a longstanding tension in the muon anomalous magnetic moment, $a_\mu=(g-2)_\mu/2$. This discrepancy was measured at the E821 experiment~\cite{Muong-2:2006rrc} in 2006, which was recently confirmed by the E989 experiment~\cite{Muong-2:2021ojo}. The combined result yields a $4.2\sigma$ discrepancy with the SM prediction,
\begin{align}
\Delta a_\mu=a^\mathrm{exp.}_\mu-a^\mathrm{SM}_\mu= (2.51\pm 0.59)\times 10^{-9},    
\end{align}
hinting towards physics beyond the Standard Model that violates lepton flavor universality. Interestingly, the example model that we have discussed in the previous sections contains a scalar LQ, $\phi\sim (\overline 3,1,1/3)$, and it can address the tension in the muon anomalous magnetic moment (for an incomplete reference list, see, for example, refs.~\cite{Kowalska:2018ulj,Bigaran:2020jil,Dorsner:2020aaz,Khasianevich:2023duu}). Since $\phi$ couples to both the left- and the right-handed up-type quarks, it is possible to have a chiral enhancement in the loop to provide adequate new physics contributions to $(g-2)_\mu$.

To compute the $(g-2)_\mu$, we work in the up-type quark mass diagonal basis (for details, see ref.~\cite{Dorsner:2020aaz} and references therein), where the CKM matrix is associated with the down-type quarks. In this basis, the Yukawa couplings of the LQ, cfr.~eq.~\eqref{LagS1}, take the following form: 
\begin{align}
\mathcal{L}=& (-V^Ty^L)_{ij} \overline{d^c_L}_i \phi^{1/3} \hat{\nu}_{Lj} 
+ (y^L)_{i j} \overline{u^c_L}_i \phi^{1/3} \ell_{Lj} + y^R_{ij}\overline{u^c_R}_i \phi^{1/3} \ell_{R j} + \text{h.c.} \, , \label{eq:lagSS1}
\end{align}
where $V$  represents the CKM matrix; for its entries, we use the PDG values~\cite{ParticleDataGroup:2020ssz}. 
With the couplings  as given above, additional contributions to the $(g-2)_\mu$ are generated at the one-loop level, which can be expressed as follows~\cite{Dorsner:2020aaz}:
\begin{align}
&\Delta a_\mu=-\frac{3 m^2_\mu}{8\pi^2 m^2_\phi}
\sum_q\left[ 
\frac{m_q}{m_\mu} Re\left( y^L_{q\mu}y^{R,*}_{q\mu} \right)
\left( \frac{7}{6}+\frac{2}{3}\ln x_q \right) - \frac{1}{12} \left( |y^R_{q\mu}|^2+|y^L_{q\mu}|^2   \right) 
\right], \label{eq:g-2}
\end{align}
here we have defined, $x_q=m^2_q/m^2_\phi$, and the index $q$ runs over $u,c,t$-quarks.  A sufficient new physic contribution to the $(g-2)_\mu$ can only be provided if a top-quark or charm-quark mass flip occurs inside the loop. Due to the very small mass, the contribution from the up-quark can be fully neglected. This is why we examine two separate scenarios, (i) one with   
$y^{L,R}_{t\mu}\neq 0$ (top-quark mass flip) and (ii) another with $y^{L,R}_{c\mu}\neq 0$ (charm-quark mass flip).  
Moreover, in eq.~\eqref{eq:g-2}, the first term corresponds to the chirality-flipping contribution, hence it dominates over the second term, which can be safely neglected. Then, the $(g-2)_\mu$  becomes approximately proportional to the ratio $y^{L}_{q\mu} y^{R}_{q\mu} m_q/m^2_\phi$. Since $m_t/m_c\sim 136$, with $y^{L}_{q\mu} y^{R}_{q\mu}\lesssim \mathcal{O}(1)$, the experimentally measured deviation in the muon anomalous magnetic moment can be incorporated for $m_\phi \lesssim 5$ TeV and $m_\phi \lesssim 60$ TeV for the charm- and top-quark scenarios, respectively.   

\begin{figure}[t!]
\centering
\includegraphics[width=0.47\textwidth]{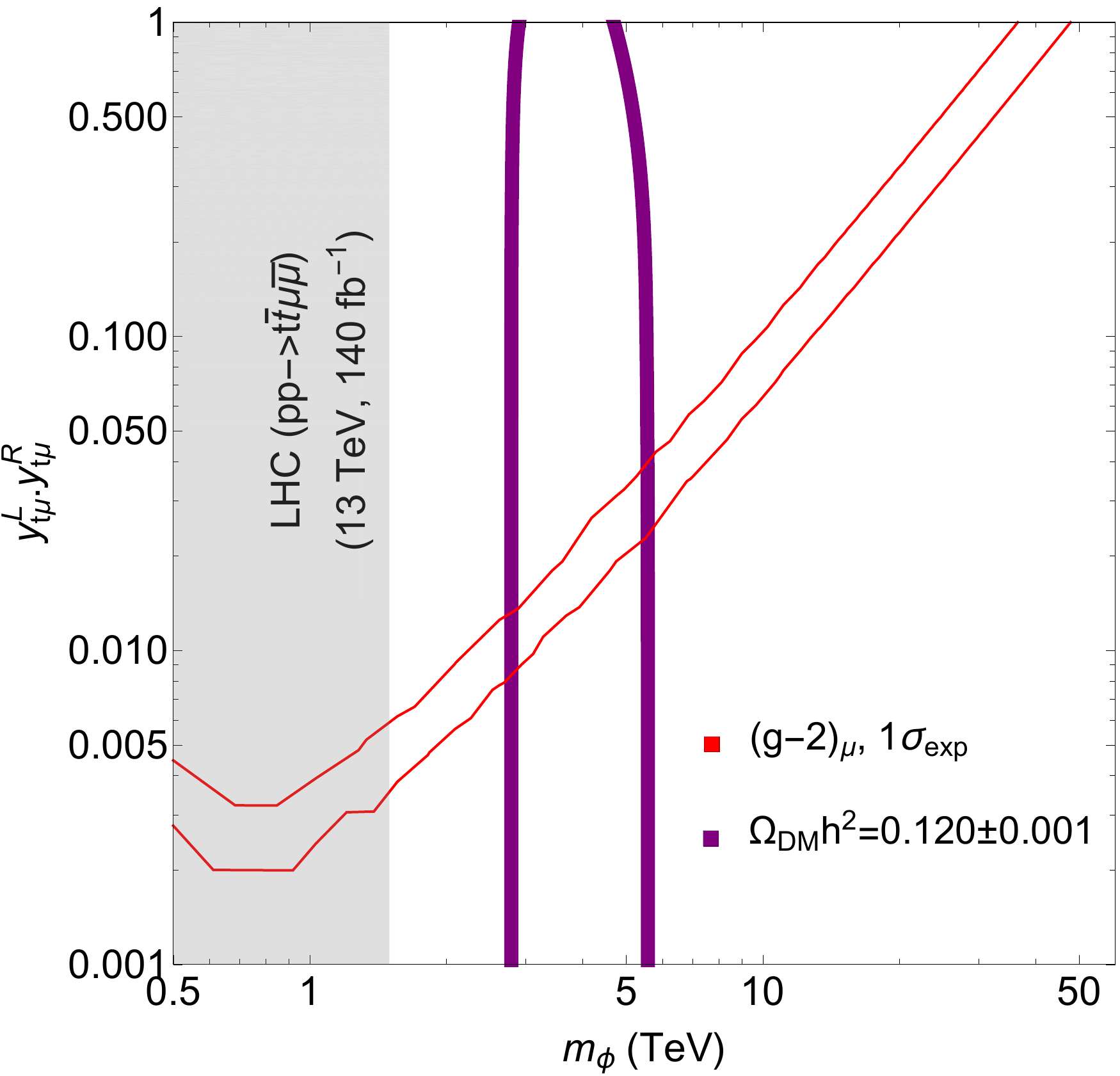}
\caption{Region inside the red lines ($1\sigma$) corresponds to the required new-physics contributions to the $(g-2)_\mu$ via top-quark mass flip in the loop in the example model with $\phi=(\overline 3,1,1/3)$. For simplicity, here we quoted the most stringent collider bound (gray-shaded area), which can be relaxed depending on the model/analysis details, see discussion in the text. The purple band gives the observed DM energy density for $m_\chi=2.5$ TeV and $\Delta m /m_\chi=10^{-2}$.} \label{fig:g-2:top}
\end{figure}

Note that due to the appearance of the CKM matrix in eq.~\eqref{eq:lagSS1}, all three generations of down-type quark couple to the (muon) neutrino. Depending on the scenario, additional interactions of these types may lead to uncontrollable flavor-violating processes (for example, an additional interaction with the electron may mediate dangerous $\mu\to e \gamma$ processes~\cite{Dorsner:2020aaz}). Owing to the requirement of small couplings for TeV scale LQ in the top-quark mediated scenario,  flavor-violating processes (for example, $B\to K^{(*)}\nu\overline \nu$) are well under control. This scenario with top-quark mass flip is illustrated in   figure~\ref{fig:g-2:top}. 

\begin{figure}[th!]
\centering
\includegraphics[scale=0.4]
{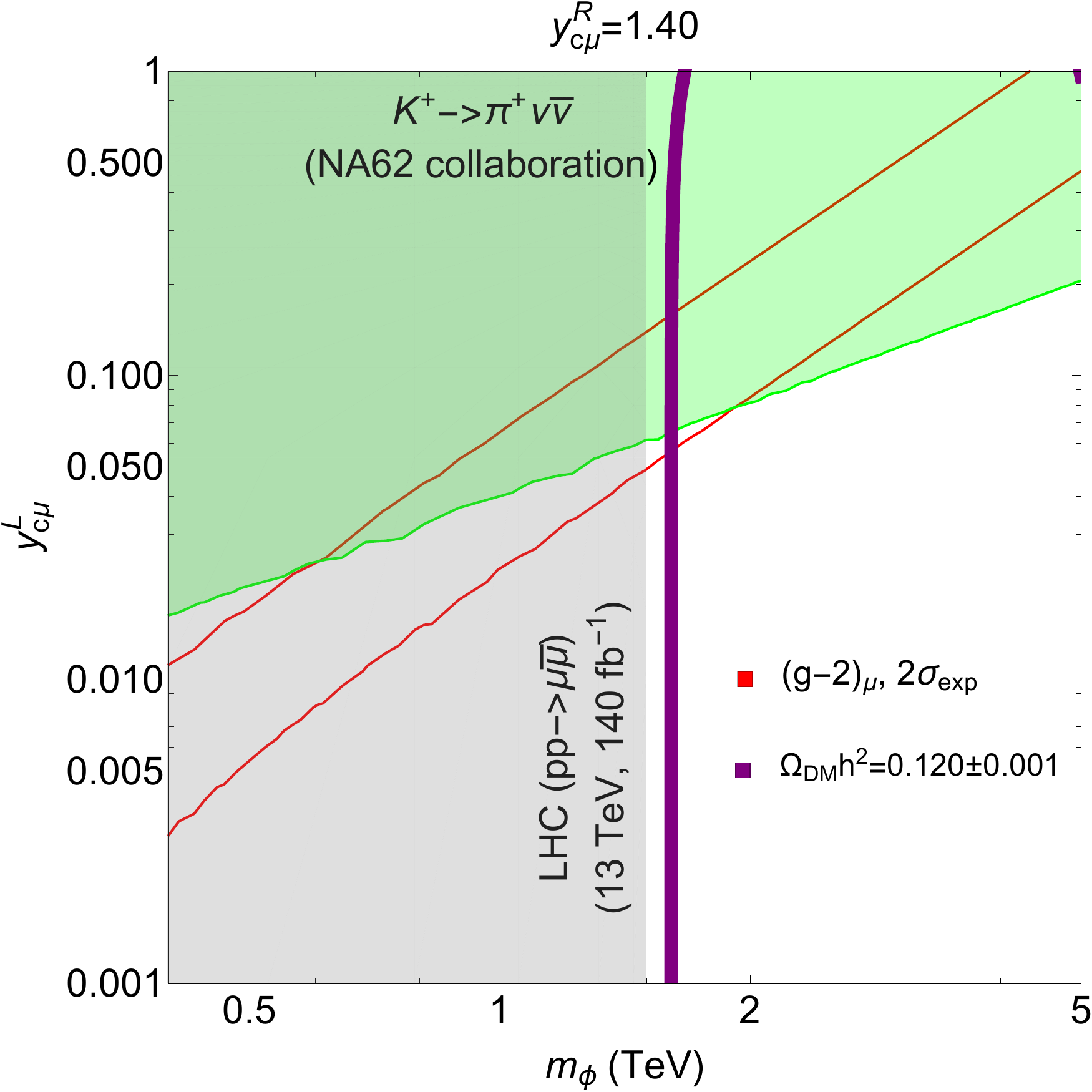}
\includegraphics[scale=0.2875]
{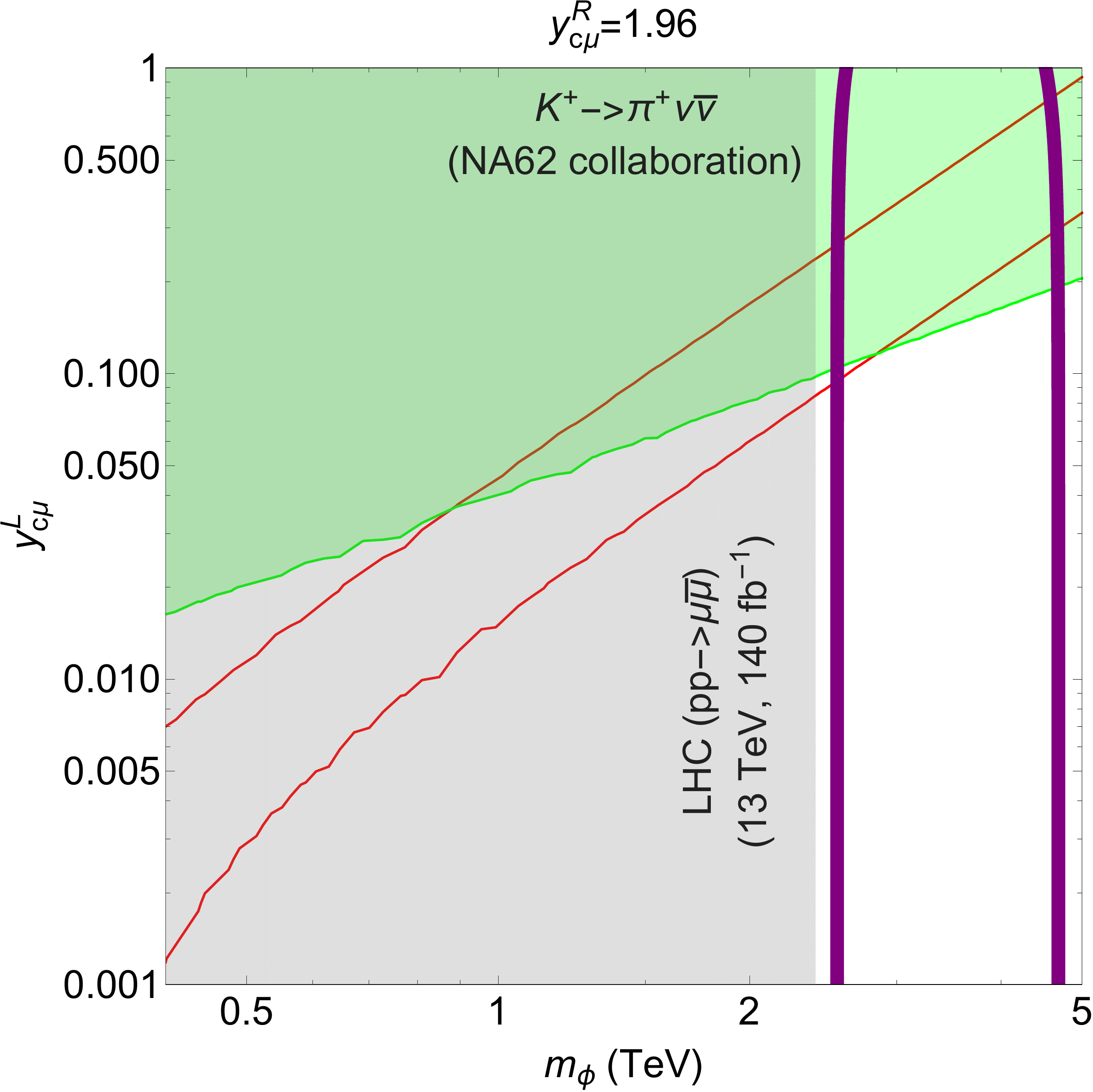}
\caption{Region inside the red lines corresponds to the required new-physics contributions to the $(g-2)_\mu$ via charm-quark mass flip in the loop in the example model with $\phi=(\overline 3,1,1/3)$.  The green-shaded area corresponds to the experimental result from NA62~\cite{NA62}, whereas the gray-shaded area to non-resonant dilepton searches at the LHC \cite{ATLAS:2020zms,CMS-PAS-EXO-19-019}. The purple band gives the observed DM energy density.} \label{fig:g-2:charm}
\end{figure}

The charm-quark mediated case, however, suffers from large flavor-violating processes. This is due to the requirement of somewhat larger couplings.  The scalar LQ couples to strange- and down-quark with almost the same strength, where the latter coupling is Cabbibo suppressed. As a result, the leptoquark mediates kaon decays of the type $K^+\to \pi^+\nu\overline \nu$,  which rules out a large part of the parameter space as shown in figure~\ref{fig:g-2:charm} (green-shaded area). The experimental result from NA62~\cite{NA62} that corresponds to $\textrm{BR}\left(K^+\to \pi^+\nu\overline \nu\right) < 1.85\times 10^{-10}$ puts strong constraint only on $y^{L}_{c\mu}$ for our scenario, which we compute following Ref.~\cite{Mandal:2019gff}.  Therefore, in order to get the correct $(g-2)_\mu$ value as observed in the experiments, a large $y^{R}_{c\mu}$ is typically required. Interestingly, such a large value of $y^{R}_{c\mu}$ is also highly constrained~\cite{Angelescu:2021lln,Julio:2022bue} from non-resonant dilepton searches at the LHC~\cite{ATLAS:2020zms,CMS-PAS-EXO-19-019}, to be discussed below, see section~\ref{sec:collider_constraints}. Once these two constraints are imposed, a tiny portion of the parameter space remains consistent with the experimentally observed $(g-2)_\mu$ only at the $2\sigma$ C.L. as depicted in figure~\ref{fig:g-2:charm}.   

In order to highlight the interplay between the explanation of the observed values $(g-2)_\mu$ and dark matter relic density, we superimpose the curves that reproduce $\Omega_{\textrm{DM}}h^2 = 0.1200\pm 0.001$ in figures~\ref{fig:g-2:top} and \ref{fig:g-2:charm}.To this aim, we consider the Majorana dark matter option.  As regards the charm case, the quite small available parameter space demands a careful choice of the DM mass once the portal coupling $y$ and the mass splitting are fixed; we choose $y=1$ and $\Delta m /m_\chi=10^{-2}$. For the right-handed Yukawa value $y^R_{c \mu}=1.40$ ($y^R_{c \mu}=1.96$), we find the corresponding DM mass to be $m_\chi=2.84$ (2.64) TeV in order to lie in the still viable window. Changing the DM mass more than about $5\%$ is sufficient for loosing the interplay (one could however tune again the mass splitting and $y$). The top quark scenario is much less restrictive in this respect and we show an exemplary case, which lies beyond the most stringent LHC limit but it is still in the TeV range of the leptoquark mass. Here, the DM is $m_\chi=2.5$ TeV, whereas we keep the same values for $y$ and $\Delta m /m_\chi$ as in the charm case.

Before concluding the discussion about the muon anomalous magnetic moment, we point out that recent lattice determinations~\cite{Borsanyi:2020mff,Ce:2022kxy,Alexandrou:2022amy} of the hadronic vacuum
polarization give a SM prediction that  agrees with the experimental result, however, it is in tension
with the previous calculations based on dispersive methods~\cite{Borsanyi:2020mff}. Forthcoming experiments will be able to shed light on this unresolved issue.

\subsection{Collider implications}
\label{sec:collider_constraints}
\textbf{LHC constraints:}-- Since the DM in our framework is a SM singlet, it cannot be directly produced at the LHC. However, the corresponding productions of the LQ and the DVLF are unsuppressed since they carry color charge. Relevant bounds on these masses from LHC searches are discussed in the following.

Leptoquarks can be pair produced at LHC via gluon-fusion $pp\to \phi\phi^\dagger$~\cite{Diaz:2017lit,Dorsner:2018ynv}. After production, each of these LQs would decay into a quark and a lepton. Several dedicated searches for LQ pairs have
been carried out by  ATLAS and CMS Collaborations for different final states with ($pp\to q\overline q \nu \overline \nu$) or without ($pp\to q\overline q \ell \overline \ell$) neutrinos. The LHC limits
on LQ mass depend on the branching ratios of various
modes.  For the top-quark mass flip solution presented above, the strongest (weakest) constraints arise if the value of $y^R_{t\mu}$ ($y^L_{t\mu}$) is somewhat larger than $y^L_{t\mu}$ ($y^R_{t\mu}$). In this case, branching ratio to $pp\to t \overline t\mu\overline \mu$ is about unity (a half) and LHC excludes LQ masses below 1.5 (1.3) TeV~\cite{ATLAS:2020xov,Julio:2022bue}.

Note that when kinematically allowed,  in addition to the $\phi\to q\ell$, there is an additional decay mode, namely,  $\phi\to \psi\chi$ (see eq.~\eqref{leptoquark_width}). As a result, the branching ratio to $q\ell$  gets modified and the bounds quoted above will be relaxed (for details, see Refs.~\cite{Baker:2015qna,Belanger:2021smw}). For pair produced LQs, a dedicated search has been performed at the LHC, where one of the LQs directly decays to jet and a muon ($\phi\to j\mu$) and the other LQ decays to $\phi\to j\mu+\slashed{E}_T$ with low-$p_T$ SM fermions via cascade decays ($\phi\to \psi\chi\to \phi^*\chi\chi\to j\mu\chi\chi$). The analysis strategy is based on the search of a peak in the LQ  invariant mass $m_{j\mu}$ distribution from the highest $p_T$ muon and jet in an event,
with the requirement of significant missing transverse momentum due to the DM particles in the final state. Since no signals above the SM background is observed, from this search, LHC rules out dark matter masses up to 600 GeV for LQ masses of order $\mathcal{O}(1.5)$ TeV~\cite{CMS:2018yiq}.

On the other hand, for the charm-quark mass flip solution, 
the most relevant LHC bound comes from the indirect high-$p_T$ searches~\cite{Eboli:1987vb}. We are interested in   the direct constraint on the coupling versus mass plane arising from the non-resonant dimuon searches at the LHC ($pp\to \mu\overline \mu$). As discussed above, we require somewhat large values of $y^R_{c\mu}$ to address the tension in the muon magnetic moment. For a LQ of mass 1 TeV, non-resonant dimuon search rules out couplings of order unity, i.e.,  $y^R_{c\mu}\leq 1.1$~\cite{Angelescu:2021lln} must be satisfied. Since the bound on the coupling depends on the mass of the LQ, this functional dependence is presented in figure~\ref{fig:g-2:charm} with varying $m_\phi$ for the two different coupling choices (gray shaded area).

DVLFs are also pair-produced at the LHC via gluon-fusion $pp\to \psi\overline \psi$. Subsequently, each DVLF decays to $\psi\to \chi\;\phi\to \chi \;q\;\ell$ leading to large MET.   Processes of these types have been searched for at the LHC that put strong bounds on the lower limit of squark masses, especially for stop and sbottom.  Depending on the exact LQ coupling and mass, as well as the value of  $\Delta m=m_\psi-m_\chi$, the bound is in between $m_\psi\gtrsim 700$ GeV and $m_\psi\gtrsim 1300$ GeV (for details, see Ref.~\cite{Belanger:2021smw}). This analysis, however, is not applicable for very small mass splitting. In such a compressed scenario, missing energy searches lose sensitivity, and conventional searches are no longer applicable. This happens typically for mass splittings $\Delta m < 5$ GeV~\cite{Schwaller:2013baa}. In fact, if the mass splitting is very small,   the DVLF becomes effectively long-lived as a result of a highly off-shell LQ. The phenomenology of DVLFs is entirely different from the one in the standard searches of these particles. These quasi-stable heavy quarks, namely, the R-hadrons~\cite{Farrar:1978xj} interact hadronically as they move through the
detector after being produced at the LHC.  The recent analysis of the ATLAS collaboration puts bounds on the mass of long-lived supersymmetric R-hadrons (squarks and gluinos), which for a state with electromagnetic charge $\pm 1/3$ (sbottom) is  1250 GeV~\cite{ATLAS:2019gqq}.  This search is quite model-independent and has been adapted to the case of
vectorlike fermion, see Ref.~\cite{Criado:2019mvu}, which finds a lower bound of $m_\psi\gtrsim 1500$ GeV.

It is important to point out that the exact LHC limits depend on the details of the multidimensional parameter space, which is beyond the scope of this work.

\textbf{Muon collider probes:}-- As discussed above, the observed large tension in the muon anomalous magnetic moment is an indication that the new physics couples strongly with the muon and not to the other lepton generations. Consequently, muon colliders are the perfect machines to test such muon-philic new physics scenarios~\cite{Buttazzo:2020ibd}.  As outlined above, the LQ can reside in the multi-TeV range, which is beyond the reach of LHC,  and yet provide the required new physics contribution to reproduce muon $g-2$ and play an important role in determining the dark matter relic abundance.  In such a scenario, by integrating out the heavy field in obtaining the effective field theory, one gets the scattering process $\mu\overline \mu\to c\overline c/ t\overline t$ depending on the charm-philic/top-philic nature of the LQ.  Then, a probe of $\Delta a_\mu$ is obtained via  computing the number of events and 
requiring a statistically significant deviation from the SM $\mu\overline \mu\to c\overline c/ t\overline t$ background. By performing extensive analysis  for the relevant  semi-leptonic
operator involving charm-quark (top-quark), ref.~\cite{Buttazzo:2020ibd} showed that muon-philic scenario  can be probed
already at $\sqrt{s}=1$ TeV, while the top-philic case can be probed at $\sqrt{s}=10$ TeV. Such a high-energy determination of $\Delta a_\mu$ is a unique feature of muon colliders.

\section{Conclusions}
\label{sec:conclusions}
In this work, we considered a class of dark matter models where the DM candidate does not interact directly with Standard Model particles. The sole interaction of the dark matter, a SM singlet fermion, is through a Yukawa coupling with a scalar mediator and a Dirac fermion. The latter is assumed to carry a dark charge and to be heavier than the DM particle to ensure the DM candidate's stability. In order to trigger DM annihilations in the early universe via thermal freeze-out, both the mediator and the fermionic partner carry some charges under the SM gauge group. We focused on the case of QCD colored states and the scalar mediator being a leptoquark. This setup provides interesting phenomenological consequences despite the absence of direct interaction between the dark matter and the visible sector—most notably, collider implications and a connection with the anomalous magnetic moment of the muon. 

The colored DVLFs play a crucial role for the annihilations of particles of the dark sector, that determine the relic energy density. Indeed, in order to avoid overclosing the universe, coannihilations of nearly degenerated colored partners have been shown to be a necessary ingredient for the model at hand.  In this regime, it is important to scrutinize relevant non-perturbative effects.  One of the  main objectives of this paper is to assess such effects for a more reliable estimation of the dark matter energy density.

Non-relativistic heavy DVLF pairs are affected by repeated soft-gluon exchange in two ways. First, above-threshold scattering states experience Sommerfeld effects. We show that they play a rather marginal role in the case of $\psi \bar{\psi}$ annihilations because of a competing enhancement and suppression in the attractive color-singlet and repulsive color-octet channels, that contribute at the same order in the velocity expansion (see eqs.~\eqref{cross_psi_psibar_QCD} and \eqref{psi_psibar_leptoquarks}). On the contrary, a more prominent effect is found for $\psi \psi$ and $\bar{\psi}\bar{\psi}$ annihilations because of an enhancing color-triplet Sommerfeld
factor. The corresponding cross section is not diminished by a suppression factor from the color-sextet repulsive channel, which only appears at higher order in the velocity expansion. This latter situation only applies to the Majorana dark matter option.  Second, repeated gluon-exchange in the attractive color-singlet and triplet channels, may sustain bound states. During the freeze-out in the early universe, bound-state formation for DVLF pairs and their subsequent decays into SM particles, works as an additional channel to effectively deplete dark sector particles. We take into account the bound-state formation process via gluon radiation. For the Majorana fermion scenario, we have obtained the bound-state formation cross section for the process $(\psi \psi)_{[6]}^n \to \mathcal{B}^n_{[3]} + g$ in the framework of pNRQCD, and computed the corresponding electric-dipole matrix elements in full generality (see appendix~\ref{sec:app_electri_dipole}). Our result can be also be useful for other simplified models that feature real scalar dark matter coannihilating with vector-like colored fermions, e.g. \cite{Giacchino:2015hvk,Belanger:2018sti}. 

We have assessed the impact of non-perturbative effects depending on the nature of the dark matter fermion, which is rarely pursued in the literature. As a general observation, Sommerfeld and bound-state effects are more relevant for the Majorana option. This is due to additional DVLF pair annihilation channels, namely $\psi \psi \to \phi \phi$ and the complex conjugate process. For this scenario, bound-state formation and decays from both color-singlet and color-triplet pairs boost the annihilations of the dark sector particles. The inclusion of non-perturbative effects, especially bound-state formation, has a sizeable impact on the dark matter mass that is compatible with the observed energy density. For the smallest mass splitting considered in this work, we find that $m_\chi$ is shifted from $3.4$ TeV to $5.6$ TeV (see figure~\ref{fig:Majorana_Mchi_versus_OmegaDM_a}). Moreover, as shown in figure~\ref{fig:Omega_Mphi_plane_a}, non-perturbative effects open new mass regions for the leptoquark, which would be deemed excluded otherwise.   Our findings motivate further investigations and a more comprehensive inclusion of bound-state effects for the models' class of this work (such as the complementary bound-state formation process via $2 \to 2$ scatterings with light plasma constituents and more excited states).  

Despite we have considered the freeze-out option in this work, some comments can be made on the conversion-driven freeze-out. Here, much smaller portal coupling $y$ are needed, which makes the Sommerfeld enhancement due to the attractive triplet channel, as well as the corresponding bound-state effects  practically irrelevant.  However, bound-state effects from $(\psi \psi)_{[8]}^n \to \mathcal{B}^n_{[1]} + g$, which are independent of the Yukawa coupling, can be relevant for estimating the thermal abundance of the DVLF partner both in the Dirac and Majorana option. We leave their inclusion in the conversion-driven freeze-out for future investigation on the subject.

Although dark matter direct and indirect detection is challenging in this setup, because of no coupling between the DM with the SM particles, the example model that we studied still has important phenomenological consequences. The mediator chosen is a scalar leptoquark, which can be directly searched for at colliders. For masses of order TeV, LHC already puts strong constraints on the LQ couplings. Interestingly, in addition to acting as a mediator between the visible and dark sectors, the leptoquark can also address the longstanding tension in the muon's anomalous magnetic moment. In such a scenario, leptoquark masses, even up to 100 TeV, can be probed in future muon colliders. Since the coannihilation partner must have identical quantum numbers as the mediator within this framework, it can also be efficiently produced in LHC and may leave exciting signatures. In view of the fact that coannihilations play a crucial role in achieving a dark matter abundance compatible with observations, which requires nearly degenerate states, detecting the coannihilation partner at LHC will already provide information about the mass of the dark matter.

\section*{Acknowledgements}
S.B. thanks Mikko Laine and Gramos Qerimi for useful discussions.
The work of S.B. is supported by the Swiss National Science Foundation (SNSF) under the Ambizione grant PZ00P2\_185783.

\appendix 
\numberwithin{equation}{section}

\section{$2 \to 2$ annihilation cross sections}
\label{sec:app_annihilations_channels}
In this section we provide the annihilation cross sections of the relevant process. Our main focus is on the cross sections that involve one or two massive leptoquarks in the final state. In this case, we give the exact expression without performing the velocity expansion, at leading order in the couplings. Each cross section is given as a function of the Maldestam variable $s$ and we perform the average over the degrees of freedom of the incoming states. Such quantity is the one that can be readily inserted in the standard form of the thermally average cross section~\cite{Gondolo:1990dk,Davidson:2008bu} 
\begin{eqnarray}
&&\langle \sigma_{ a b  \to c d } \, v_{\textrm{rel}} \rangle = \frac{1}{8 T \, m_a^2 m_b^2 \mathcal{K}_2\left( \frac{m_a}{T} \right)  \mathcal{K}_2\left( \frac{m_b}{T} \right)} 
\nonumber
\\
&& \hspace{4.8 cm} \times
\int_{s_{\textrm{min}}}^\infty ds \, s^{3/2} \, \lambda \left( 1, \frac{m_a^2}{s},  \frac{m_b^2}{s} \right) \mathcal{K}_1 \left( \frac{\sqrt{s}}{T} \right) \, \sigma_{ a b  \to c d }(s) \, ,
\nonumber
\\
\end{eqnarray}
where $T$ is the temperature, $\mathcal{K}_1(x)$ and $\mathcal{K}_2(x)$ are the modified Bessel function of the first and second kind, $\lambda(x,y,z)=(x-y-z)^2-4 x y z$ and $s_{\textrm{min}}=\textrm{max}[(m_a+m_b)^2,(m_c+m_d)^2]$. We remark that the analytical $2 \to 2$ cross sections have been checked against the model implementation in MadGraph \cite{Alwall:2014hca}. In the main body of the paper, in eq.~\eqref{co_cross}, we have abbreviated the cross sections by indicating only the incoming states $\sigma_{a b} v_{\textrm{rel}}$. 
\subsection*{$\chi \chi$ annihilations}
The pair annihilation cross section of Majorana dark matter fermions into leptoquark $\phi \phi^\dagger$ pairs is
\begin{eqnarray}
     \sigma_{\chi \chi \to \phi \phi^\dagger}(s) &=& \frac{y^4 N}{16 \pi \, s} \sqrt{\frac{s-4m_\phi^2}{s-4 m_\chi^2} } \left\lbrace \frac{1}{A^2-B^2} \left[ s \left( s- 4 m_\phi^2 \right) + 4  (m_\chi + m_\psi)^2 \left( s- 4 m^2_\chi \right)  \right. \right.
     \nonumber\\
  &&   \left. \left. -32 A m_\chi  (m_\chi + m_\psi) -12 A^2 + 8 B^2 \right] +  \frac{1}{AB}\left[8 A m_\chi (m_\chi + m_\psi)  +  A^2  \right. \right.
  \nonumber \\
 && \left. \left.  -s \left( s- 4 m_\phi^2 \right) + 4  (m_\chi + m_\psi)^2 \left( s- 4 m_\chi^2 \right) \right]\tanh^{-1} \left( \frac{B}{A}\right) \right\rbrace \, ,
\end{eqnarray}
where $N$ is the number of colors, and we have defined the following auxiliary quantities in order to write the cross section more compactly 
\begin{eqnarray}
    && A=2 m_\psi^2 + \frac{1}{2} \left( s-4 m_\phi^2 \right) + \frac{1}{2} \left( s- 4 m_\chi^2 \right) \, , 
     \\
     && B = \sqrt{s-4 m_\phi^2} \sqrt{s- 4 m_\chi^2} \, .
\end{eqnarray}
The pair annihilation cross section of Dirac dark matter fermions into $\phi \phi^\dagger$ pairs reads instead
\begin{eqnarray}
     \sigma_{\chi \bar{\chi} \to \phi \phi^\dagger}(s) &=& \frac{y^4 N}{32 \pi \, s} \sqrt{\frac{s-4m_\phi^2}{s-4 m_\chi^2} } \left\lbrace \frac{1}{A^2-B^2} \left[ s \left( s- 4 m_\phi^2 \right) + 4  (m_\chi + m_\psi)^2 \left( s- 4 m^2_\chi \right)  \right. \right.
     \nonumber\\
  &&   \left. \left. -8 A m_\chi  (m_\chi + m_\psi) -2 A^2 +  B^2 \right] +  \frac{1}{B}\left[4  m_\chi (m_\chi + m_\psi)  +  A  \right]\tanh^{-1} \left( \frac{B}{A}\right) \right\rbrace \, .
  \nonumber
  \\
\end{eqnarray}
Upon expanding in the non-relativistic velocity, which is a good approximation sufficiently away from the opening of the threshold, the cross section times the relative velocity reads
\begin{eqnarray}
\sigma_{\chi \chi \to \phi \phi^\dagger} v_{\textrm{rel}} &=& \frac{N \, y^4 v_{\textrm{rel}}^2}{16 \pi  (m_\psi^2+m_\chi^2-m_\phi^2)^4 }  \left[ m_\chi^2(m_\chi^2-m_\phi^2)^2 + m_\psi^4 (3m_\chi^2-2m_\phi^2) \right. 
\nonumber \\ 
&& \left.    +\frac{2}{3} m_\chi m_\psi (m_\chi^2-m_\phi^2)^2 + \frac{m_\psi^2}{3} (3 m_\phi^4 - 8 m_\phi^2 m_\chi^2 +5 m_\chi^4)
 \right. 
\nonumber \\
&& \left.  + \frac{8}{3} m_\chi m_\psi^3 (m_\chi^2-m_\phi^2) + 2 m_\chi m_\psi^5 + m_\psi^6 \right]\left( 1- \frac{m_\phi^2}{m_\chi^2}\right)^{\frac{1}{2}} \, , 
\label{chi_chi_ann_exp}
\end{eqnarray}
for the Majorana case,   whereas for the Dirac case we find 
\begin{eqnarray}
 \sigma_{\chi \bar{\chi} \to \phi \phi^\dagger} v_{\textrm{rel}} = \frac{N \, y^4 m_\chi^2}{16 \pi  (m_\psi^2+m_\chi^2-m_\phi^2)^2 } \left( 1- \frac{m_\phi^2}{m_\chi^2}\right)^{\frac{3}{2}} \,  \, .
 \label{chi_barchi_ann_exp}
\end{eqnarray}
We have kept here the corresponding leading terms in the velocity expansion.   
\subsection*{$\chi \psi$ annihilations}
In this case there are two class of processes: (i) annihilation processes into lepton and quark pairs via a s-channel leptoquark exchange; (ii) annihilations into a gluon and a leptoquark. In the latter case, the unstable leptoquark decays in turn into a lepton and a quark. 

The coannihilations into a right-handed lepton and quark reads
\begin{eqnarray}
\sigma _{ \chi \psi \to u^c \bar{\ell}}(s) = \frac{y^2 |y_R|^2}{32 \pi} \frac{ s-(m_\chi+m_\psi)^2}{\left[ (s-m_\phi^2)^2+ m_\phi^2 \Gamma_\phi^2\right] \sqrt{\lambda(1,m_\chi^2/s,m_\psi^2/s)}} \, ,
\label{resonant_right_handed}
\end{eqnarray}
whereas for final states left-handed SM quarks and leptons doublets, we find 
\begin{eqnarray}
\sigma _{ \chi \psi \to Q^c \bar{L}} (s) = \frac{y^2 |y_L|^2}{16 \pi} \frac{ s-(m_\chi+m_\psi)^2}{\left[ (s-m_\phi^2)^2+ m_\phi^2 \Gamma_\phi^2\right] \sqrt{\lambda(1,m_\chi^2/s,m_\psi^2/s)}} \, , 
\end{eqnarray}
where the factor of 2 simply originates from the SU(2) multiplicity. 

For process $\chi \psi \to \phi g $ the analytical expression of the cross section turns out to be quite lengthy. We list the squared amplitude of the $s$ and $t$-channels, as well as the interference term. One can easily obtain the cross section by incorporating the flux factor, namely $2 s \sqrt{\lambda(1,m_\chi^2/s,m_\psi^2/s)}$,  the two-body final state (for a massive letpoquark and a massless gluon) and performing the spin and color averages. The $s$-channel  squared amplitude is
\begin{eqnarray}
    |\mathcal{M}^s_{\chi \psi \to \phi g }|^2 = -N C_F \frac{(s+m_\phi^2)(s-(m_\psi+m_\chi)^2)}{(s-m_\phi^2)^2+m_\phi^2 \Gamma_\phi^2} \, ,
\end{eqnarray}
the $t$-channel squared amplitude reads
\begin{eqnarray}
   && |\mathcal{M}^t_{\chi \psi \to \phi g }|^2 =8 N C_F \times\nonumber
   \\
   &&\left\lbrace \frac{2}{s} (m_\psi+m_\chi) \left[ (s-m_\phi^2)(m_\psi-m_\chi) + 2 m_\psi s \right] \left[ \frac{C}{C^2-D^2}-\frac{\tanh^{-1} \left( \frac{D}{C}\right)}{D}\right] \right.
    \nonumber \\
   && \left. -    \frac{2C^2-D^2}{C^2-D^2} - 2 \frac{D}{C} \tanh^{-1} \left( \frac{D}{C}\right) + \frac{1}{s^2(C^2-D^2)} \left[ m_\phi^4(s^2-(m^2_\psi-m^2_\chi)^2) \right. \right. 
   \nonumber \\
  && \left.  \left. - 2 s m_\phi^2 (s^2+2sm_\psi(m_\psi-m_\chi)-3 m_\psi^4-2 m_\psi^3 m_\chi + 4 m_\psi^2 m_\chi^2 + 2 m_\psi m_\chi^3 -m_\chi^4 ) \right. \right. 
  \nonumber \\
  && \left.  \left. + s^2(s^2-4s m_\psi(m_\psi+m_\chi)+3 m_\psi^4+12 m_\psi^3 m_\chi + 14 m_\psi^2 m_\chi^2 + 4 m_\psi m_\chi^3 -m_\chi^4 ) \right] \right\rbrace \, ,
  \nonumber
  \\
\end{eqnarray}
and the interference terms is 
\begin{eqnarray}
  &&  2 \textrm{Re} \left( \mathcal{M}^s_{\chi \psi \to \phi g} \mathcal{M}_{\chi \psi \to \phi g} ^{t\, \dagger}\right)= \frac{4   N C_F  (s-m_\phi^2)}{(s-m_\phi^2)^2+m_\phi^2 \Gamma_\phi^2} \left\lbrace 2 (s+(m_\psi+m_\chi)^2) + \frac{2}{s D} \tanh^{-1} \left( \frac{D}{C}\right) \times  \right. 
    \nonumber \\
  &&  \left. \left[ (s-(m_\psi+m_\chi)^2) (s(s+3m_\psi^2-3m_\chi^2)+m_\phi^2(3s + m_\psi^2-m_\chi^2))-sA (s+(m_\psi+m_\chi)^2)\right]
    \right\rbrace  \, .
    \nonumber
    \\
\end{eqnarray}
where the auxiliary functions are, in this case, as follows
\begin{eqnarray}
   && C= (s + m_\psi^2 - m_\chi^2) \left( 1 - \frac{m_\phi^2}{s} \right) \, , 
    \\
   && D= (s-m_\phi^2) \sqrt{\lambda(1,m_\chi^2/s,m_\psi^2/s)}\, .
\end{eqnarray}
\subsection*{$\psi \bar{\psi} $ annihilations}
Particle-antiparticle annihilation of the DVLF are divided in two classes. On the one hand, there are $2 \to 2$ annihilation processes directly into light SM states, namely gluons and quarks. On the other hand, annihilation into leptoquark pairs are also viable. The latter induce a four-body final state which is relevant above the leptoquark threshold. For the first class, the velocity expansion works fine and one can readily extract the cross section from the matching coefficients of NRQCD \cite{Bodwin:1994jh} (see results in the body of the paper, cfr.~eq.~\eqref{cross_psi_psibar_QCD}). As for the annihilation into leptoquark pairs, without performing the velocity expansion, we obtain 
\begin{eqnarray}
     && \sigma_{ \psi \bar{\psi} \to \phi \phi^\dagger }(s) = \frac{y^4 }{ 32 \, \pi \,s }  \sqrt{\frac{s-4m_\phi^2}{s-4 m_\chi^2} } \left\lbrace \frac{1}{E^2-F^2} \left[  s \left( s+4 m_\phi^2 \right) -4  (m_\psi+m_\chi)^2 \left( s + 4 m_\psi^2 \right) \right. \right.
     \nonumber\\
     &&\left. \left. \hspace{1 cm} -8 E m_\psi(m_\psi+2 m_\chi) - 2 E^2 -F^2\right] +  \frac{2}{F}  \left[E+4 m_\psi(m_\psi+m_\chi) \right]   \tanh^{-1} \left( \frac{F}{E}\right) \right\rbrace 
     \nonumber
     \\
     &+&\frac{\pi \alpha_s^2}{6 s^3}  \frac{C_{\textrm{F}}}{N}  \sqrt{\frac{s-4m_\phi^2}{s-4 m_\chi^2} }  \left( s+ 2 m_\psi^2 \right) \left( s- 4 m_\phi^2 \right) 
     \nonumber
     \\
     &-&\frac{\alpha_s y^2}{8 s^2} \frac{C_{\textrm{F}}}{N}\sqrt{\frac{s-4m_\phi^2}{s-4 m_\chi^2} }  \left\lbrace E+4 m_\psi^2 + \frac{ \tanh^{-1} \left( \frac{F}{E}\right)}{ F} \left[ s \left( s- 4 m_\phi^2 \right) - 4 m_\psi^2 E -E^2\right]\right\rbrace \, , 
     \nonumber
     \\
     \label{cross_unEX_psi_bar_psi}
\end{eqnarray}
where 
\begin{eqnarray}
    && E=2 m_\chi^2 + \frac{1}{2} \left( s-4 m_\phi^2 \right) + \frac{1}{2} \left( s- 4 m_\psi^2 \right) \, , 
    \label{coeff_C}
     \\
     && F = \sqrt{s-4 m_\phi^2} \sqrt{s- 4 m_\psi^2} \, .
       \label{coeff_D}
\end{eqnarray}
The result in eq.~\eqref{cross_unEX_psi_bar_psi} enters the total cross section of both scenarios, namely a dark matter Dirac or Majorana fermion. Away from the leptoquark mass threshold, the expansion of the cross section in eq.~\eqref{cross_unEX_psi_bar_psi} gives the result in eq.~\eqref{psi_psibar_leptoquarks}, when decomposed in the corresponding color singlet and color octet contributions. 
\subsection*{$\psi \psi $ and $\bar{\psi} \bar{\psi}$ annihilations}
When the dark matter fermion is assumed to be Majorana, there are  additional annihilation channels for particle-particle ($\psi \psi$) and antiparticle-antiparticle ($\bar{\psi} \bar{\psi}$) DVLF pairs. In this case, a $t$ and $u$-channel mediated by the exchange of a Majorana fermion are possible. The cross section for the process reads 
\begin{eqnarray}
     && \sigma_{ \psi \psi \to \phi \phi }(s) = \frac{y^4}{32 \pi \, N \, s} \sqrt{\frac{s-4m_\phi^2}{s-4 m_\psi^2} } \left\lbrace \frac{1}{C^2-D^2} \left[ N s \left( s- 4 m_\phi^2 \right) + 4 N  (m_\chi + m_\psi)^2 \left( s- 4 m^2_\psi \right)  \right. \right.
     \nonumber\\
  &&   \left. \left. -8 N C m_\psi  (m_\chi + m_\psi) -2 C^2 (N+1) +  D^2 (N+2) \right] +  \frac{1}{AB}\left[8 N C m_\psi (m_\chi + m_\psi)     \right. \right.
  \nonumber \\
 && \left. \left. +C^2(2N+1)  -s \left( s- 4 m_\phi^2 \right) + 4  (m_\chi + m_\psi)^2 \left( s- 4 m_\psi^2 \right) \right]\tanh^{-1} \left( \frac{D}{C}\right) \right\rbrace \, ,
\end{eqnarray}
where the auxiliary coefficeints $E$ and $F$ can be read off eqs.~\eqref{coeff_C} and \eqref{coeff_D}, and $ \sigma_{ \bar{\psi} \bar{\psi} \to \phi^\dagger \phi^\dagger }(s) =  \sigma_{ \psi \psi \to \phi \phi }(s)$.
\section{Electric-dipole matrix element}
\label{sec:app_electri_dipole}
In this section we provide the derivation of the electric dipole matrix element for the transition between color-sextet scattering states and color-antitriplet bound states.
We give the result for a generic bound state and by choosing $\bm{p}$ along the $z$-direction.  We shall derive a general expression using the notation and decomposition of the scattering and bound-state wave functions following the derivation for the octet-singlet electric dipole (see ref.\cite{Biondini:2023zcz}), which was in turn based on refs.~\cite{gordon,stobbe}.
The necessary ingredients are the wavefunctions of the Coulombic scattering and bound states.
The Coulomb wavefunction for a DVLF scattering state of positive energy $\bm{p}^2/M$ reads, when expanded into partial waves $\Psi_{\bm{p} \ell}(\bm{r})=\langle\bm{r}|\bm{p}_{[6]}\ell\rangle$ as  
\begin{eqnarray}
  \Psi_{\bm{p}}(\bm{r}) &=& \sum_{\ell=0}^{\infty} \Psi_{\bm{p}_{[6]} \ell}(\bm{r}) 
= \sqrt{\frac{\frac{\pi (N-1)}{N} \frac{\alpha_s}{v_{\textrm{rel}}}}{1-e^{\frac{\pi (N-1)}{N}\frac{\alpha_s}{v_{\textrm{rel}}}}}}\sum_{\ell=0}^{\infty} e^{i\frac{\pi}{2}\ell}\frac{(2pr)^{\ell}}{(2\ell+1)!}(2\ell+1)P_{\ell}(\cos\theta)e^{ipr} \nonumber \\
&&\times ~_{1}F_{1}\left(\ell+1+i \frac{N-1}{2N}\frac{\alpha_s}{v_{\textrm{rel}}},2\ell+2,-2ipr\right)\prod \limits_{\kappa=1}^{\ell}\sqrt{\kappa^{2} + \left(\frac{N-1}{2N}\frac{\alpha_s}{v_{\textrm{rel}}}\right)^{2}}~,
\label{coulomb_wave}
\end{eqnarray}
where $P_\ell(x)$ are Legendre polynomials and $_{1}F_{1}\left(  a,b,c\right)$ is the confluent hypergeometric function.
The Coulomb wavefunction for a bound state made of two DVLF particles, namely $\psi \psi$, of quantum numbers $n$, $\ell$ and $m$, Bohr radius $\tilde{a}_0=2/(m_\psi C_a \alpha_s)$ and negative binding energy $E^b_n=-m_\psi C_a^2 \alpha_s^2/(4n^2)$ reads
\begin{equation}
\Psi_{[3],n\ell m}(\bm{r}) = \langle\bm{r}|n\ell m \rangle = R_{n \ell}(r)Y_{\ell m}(\Omega) \,,
\label{boundwave}
\end{equation}
with $Y_{\ell m}(\Omega)$ being the spherical harmonics and the radial functions given by
\begin{equation}
\begin{aligned}
  R_{n \ell}(r) = \frac{1}{(2\ell+1)!}\sqrt{\left(\frac{2}{n \tilde{a}_0}\right)^3\frac{(n+\ell)!}{2n(n-\ell-1)!}}
  \left(\frac{2r}{n \tilde{a}_0}\right)^\ell e^{-\frac{r}{n \tilde{a}_0}}~_{1}F_{1}\left(\ell+1-n,2\ell+2,\frac{2r}{n \tilde{a}_0}\right) \,.
\end{aligned}
\end{equation}

The electric-dipole matrix element is
\begin{equation}
\begin{aligned}
&\langle n\ell m_{[3]}|\bm{r}|\bm{p}_{[6]}\rangle = \int d^3r\,\bm{r}\,\Psi_{n\ell m}^*(\bm{r})\Psi_{\bm{p}}(\bm{r}) \\
&~~~= \mathcal{N}\left[\sqrt{\ell(\ell+1)}(\delta_{m,1}-\delta_{m,-1})\bm{e}_x - i\sqrt{\ell(\ell+1)}(\delta_{m,1}+\delta_{m,-1})\bm{e}_y + 2(\ell+1)\delta_{m,0}\bm{e}_z\right]XG_1 \\
&~~~~~+\mathcal{N}\left[-\sqrt{\ell(\ell+1)}(\delta_{m,1}-\delta_{m,-1})\bm{e}_x + i\sqrt{\ell(\ell+1)}(\delta_{m,1}+\delta_{m,-1})\bm{e}_y + 2\ell\delta_{m,0}\bm{e}_z\right]YG_2 \, ,
\label{matrixelement_su(N)}
\end{aligned}
\end{equation}
where
\begin{align}
\mathcal{N} \equiv&  \frac{i^{\ell+3}(-1)^{n-\ell}}{(2\ell+1)!}\sqrt{\left(\frac{2}{n\tilde{a}_0}\right)^3\frac{(n+\ell)!}{2n(n-\ell-1)!}}\left(\frac{2}{n\tilde{a}_0}\right)^\ell
     \sqrt{\frac{\frac{\pi (N-1)}{N} \frac{\alpha_s}{v_{\textrm{rel}}}}{e^{\frac{\pi (N-1)}{N}\frac{\alpha_s}{v_{\textrm{rel}}}}-1}} \nonumber\\
  &\times \sqrt{\frac{\pi}{2\ell+1}}\frac{1}
    {\left[m_\psi^2v_{\textrm{rel}}^2\left(1+\frac{C_a^2\alpha_s^2}{n^2v_{\textrm{rel}}^2}\right)\right]^\ell}\,
    e^{-2\left[i(\ell+1-n) - \frac{\alpha_s(N-1)}{2 N v_{\textrm{rel}}} \right] \arccot{\left(\frac{C_a\alpha_s}{nv_{\textrm{rel}}}\right)}} ,
\end{align}
\begin{align}
X &\equiv \frac{i(m_\psi v_{\textrm{rel}})^{\ell+1}2^{2\ell+4}}{m_\psi^5v_{\textrm{rel}}^5\left(1+\frac{C_a^2\alpha_s^2}{n^2v_{\textrm{rel}}^2}\right)^2}
     e^{-2i\arccot{\left(\frac{C_a\alpha_s}{nv_{\textrm{rel}}}\right)}}\prod \limits_{\kappa=1}^{\ell+1}\sqrt{\kappa^{2} + \left(\frac{\alpha_s (N-1)}{2N v_{\textrm{rel}}}\right)^2} ,
\end{align}
\begin{align}
Y &\equiv \frac{n\ell(2\ell+1)(m_\psi v_{\textrm{rel}})^{\ell-1}2^{2\ell+3}}{C_a\alpha_s m_\psi^3v_{\textrm{rel}}^2\left(1+\frac{C_a^2\alpha_s^2}{n^2v_{\textrm{rel}}^2}\right)}
     \prod \limits_{\kappa=1}^{\ell-1}\sqrt{\kappa^{2} + \left(\frac{\alpha_s(N-1)}{2N v_{\textrm{rel}}}\right)^2} ,
\end{align}
\begin{align}
G_1 &\equiv \left(1+i\frac{\alpha_s}{v_{\textrm{rel}}}\right)\,_{2}F_{1}\left(\ell+2+i\frac{\alpha_s(N-1)}{2N v_{\textrm{rel}}},\ell+1-n,2\ell+2,\frac{-4i C_a \alpha_s}{nv_{\textrm{rel}}
      \left(1-i\frac{C_a \alpha_s}{nv_{\textrm{rel}}}\right)^2}\right) \nonumber\\
& -i\frac{2 \alpha_s}{v_{\textrm{rel}}}e^{2i\arccot{\left(\frac{C_a\alpha_s}{nv_{\textrm{rel}}}\right)}}\,_{2}F_{1}\left(\ell+1+i\frac{\alpha_s(N-1)}{2N v_{\textrm{rel}}},\ell+1-n,2\ell+2,
  \frac{-4i C_a \alpha_s}{nv_{\textrm{rel}}\left(1-i\frac{C_a\alpha_s}{nv_{\textrm{rel}}}\right)^2}\right) \nonumber\\
& -\left(1-i\frac{ \alpha_s}{v_{\textrm{rel}}}\right)e^{4i\arccot{\left(\frac{C_a\alpha_s}{nv_{\textrm{rel}}}\right)}}\,_{2}F_{1}\left(\ell+i\frac{\alpha_s (N-1)}{2N v_{\textrm{rel}}},
  \ell+1-n,2\ell+2,\frac{-4i C_a \alpha_s }{nv_{\textrm{rel}}\left(1-i\frac{C_a\alpha_s}{nv_{\textrm{rel}}}\right)^2}\right), 
\end{align} 
\begin{align}
G_2 &\equiv \left(1-\frac{ n }{C_a}\right)\,_{2}F_{1}\left(\ell+1-n,\ell+i\frac{\alpha_s(N-1)}{2N v_{\textrm{rel}}},2\ell,
    \frac{-4i C_a \alpha_s }{nv_{\textrm{rel}}\left(1-i\frac{C_a\alpha_s}{nv_{\textrm{rel}}}\right)^2}\right) \nonumber\\
&+\frac{2n}{C_a}e^{2i\arccot{\left(\frac{C_a\alpha_s}{nv_{\textrm{rel}}}\right)}}\,_{2}F_{1}\left(\ell-n,\ell+i\frac{\alpha_s(N-1)}{2N v_{\textrm{rel}}},2\ell,
  \frac{-4iC_a\alpha_s }{nv_{\textrm{rel}}\left(1-i\frac{C_a \alpha_s}{nv_{\textrm{rel}}}\right)^2}\right) \nonumber\\
&-\left(1+\frac{ n}{C_a}\right)e^{4i\arccot{\left(\frac{C_a\alpha_s}{nv_{\textrm{rel}}}\right)}}\,_{2}F_{1}\left(\ell-1-n,\ell+i\frac{\alpha_s(N-1)}{2N v_{\textrm{rel}}},2\ell,
  \frac{-4iC_a\alpha_s }{nv_{\textrm{rel}}\left(1-i\frac{C_a\alpha_s}{nv_{\textrm{rel}}}\right)^2}\right) \, .
\end{align}
In the dipole matrix element the natural renormalization scale of the coupling $\alpha_s$ is $\mu_{\textrm{s}}$, which is of the order of the soft scale. We do not distinguish the soft scale between singlet and triplet bound states.

In the case of the lowest-lying $1\textrm{S}$ bound state, the wavefunction reads $\langle \bm{r}|1\textrm{S}{[3]}\rangle = R_{[3],10}(r)/(4\pi)$.
In this case, only scattering states in the partial wave $\ell=1$ contribute, and the corresponding wavefunction is 
$\langle \bm{r}|\bm{p}_{[6]}1\,\rangle = \Psi_{\bm{p}_{[6]}1}(\bm{r})$, where the radial Coulombic wave function is $R_{[3],10}(r)=2 e^{-r/\tilde{a}_0}/\tilde{a}_0^{3/2}$. The squared matrix element for the ground state reads 
\begin{eqnarray}
    \left|\langle 1\textrm{S}_{[3]} | \boldsymbol{r} | \boldsymbol{p}_{[\bm{6}]} \rangle\right|^2 = \frac{\pi^2}{m_\psi^5}  \frac{2^{13}  C^3_a  \, \alpha_s^6}{N \, v_{\hbox{\scriptsize rel}}^{11}}  \frac{(C_a+1)\left( C_a-\frac{1}{N}\right)}{\left[ 1+ \left(\frac{C_a\alpha_s}{v_{\textrm{rel}}}\right)^2\right]^6} \left[ 1+\left(\frac{\alpha_s (N-1)}{2Nv_{\textrm{rel}}}\right)^2 \right]  \frac{e^{\frac{2\alpha_s(N-1)}{Nv_{\textrm{rel}}} \hbox{\scriptsize arccot} \frac{C_a\alpha_s}{v_{\textrm{rel}}} }}{e^{\frac{\pi(N-1)}{N} \frac{\alpha_s}{v_{\textrm{rel}}}}-1} \, ,
    \nonumber
    \\
\end{eqnarray}
which can be inserted in eq.\eqref{bsf_sextet_to_bound_1S} to provide the explicit result for the bound-state formation cross section. 

\bibliographystyle{style}
\bibliography{references}

\providecommand{\href}[2]{#2}\begingroup\raggedright\begin{thebibliography}{100}

\bibitem{Bertone:2004pz}
G.~Bertone, D.~Hooper, and J.~Silk, ``{Particle dark matter: Evidence,
  candidates and constraints},''
  \href{http://dx.doi.org/10.1016/j.physrep.2004.08.031}{{\em Phys. Rept.}
  {\bfseries 405} (2005) 279--390},
  \href{http://arxiv.org/abs/hep-ph/0404175}{{\ttfamily arXiv:hep-ph/0404175}}.

\bibitem{Feng:2010gw}
J.~L. Feng, ``{Dark Matter Candidates from Particle Physics and Methods of
  Detection},''
  \href{http://dx.doi.org/10.1146/annurev-astro-082708-101659}{{\em Ann. Rev.
  Astron. Astrophys.} {\bfseries 48} (2010) 495--545},
  \href{http://arxiv.org/abs/1003.0904}{{\ttfamily arXiv:1003.0904
  [astro-ph.CO]}}.

\bibitem{Planck:2018vyg}
{\bfseries Planck} Collaboration, N.~Aghanim {\em et~al.}, ``{Planck 2018
  results. VI. Cosmological parameters},''
  \href{http://dx.doi.org/10.1051/0004-6361/201833910}{{\em Astron. Astrophys.}
  {\bfseries 641} (2020) A6}, \href{http://arxiv.org/abs/1807.06209}{{\ttfamily
  arXiv:1807.06209 [astro-ph.CO]}}. [Erratum: Astron.Astrophys. 652, C4
  (2021)].

\bibitem{Arcadi:2017kky}
G.~Arcadi, M.~Dutra, P.~Ghosh, M.~Lindner, Y.~Mambrini, M.~Pierre, S.~Profumo,
  and F.~S. Queiroz, ``{The waning of the WIMP? A review of models, searches,
  and constraints},''
  \href{http://dx.doi.org/10.1140/epjc/s10052-018-5662-y}{{\em Eur. Phys. J. C}
  {\bfseries 78} no.~3, (2018) 203},
  \href{http://arxiv.org/abs/1703.07364}{{\ttfamily arXiv:1703.07364
  [hep-ph]}}.

\bibitem{McDonald:2001vt}
J.~McDonald, ``{Thermally generated gauge singlet scalars as selfinteracting
  dark matter},'' \href{http://dx.doi.org/10.1103/PhysRevLett.88.091304}{{\em
  Phys. Rev. Lett.} {\bfseries 88} (2002) 091304},
  \href{http://arxiv.org/abs/hep-ph/0106249}{{\ttfamily arXiv:hep-ph/0106249}}.

\bibitem{Hall:2009bx}
L.~J. Hall, K.~Jedamzik, J.~March-Russell, and S.~M. West, ``{Freeze-In
  Production of FIMP Dark Matter},''
  \href{http://dx.doi.org/10.1007/JHEP03(2010)080}{{\em JHEP} {\bfseries 03}
  (2010) 080}, \href{http://arxiv.org/abs/0911.1120}{{\ttfamily arXiv:0911.1120
  [hep-ph]}}.

\bibitem{Bernal:2017kxu}
N.~Bernal, M.~Heikinheimo, T.~Tenkanen, K.~Tuominen, and V.~Vaskonen, ``{The
  Dawn of FIMP Dark Matter: A Review of Models and Constraints},''
  \href{http://dx.doi.org/10.1142/S0217751X1730023X}{{\em Int. J. Mod. Phys.}
  {\bfseries A32} no.~27, (2017) 1730023},
\href{http://arxiv.org/abs/1706.07442}{{\ttfamily arXiv:1706.07442 [hep-ph]}}.

\bibitem{Donoghue:1994dn}
J.~F. Donoghue, ``{General relativity as an effective field theory: The leading
  quantum corrections},''
  \href{http://dx.doi.org/10.1103/PhysRevD.50.3874}{{\em Phys. Rev. D}
  {\bfseries 50} (1994) 3874--3888},
  \href{http://arxiv.org/abs/gr-qc/9405057}{{\ttfamily arXiv:gr-qc/9405057}}.

\bibitem{Choi:1994ax}
S.~Y. Choi, J.~S. Shim, and H.~S. Song, ``{Factorization and polarization in
  linearized gravity},'' \href{http://dx.doi.org/10.1103/PhysRevD.51.2751}{{\em
  Phys. Rev. D} {\bfseries 51} (1995) 2751--2769},
  \href{http://arxiv.org/abs/hep-th/9411092}{{\ttfamily arXiv:hep-th/9411092}}.

\bibitem{Holstein:2006bh}
B.~R. Holstein, ``{Graviton Physics},''
  \href{http://dx.doi.org/10.1119/1.2338547}{{\em Am. J. Phys.} {\bfseries 74}
  (2006) 1002--1011}, \href{http://arxiv.org/abs/gr-qc/0607045}{{\ttfamily
  arXiv:gr-qc/0607045}}.

\bibitem{Garny:2015sjg}
M.~Garny, M.~Sandora, and M.~S. Sloth, ``{Planckian Interacting Massive
  Particles as Dark Matter},''
  \href{http://dx.doi.org/10.1103/PhysRevLett.116.101302}{{\em Phys. Rev.
  Lett.} {\bfseries 116} no.~10, (2016) 101302},
  \href{http://arxiv.org/abs/1511.03278}{{\ttfamily arXiv:1511.03278
  [hep-ph]}}.

\bibitem{Mambrini:2021zpp}
Y.~Mambrini and K.~A. Olive, ``{Gravitational Production of Dark Matter during
  Reheating},'' \href{http://dx.doi.org/10.1103/PhysRevD.103.115009}{{\em Phys.
  Rev. D} {\bfseries 103} no.~11, (2021) 115009},
  \href{http://arxiv.org/abs/2102.06214}{{\ttfamily arXiv:2102.06214
  [hep-ph]}}.

\bibitem{Barman:2021ugy}
B.~Barman and N.~Bernal, ``{Gravitational SIMPs},''
  \href{http://dx.doi.org/10.1088/1475-7516/2021/06/011}{{\em JCAP} {\bfseries
  06} (2021) 011}, \href{http://arxiv.org/abs/2104.10699}{{\ttfamily
  arXiv:2104.10699 [hep-ph]}}.

\bibitem{Fong:2022cmq}
C.~S. Fong, M.~H. Rahat, and S.~Saad, ``{BBN photodisintegration constraints on
  gravitationally produced vector bosons},''
  \href{http://dx.doi.org/10.1007/JHEP11(2022)067}{{\em JHEP} {\bfseries 11}
  (2022) 067}, \href{http://arxiv.org/abs/2206.02802}{{\ttfamily
  arXiv:2206.02802 [hep-ph]}}.

\bibitem{Sommerfeld}
A.~Sommerfeld, ``{\"Uber die Beugung und Bremsung der Elektronen},'' {\em Ann.
  Phys.(1931)} {\bfseries 403} (1931) .

\bibitem{Hisano:2006nn}
J.~Hisano, S.~Matsumoto, M.~Nagai, O.~Saito, and M.~Senami, ``{Non-perturbative
  effect on thermal relic abundance of dark matter},''
  \href{http://dx.doi.org/10.1016/j.physletb.2007.01.012}{{\em Phys. Lett. B}
  {\bfseries 646} (2007) 34--38},
  \href{http://arxiv.org/abs/hep-ph/0610249}{{\ttfamily arXiv:hep-ph/0610249}}.

\bibitem{Cirelli:2007xd}
M.~Cirelli, A.~Strumia, and M.~Tamburini, ``{Cosmology and Astrophysics of
  Minimal Dark Matter},''
  \href{http://dx.doi.org/10.1016/j.nuclphysb.2007.07.023}{{\em Nucl. Phys.}
  {\bfseries B787} (2007) 152--175},
\href{http://arxiv.org/abs/0706.4071}{{\ttfamily arXiv:0706.4071 [hep-ph]}}.

\bibitem{Feng:2009mn}
J.~L. Feng, M.~Kaplinghat, H.~Tu, and H.-B. Yu, ``{Hidden Charged Dark
  Matter},'' \href{http://dx.doi.org/10.1088/1475-7516/2009/07/004}{{\em JCAP}
  {\bfseries 0907} (2009) 004},
\href{http://arxiv.org/abs/0905.3039}{{\ttfamily arXiv:0905.3039 [hep-ph]}}.

\bibitem{vonHarling:2014kha}
B.~von Harling and K.~Petraki, ``{Bound-state formation for thermal relic dark
  matter and unitarity},''
  \href{http://dx.doi.org/10.1088/1475-7516/2014/12/033}{{\em JCAP} {\bfseries
  1412} (2014) 033},
\href{http://arxiv.org/abs/1407.7874}{{\ttfamily arXiv:1407.7874 [hep-ph]}}.

\bibitem{Cirelli:2008id}
M.~Cirelli, R.~Franceschini, and A.~Strumia, ``{Minimal Dark Matter predictions
  for galactic positrons, anti-protons, photons},''
  \href{http://dx.doi.org/10.1016/j.nuclphysb.2008.03.013}{{\em Nucl. Phys.}
  {\bfseries B800} (2008) 204--220},
\href{http://arxiv.org/abs/0802.3378}{{\ttfamily arXiv:0802.3378 [hep-ph]}}.

\bibitem{Cirelli:2009uv}
M.~Cirelli and A.~Strumia, ``{Minimal Dark Matter: Model and results},''
  \href{http://dx.doi.org/10.1088/1367-2630/11/10/105005}{{\em New J. Phys.}
  {\bfseries 11} (2009) 105005},
\href{http://arxiv.org/abs/0903.3381}{{\ttfamily arXiv:0903.3381 [hep-ph]}}.

\bibitem{Feng:2010zp}
J.~L. Feng, M.~Kaplinghat, and H.-B. Yu, ``{Sommerfeld Enhancements for Thermal
  Relic Dark Matter},''
  \href{http://dx.doi.org/10.1103/PhysRevD.82.083525}{{\em Phys. Rev.}
  {\bfseries D82} (2010) 083525},
\href{http://arxiv.org/abs/1005.4678}{{\ttfamily arXiv:1005.4678 [hep-ph]}}.

\bibitem{deSimone:2014pda}
A.~De~Simone, G.~F. Giudice, and A.~Strumia, ``{Benchmarks for Dark Matter
  Searches at the LHC},'' \href{http://dx.doi.org/10.1007/JHEP06(2014)081}{{\em
  JHEP} {\bfseries 06} (2014) 081},
\href{http://arxiv.org/abs/1402.6287}{{\ttfamily arXiv:1402.6287 [hep-ph]}}.

\bibitem{Beneke:2014gja}
M.~Beneke, C.~Hellmann, and P.~Ruiz-Femenia, ``{Non-relativistic pair
  annihilation of nearly mass degenerate neutralinos and charginos III.
  Computation of the Sommerfeld enhancements},''
  \href{http://dx.doi.org/10.1007/JHEP05(2015)115}{{\em JHEP} {\bfseries 05}
  (2015) 115}, \href{http://arxiv.org/abs/1411.6924}{{\ttfamily arXiv:1411.6924
  [hep-ph]}}.

\bibitem{Beneke:2014hja}
M.~Beneke, C.~Hellmann, and P.~Ruiz-Femenia, ``{Heavy neutralino relic
  abundance with Sommerfeld enhancements - a study of pMSSM scenarios},''
  \href{http://dx.doi.org/10.1007/JHEP03(2015)162}{{\em JHEP} {\bfseries 03}
  (2015) 162}, \href{http://arxiv.org/abs/1411.6930}{{\ttfamily arXiv:1411.6930
  [hep-ph]}}.

\bibitem{Ibarra:2015nca}
A.~Ibarra, A.~Pierce, N.~R. Shah, and S.~Vogl, ``{Anatomy of Coannihilation
  with a Scalar Top Partner},''
  \href{http://dx.doi.org/10.1103/PhysRevD.91.095018}{{\em Phys. Rev. D}
  {\bfseries 91} no.~9, (2015) 095018},
  \href{http://arxiv.org/abs/1501.03164}{{\ttfamily arXiv:1501.03164
  [hep-ph]}}.

\bibitem{Ellis:2014ipa}
J.~Ellis, K.~A. Olive, and J.~Zheng, ``{The Extent of the Stop Coannihilation
  Strip},'' \href{http://dx.doi.org/10.1140/epjc/s10052-014-2947-7}{{\em Eur.
  Phys. J.} {\bfseries C74} (2014) 2947},
\href{http://arxiv.org/abs/1404.5571}{{\ttfamily arXiv:1404.5571 [hep-ph]}}.

\bibitem{Liew:2016hqo}
S.~P. Liew and F.~Luo, ``{Effects of QCD bound states on dark matter relic
  abundance},'' \href{http://dx.doi.org/10.1007/JHEP02(2017)091}{{\em JHEP}
  {\bfseries 02} (2017) 091},
\href{http://arxiv.org/abs/1611.08133}{{\ttfamily arXiv:1611.08133 [hep-ph]}}.

\bibitem{Mitridate:2017izz}
A.~Mitridate, M.~Redi, J.~Smirnov, and A.~Strumia, ``{Cosmological Implications
  of Dark Matter Bound States},''
  \href{http://dx.doi.org/10.1088/1475-7516/2017/05/006}{{\em JCAP} {\bfseries
  1705} no.~05, (2017) 006},
\href{http://arxiv.org/abs/1702.01141}{{\ttfamily arXiv:1702.01141 [hep-ph]}}.

\bibitem{Garny:2021qsr}
M.~Garny and J.~Heisig, ``{Bound-state effects on dark matter coannihilation:
  Pushing the boundaries of conversion-driven freeze-out},''
  \href{http://dx.doi.org/10.1103/PhysRevD.105.055004}{{\em Phys. Rev. D}
  {\bfseries 105} no.~5, (2022) 055004},
  \href{http://arxiv.org/abs/2112.01499}{{\ttfamily arXiv:2112.01499
  [hep-ph]}}.

\bibitem{Harz:2018csl}
J.~Harz and K.~Petraki, ``{Radiative bound-state formation in unbroken
  perturbative non-Abelian theories and implications for dark matter},''
  \href{http://dx.doi.org/10.1007/JHEP07(2018)096}{{\em JHEP} {\bfseries 07}
  (2018) 096}, \href{http://arxiv.org/abs/1805.01200}{{\ttfamily
  arXiv:1805.01200 [hep-ph]}}.

\bibitem{Biondini:2018ovz}
S.~Biondini and S.~Vogl, ``{Coloured coannihilations: Dark matter phenomenology
  meets non-relativistic EFTs},''
  \href{http://dx.doi.org/10.1007/JHEP02(2019)016}{{\em JHEP} {\bfseries 02}
  (2019) 016},
\href{http://arxiv.org/abs/1811.02581}{{\ttfamily arXiv:1811.02581 [hep-ph]}}.

\bibitem{Biondini:2019int}
S.~Biondini and S.~Vogl, ``{Scalar dark matter coannihilating with a coloured
  fermion},'' \href{http://dx.doi.org/10.1007/JHEP11(2019)147}{{\em JHEP}
  {\bfseries 11} (2019) 147},
\href{http://arxiv.org/abs/1907.05766}{{\ttfamily arXiv:1907.05766 [hep-ph]}}.

\bibitem{Becker:2022iso}
M.~Becker, E.~Copello, J.~Harz, K.~A. Mohan, and D.~Sengupta, ``{Impact of
  Sommerfeld effect and bound state formation in simplified t-channel dark
  matter models},'' \href{http://dx.doi.org/10.1007/JHEP08(2022)145}{{\em JHEP}
  {\bfseries 08} (2022) 145}, \href{http://arxiv.org/abs/2203.04326}{{\ttfamily
  arXiv:2203.04326 [hep-ph]}}.

\bibitem{Baker:2015qna}
M.~J. Baker {\em et~al.}, ``{The Coannihilation Codex},''
  \href{http://dx.doi.org/10.1007/JHEP12(2015)120}{{\em JHEP} {\bfseries 12}
  (2015) 120}, \href{http://arxiv.org/abs/1510.03434}{{\ttfamily
  arXiv:1510.03434 [hep-ph]}}.

\bibitem{Belanger:2021smw}
G.~Belanger {\em et~al.}, ``{Leptoquark manoeuvres in the dark: a simultaneous
  solution of the dark matter problem and the $ {R}_{D^{\left(\ast \right)}} $
  anomalies},'' \href{http://dx.doi.org/10.1007/JHEP02(2022)042}{{\em JHEP}
  {\bfseries 02} (2022) 042}, \href{http://arxiv.org/abs/2111.08027}{{\ttfamily
  arXiv:2111.08027 [hep-ph]}}.

\bibitem{Manzari:2022iyn}
C.~A. Manzari and S.~Profumo, ``{A Flavour Inspired Model for Dark Matter},''
  \href{http://arxiv.org/abs/2206.06768}{{\ttfamily arXiv:2206.06768
  [hep-ph]}}.

\bibitem{Carpenter:2022lhj}
L.~M. Carpenter, T.~Murphy, and T.~M.~P. Tait, ``{Distinctive signals of
  frustrated dark matter},''
  \href{http://dx.doi.org/10.1007/JHEP09(2022)175}{{\em JHEP} {\bfseries 09}
  (2022) 175}, \href{http://arxiv.org/abs/2205.06824}{{\ttfamily
  arXiv:2205.06824 [hep-ph]}}.

\bibitem{Buchmuller:1986zs}
W.~Buchmuller, R.~Ruckl, and D.~Wyler, ``{Leptoquarks in Lepton - Quark
  Collisions},'' \href{http://dx.doi.org/10.1016/0370-2693(87)90637-X}{{\em
  Phys. Lett. B} {\bfseries 191} (1987) 442--448}. [Erratum: Phys.Lett.B 448,
  320--320 (1999)].

\bibitem{Gondolo:1990dk}
P.~Gondolo and G.~Gelmini, ``{Cosmic abundances of stable particles: Improved
  analysis},'' \href{http://dx.doi.org/10.1016/0550-3213(91)90438-4}{{\em Nucl.
  Phys. B} {\bfseries 360} (1991) 145--179}.

\bibitem{Griest:1990kh}
K.~Griest and D.~Seckel, ``{Three exceptions in the calculation of relic
  abundances},'' \href{http://dx.doi.org/10.1103/PhysRevD.43.3191}{{\em Phys.
  Rev. D} {\bfseries 43} (1991) 3191--3203}.

\bibitem{Edsjo:1997bg}
J.~Edsjo and P.~Gondolo, ``{Neutralino relic density including
  coannihilations},'' \href{http://dx.doi.org/10.1103/PhysRevD.56.1879}{{\em
  Phys. Rev.} {\bfseries D56} (1997) 1879--1894},
\href{http://arxiv.org/abs/hep-ph/9704361}{{\ttfamily arXiv:hep-ph/9704361
  [hep-ph]}}.

\bibitem{Garny:2017rxs}
M.~Garny, J.~Heisig, B.~L\"ulf, and S.~Vogl, ``{Coannihilation without chemical
  equilibrium},'' \href{http://dx.doi.org/10.1103/PhysRevD.96.103521}{{\em
  Phys. Rev. D} {\bfseries 96} no.~10, (2017) 103521},
  \href{http://arxiv.org/abs/1705.09292}{{\ttfamily arXiv:1705.09292
  [hep-ph]}}.

\bibitem{DAgnolo:2017dbv}
R.~T. D'Agnolo, D.~Pappadopulo, and J.~T. Ruderman, ``{Fourth Exception in the
  Calculation of Relic Abundances},''
  \href{http://dx.doi.org/10.1103/PhysRevLett.119.061102}{{\em Phys. Rev.
  Lett.} {\bfseries 119} no.~6, (2017) 061102},
  \href{http://arxiv.org/abs/1705.08450}{{\ttfamily arXiv:1705.08450
  [hep-ph]}}.

\bibitem{Denner:2005fg}
A.~Denner, S.~Dittmaier, M.~Roth, and L.~H. Wieders, ``{Electroweak corrections
  to charged-current e+ e- ---\ensuremath{>} 4 fermion processes: Technical
  details and further results},''
  \href{http://dx.doi.org/10.1016/j.nuclphysb.2011.09.001}{{\em Nucl. Phys. B}
  {\bfseries 724} (2005) 247--294},
  \href{http://arxiv.org/abs/hep-ph/0505042}{{\ttfamily arXiv:hep-ph/0505042}}.
  [Erratum: Nucl.Phys.B 854, 504--507 (2012)].

\bibitem{Denner:1999gp}
A.~Denner, S.~Dittmaier, M.~Roth, and D.~Wackeroth, ``{Predictions for all
  processes e+ e- ---\ensuremath{>} 4 fermions + gamma},''
  \href{http://dx.doi.org/10.1016/S0550-3213(99)00437-X}{{\em Nucl. Phys. B}
  {\bfseries 560} (1999) 33--65},
  \href{http://arxiv.org/abs/hep-ph/9904472}{{\ttfamily arXiv:hep-ph/9904472}}.

\bibitem{Berdine:2007uv}
D.~Berdine, N.~Kauer, and D.~Rainwater, ``{Breakdown of the Narrow Width
  Approximation for New Physics},''
  \href{http://dx.doi.org/10.1103/PhysRevLett.99.111601}{{\em Phys. Rev. Lett.}
  {\bfseries 99} (2007) 111601},
  \href{http://arxiv.org/abs/hep-ph/0703058}{{\ttfamily arXiv:hep-ph/0703058}}.

\bibitem{Alwall:2014hca}
J.~Alwall, R.~Frederix, S.~Frixione, V.~Hirschi, F.~Maltoni, O.~Mattelaer,
  H.~S. Shao, T.~Stelzer, P.~Torrielli, and M.~Zaro, ``{The automated
  computation of tree-level and next-to-leading order differential cross
  sections, and their matching to parton shower simulations},''
  \href{http://dx.doi.org/10.1007/JHEP07(2014)079}{{\em JHEP} {\bfseries 07}
  (2014) 079}, \href{http://arxiv.org/abs/1405.0301}{{\ttfamily arXiv:1405.0301
  [hep-ph]}}.

\bibitem{Laine:2022ner}
M.~Laine, ``{Resonant s-channel dark matter annihilation at NLO},''
  \href{http://dx.doi.org/10.1007/JHEP01(2023)157}{{\em JHEP} {\bfseries 01}
  (2023) 157}, \href{http://arxiv.org/abs/2211.06008}{{\ttfamily
  arXiv:2211.06008 [hep-ph]}}.

\bibitem{Biondini:2018pwp}
S.~Biondini and M.~Laine, ``{Thermal dark matter co-annihilating with a
  strongly interacting scalar},''
  \href{http://dx.doi.org/10.1007/JHEP04(2018)072}{{\em JHEP} {\bfseries 04}
  (2018) 072}, \href{http://arxiv.org/abs/1801.05821}{{\ttfamily
  arXiv:1801.05821 [hep-ph]}}.

\bibitem{Binder:2021vfo}
T.~Binder, A.~Filimonova, K.~Petraki, and G.~White, ``{Saha equilibrium for
  metastable bound states and dark matter freeze-out},''
  \href{http://dx.doi.org/10.1016/j.physletb.2022.137323}{{\em Phys. Lett. B}
  {\bfseries 833} (2022) 137323},
  \href{http://arxiv.org/abs/2112.00042}{{\ttfamily arXiv:2112.00042
  [hep-ph]}}.

\bibitem{Belanger:2006is}
G.~Belanger, F.~Boudjema, A.~Pukhov, and A.~Semenov, ``{MicrOMEGAs 2.0: A
  Program to calculate the relic density of dark matter in a generic model},''
  \href{http://dx.doi.org/10.1016/j.cpc.2006.11.008}{{\em Comput. Phys.
  Commun.} {\bfseries 176} (2007) 367--382},
  \href{http://arxiv.org/abs/hep-ph/0607059}{{\ttfamily arXiv:hep-ph/0607059}}.

\bibitem{Caswell:1985ui}
W.~E. Caswell and G.~P. Lepage, ``{Effective Lagrangians for Bound State
  Problems in QED, QCD, and Other Field Theories},''
\href{http://dx.doi.org/10.1016/0370-2693(86)91297-9}{{\em Phys. Lett.}
  {\bfseries 167B} (1986) 437--442}.

\bibitem{Bodwin:1994jh}
G.~T. Bodwin, E.~Braaten, and G.~P. Lepage, ``{Rigorous QCD analysis of
  inclusive annihilation and production of heavy quarkonium},''
  \href{http://dx.doi.org/10.1103/PhysRevD.55.5853}{{\em Phys. Rev. D}
  {\bfseries 51} (1995) 1125--1171},
  \href{http://arxiv.org/abs/hep-ph/9407339}{{\ttfamily arXiv:hep-ph/9407339}}.
  [Erratum: Phys.Rev.D 55, 5853 (1997)].

\bibitem{Pineda:1997bj}
A.~Pineda and J.~Soto, ``{Effective field theory for ultrasoft momenta in NRQCD
  and NRQED},'' \href{http://dx.doi.org/10.1016/S0920-5632(97)01102-X}{{\em
  Nucl. Phys. Proc. Suppl.} {\bfseries 64} (1998) 428--432},
\href{http://arxiv.org/abs/hep-ph/9707481}{{\ttfamily arXiv:hep-ph/9707481
  [hep-ph]}}.

\bibitem{Brambilla:1999xf}
N.~Brambilla, A.~Pineda, J.~Soto, and A.~Vairo, ``{Potential NRQCD: An
  Effective theory for heavy quarkonium},''
  \href{http://dx.doi.org/10.1016/S0550-3213(99)00693-8}{{\em Nucl. Phys.}
  {\bfseries B566} (2000) 275},
\href{http://arxiv.org/abs/hep-ph/9907240}{{\ttfamily arXiv:hep-ph/9907240
  [hep-ph]}}.

\bibitem{Brambilla:2005yk}
N.~Brambilla, A.~Vairo, and T.~Rosch, ``{Effective field theory Lagrangians for
  baryons with two and three heavy quarks},''
  \href{http://dx.doi.org/10.1103/PhysRevD.72.034021}{{\em Phys. Rev. D}
  {\bfseries 72} (2005) 034021},
  \href{http://arxiv.org/abs/hep-ph/0506065}{{\ttfamily arXiv:hep-ph/0506065}}.

\bibitem{Brambilla:2008cx}
N.~Brambilla, J.~Ghiglieri, A.~Vairo, and P.~Petreczky, ``{Static
  quark-antiquark pairs at finite temperature},''
  \href{http://dx.doi.org/10.1103/PhysRevD.78.014017}{{\em Phys. Rev.}
  {\bfseries D78} (2008) 014017},
\href{http://arxiv.org/abs/0804.0993}{{\ttfamily arXiv:0804.0993 [hep-ph]}}.

\bibitem{Escobedo:2008sy}
M.~A. Escobedo and J.~Soto, ``{Non-relativistic bound states at finite
  temperature (I): The Hydrogen atom},''
  \href{http://dx.doi.org/10.1103/PhysRevA.78.032520}{{\em Phys. Rev. A}
  {\bfseries 78} (2008) 032520},
  \href{http://arxiv.org/abs/0804.0691}{{\ttfamily arXiv:0804.0691 [hep-ph]}}.

\bibitem{Escobedo:2010tu}
M.~A. Escobedo and J.~Soto, ``{Non-relativistic bound states at finite
  temperature (II): the muonic hydrogen},''
  \href{http://dx.doi.org/10.1103/PhysRevA.82.042506}{{\em Phys. Rev. A}
  {\bfseries 82} (2010) 042506},
  \href{http://arxiv.org/abs/1008.0254}{{\ttfamily arXiv:1008.0254 [hep-ph]}}.

\bibitem{Kim:2016kxt}
S.~Kim and M.~Laine, ``{On thermal corrections to near-threshold
  annihilation},'' \href{http://dx.doi.org/10.1088/1475-7516/2017/01/013}{{\em
  JCAP} {\bfseries 01} (2017) 013},
  \href{http://arxiv.org/abs/1609.00474}{{\ttfamily arXiv:1609.00474
  [hep-ph]}}.

\bibitem{Binder:2020efn}
T.~Binder, B.~Blobel, J.~Harz, and K.~Mukaida, ``{Dark matter bound-state
  formation at higher order: a non-equilibrium quantum field theory
  approach},'' \href{http://dx.doi.org/10.1007/JHEP09(2020)086}{{\em JHEP}
  {\bfseries 09} (2020) 086}, \href{http://arxiv.org/abs/2002.07145}{{\ttfamily
  arXiv:2002.07145 [hep-ph]}}.

\bibitem{Binder:2018znk}
T.~Binder, L.~Covi, and K.~Mukaida, ``{Dark Matter Sommerfeld-enhanced
  annihilation and Bound-state decay at finite temperature},''
  \href{http://dx.doi.org/10.1103/PhysRevD.98.115023}{{\em Phys. Rev.}
  {\bfseries D98} no.~11, (2018) 115023},
\href{http://arxiv.org/abs/1808.06472}{{\ttfamily arXiv:1808.06472 [hep-ph]}}.

\bibitem{Biondini:2023zcz}
S.~Biondini, N.~Brambilla, G.~Qerimi, and A.~Vairo, ``{Effective field theories
  for dark matter pairs in the early universe: cross sections and widths},''
  \href{http://arxiv.org/abs/2304.00113}{{\ttfamily arXiv:2304.00113
  [hep-ph]}}.

\bibitem{Brambilla:2004jw}
N.~Brambilla, A.~Pineda, J.~Soto, and A.~Vairo, ``{Effective field theories for
  heavy quarkonium},'' \href{http://dx.doi.org/10.1103/RevModPhys.77.1423}{{\em
  Rev. Mod. Phys.} {\bfseries 77} (2005) 1423},
\href{http://arxiv.org/abs/hep-ph/0410047}{{\ttfamily arXiv:hep-ph/0410047
  [hep-ph]}}.

\bibitem{ElHedri:2016onc}
S.~El~Hedri, A.~Kaminska, and M.~de~Vries, ``{A Sommerfeld Toolbox for Colored
  Dark Sectors},'' \href{http://dx.doi.org/10.1140/epjc/s10052-017-5168-z}{{\em
  Eur. Phys. J. C} {\bfseries 77} no.~9, (2017) 622},
  \href{http://arxiv.org/abs/1612.02825}{{\ttfamily arXiv:1612.02825
  [hep-ph]}}.

\bibitem{Biondini:2021ycj}
S.~Biondini and V.~Shtabovenko, ``{Bound-state formation, dissociation and
  decays of darkonium with potential non-relativistic Yukawa theory for scalar
  and pseudoscalar mediators},''
  \href{http://dx.doi.org/10.1007/JHEP03(2022)172}{{\em JHEP} {\bfseries 03}
  (2022) 172}, \href{http://arxiv.org/abs/2112.10145}{{\ttfamily
  arXiv:2112.10145 [hep-ph]}}.

\bibitem{Brambilla:2011sg}
N.~Brambilla, M.~A. Escobedo, J.~Ghiglieri, and A.~Vairo, ``{Thermal width and
  gluo-dissociation of quarkonium in pNRQCD},''
  \href{http://dx.doi.org/10.1007/JHEP12(2011)116}{{\em JHEP} {\bfseries 12}
  (2011) 116},
\href{http://arxiv.org/abs/1109.5826}{{\ttfamily arXiv:1109.5826 [hep-ph]}}.

\bibitem{Bellac:2011kqa}
M.~L. Bellac, \href{http://dx.doi.org/10.1017/CBO9780511721700}{{\em {Thermal
  Field Theory}}}.
\newblock Cambridge Monographs on Mathematical Physics. Cambridge University
  Press, 3, 2011.

\bibitem{Laine:2016hma}
M.~Laine and A.~Vuorinen,
  \href{http://dx.doi.org/10.1007/978-3-319-31933-9}{{\em {Basics of Thermal
  Field Theory}}}, vol.~925.
\newblock Springer, 2016.
\newblock \href{http://arxiv.org/abs/1701.01554}{{\ttfamily arXiv:1701.01554
  [hep-ph]}}.

\bibitem{Kharzeev:1994pz}
D.~Kharzeev and H.~Satz, ``{Quarkonium interactions in hadronic matter},''
  \href{http://dx.doi.org/10.1016/0370-2693(94)90604-1}{{\em Phys. Lett. B}
  {\bfseries 334} (1994) 155--162},
  \href{http://arxiv.org/abs/hep-ph/9405414}{{\ttfamily arXiv:hep-ph/9405414}}.

\bibitem{Xu:1995eb}
X.-M. Xu, D.~Kharzeev, H.~Satz, and X.-N. Wang, ``{J / psi suppression in an
  equilibrating parton plasma},''
  \href{http://dx.doi.org/10.1103/PhysRevC.53.3051}{{\em Phys. Rev. C}
  {\bfseries 53} (1996) 3051--3056},
  \href{http://arxiv.org/abs/hep-ph/9511331}{{\ttfamily arXiv:hep-ph/9511331}}.

\bibitem{Grandchamp:2001pf}
L.~Grandchamp and R.~Rapp, ``{Thermal versus direct J / Psi production in
  ultrarelativistic heavy ion collisions},''
  \href{http://dx.doi.org/10.1016/S0370-2693(01)01311-9}{{\em Phys. Lett. B}
  {\bfseries 523} (2001) 60--66},
  \href{http://arxiv.org/abs/hep-ph/0103124}{{\ttfamily arXiv:hep-ph/0103124}}.

\bibitem{Grandchamp:2002wp}
L.~Grandchamp and R.~Rapp, ``{Charmonium suppression and regeneration from SPS
  to RHIC},'' \href{http://dx.doi.org/10.1016/S0375-9474(02)01027-8}{{\em Nucl.
  Phys. A} {\bfseries 709} (2002) 415--439},
  \href{http://arxiv.org/abs/hep-ph/0205305}{{\ttfamily arXiv:hep-ph/0205305}}.

\bibitem{Laine:2006ns}
M.~Laine, O.~Philipsen, P.~Romatschke, and M.~Tassler, ``{Real-time static
  potential in hot QCD},''
  \href{http://dx.doi.org/10.1088/1126-6708/2007/03/054}{{\em JHEP} {\bfseries
  03} (2007) 054}, \href{http://arxiv.org/abs/hep-ph/0611300}{{\ttfamily
  arXiv:hep-ph/0611300}}.

\bibitem{Brambilla:2013dpa}
N.~Brambilla, M.~A. Escobedo, J.~Ghiglieri, and A.~Vairo, ``{Thermal width and
  quarkonium dissociation by inelastic parton scattering},''
  \href{http://dx.doi.org/10.1007/JHEP05(2013)130}{{\em JHEP} {\bfseries 05}
  (2013) 130}, \href{http://arxiv.org/abs/1303.6097}{{\ttfamily arXiv:1303.6097
  [hep-ph]}}.

\bibitem{Biondini:2017ufr}
S.~Biondini and M.~Laine, ``{Re-derived overclosure bound for the inert doublet
  model},'' \href{http://dx.doi.org/10.1007/JHEP08(2017)047}{{\em JHEP}
  {\bfseries 08} (2017) 047}, \href{http://arxiv.org/abs/1706.01894}{{\ttfamily
  arXiv:1706.01894 [hep-ph]}}.

\bibitem{Binder:2021otw}
T.~Binder, K.~Mukaida, B.~Scheihing-Hitschfeld, and X.~Yao, ``{Non-Abelian
  electric field correlator at NLO for dark matter relic abundance and
  quarkonium transport},''
  \href{http://dx.doi.org/10.1007/JHEP01(2022)137}{{\em JHEP} {\bfseries 01}
  (2022) 137}, \href{http://arxiv.org/abs/2107.03945}{{\ttfamily
  arXiv:2107.03945 [hep-ph]}}.

\bibitem{Binder:2019erp}
T.~Binder, K.~Mukaida, and K.~Petraki, ``{Rapid bound-state formation of Dark
  Matter in the Early Universe},''
  \href{http://dx.doi.org/10.1103/PhysRevLett.124.161102}{{\em Phys. Rev.
  Lett.} {\bfseries 124} no.~16, (2020) 161102},
\href{http://arxiv.org/abs/1910.11288}{{\ttfamily arXiv:1910.11288 [hep-ph]}}.

\bibitem{Mohan:2019zrk}
K.~A. Mohan, D.~Sengupta, T.~M.~P. Tait, B.~Yan, and C.~P. Yuan, ``{Direct
  detection and LHC constraints on a $t$-channel simplified model of Majorana
  dark matter at one loop},''
  \href{http://dx.doi.org/10.1007/JHEP05(2019)115}{{\em JHEP} {\bfseries 05}
  (2019) 115}, \href{http://arxiv.org/abs/1903.05650}{{\ttfamily
  arXiv:1903.05650 [hep-ph]}}. [Erratum: JHEP 05, 232 (2023)].

\bibitem{LZ:2022ufs}
{\bfseries LZ} Collaboration, J.~Aalbers {\em et~al.}, ``{First Dark Matter
  Search Results from the LUX-ZEPLIN (LZ) Experiment},''
  \href{http://arxiv.org/abs/2207.03764}{{\ttfamily arXiv:2207.03764
  [hep-ex]}}.

\bibitem{Strigari:2009bq}
L.~E. Strigari, ``{Neutrino Coherent Scattering Rates at Direct Dark Matter
  Detectors},'' \href{http://dx.doi.org/10.1088/1367-2630/11/10/105011}{{\em
  New J. Phys.} {\bfseries 11} (2009) 105011},
  \href{http://arxiv.org/abs/0903.3630}{{\ttfamily arXiv:0903.3630
  [astro-ph.CO]}}.

\bibitem{Hess:2021cdp}
{\bfseries Hess, HAWC, VERITAS, MAGIC, H.E.S.S., Fermi-LAT} Collaboration,
  H.~Abdalla {\em et~al.}, ``{Combined dark matter searches towards dwarf
  spheroidal galaxies with Fermi-LAT, HAWC, H.E.S.S., MAGIC, and VERITAS},''
  \href{http://dx.doi.org/10.22323/1.395.0528}{{\em PoS} {\bfseries ICRC2021}
  (2021) 528}, \href{http://arxiv.org/abs/2108.13646}{{\ttfamily
  arXiv:2108.13646 [hep-ex]}}.

\bibitem{Cirelli:2010xx}
M.~Cirelli, G.~Corcella, A.~Hektor, G.~Hutsi, M.~Kadastik, P.~Panci, M.~Raidal,
  F.~Sala, and A.~Strumia, ``{PPPC 4 DM ID: A Poor Particle Physicist Cookbook
  for Dark Matter Indirect Detection},''
  \href{http://dx.doi.org/10.1088/1475-7516/2012/10/E01}{{\em JCAP} {\bfseries
  03} (2011) 051}, \href{http://arxiv.org/abs/1012.4515}{{\ttfamily
  arXiv:1012.4515 [hep-ph]}}. [Erratum: JCAP 10, E01 (2012)].

\bibitem{CTA:2020qlo}
{\bfseries CTA} Collaboration, A.~Acharyya {\em et~al.}, ``{Sensitivity of the
  Cherenkov Telescope Array to a dark matter signal from the Galactic
  centre},'' \href{http://dx.doi.org/10.1088/1475-7516/2021/01/057}{{\em JCAP}
  {\bfseries 01} (2021) 057}, \href{http://arxiv.org/abs/2007.16129}{{\ttfamily
  arXiv:2007.16129 [astro-ph.HE]}}.

\bibitem{Muong-2:2006rrc}
{\bfseries Muon g-2} Collaboration, G.~W. Bennett {\em et~al.}, ``{Final Report
  of the Muon E821 Anomalous Magnetic Moment Measurement at BNL},''
  \href{http://dx.doi.org/10.1103/PhysRevD.73.072003}{{\em Phys. Rev. D}
  {\bfseries 73} (2006) 072003},
  \href{http://arxiv.org/abs/hep-ex/0602035}{{\ttfamily arXiv:hep-ex/0602035}}.

\bibitem{Muong-2:2021ojo}
{\bfseries Muon g-2} Collaboration, B.~Abi {\em et~al.}, ``{Measurement of the
  Positive Muon Anomalous Magnetic Moment to 0.46 ppm},''
  \href{http://dx.doi.org/10.1103/PhysRevLett.126.141801}{{\em Phys. Rev.
  Lett.} {\bfseries 126} no.~14, (2021) 141801},
  \href{http://arxiv.org/abs/2104.03281}{{\ttfamily arXiv:2104.03281
  [hep-ex]}}.

\bibitem{Kowalska:2018ulj}
K.~Kowalska, E.~M. Sessolo, and Y.~Yamamoto, ``{Constraints on charmphilic
  solutions to the muon g-2 with leptoquarks},''
  \href{http://dx.doi.org/10.1103/PhysRevD.99.055007}{{\em Phys. Rev. D}
  {\bfseries 99} no.~5, (2019) 055007},
  \href{http://arxiv.org/abs/1812.06851}{{\ttfamily arXiv:1812.06851
  [hep-ph]}}.

\bibitem{Bigaran:2020jil}
I.~Bigaran and R.~R. Volkas, ``{Getting chirality right: Single scalar
  leptoquark solutions to the $(g-2)_{e,\mu}$ puzzle},''
  \href{http://dx.doi.org/10.1103/PhysRevD.102.075037}{{\em Phys. Rev. D}
  {\bfseries 102} no.~7, (2020) 075037},
  \href{http://arxiv.org/abs/2002.12544}{{\ttfamily arXiv:2002.12544
  [hep-ph]}}.

\bibitem{Dorsner:2020aaz}
I.~Dor\v{s}ner, S.~Fajfer, and S.~Saad, ``{$\mu \to e \gamma$ selecting scalar
  leptoquark solutions for the $(g-2)_{e,\mu}$ puzzles},''
  \href{http://dx.doi.org/10.1103/PhysRevD.102.075007}{{\em Phys. Rev. D}
  {\bfseries 102} no.~7, (2020) 075007},
  \href{http://arxiv.org/abs/2006.11624}{{\ttfamily arXiv:2006.11624
  [hep-ph]}}.

\bibitem{Khasianevich:2023duu}
U.~Khasianevich, D.~Stoeckinger, H.~Stoeckinger-Kim, and J.~Wuensche,
  ``{Constraint on scalar leptoquark from low energy leptonic observables},''
  \href{http://arxiv.org/abs/2305.05016}{{\ttfamily arXiv:2305.05016
  [hep-ph]}}.

\bibitem{ParticleDataGroup:2020ssz}
{\bfseries Particle Data Group} Collaboration, P.~A. Zyla {\em et~al.},
  ``{Review of Particle Physics},''
  \href{http://dx.doi.org/10.1093/ptep/ptaa104}{{\em PTEP} {\bfseries 2020}
  no.~8, (2020) 083C01}.

\bibitem{NA62}
G.~Ruggiero, ``{(NA62) talk at Kaon 2019, Perugia (Italy), 10 September
  2019},''.

\bibitem{Mandal:2019gff}
R.~Mandal and A.~Pich, ``{Constraints on scalar leptoquarks from lepton and
  kaon physics},'' \href{http://dx.doi.org/10.1007/JHEP12(2019)089}{{\em JHEP}
  {\bfseries 12} (2019) 089}, \href{http://arxiv.org/abs/1908.11155}{{\ttfamily
  arXiv:1908.11155 [hep-ph]}}.

\bibitem{Angelescu:2021lln}
A.~Angelescu, D.~Be\v{c}irevi\'c, D.~A. Faroughy, F.~Jaffredo, and
  O.~Sumensari, ``{Single leptoquark solutions to the B-physics anomalies},''
  \href{http://dx.doi.org/10.1103/PhysRevD.104.055017}{{\em Phys. Rev. D}
  {\bfseries 104} no.~5, (2021) 055017},
  \href{http://arxiv.org/abs/2103.12504}{{\ttfamily arXiv:2103.12504
  [hep-ph]}}.

\bibitem{Julio:2022bue}
J.~Julio, S.~Saad, and A.~Thapa, ``{Marriage between neutrino mass and flavor
  anomalies},'' \href{http://dx.doi.org/10.1103/PhysRevD.106.055003}{{\em Phys.
  Rev. D} {\bfseries 106} no.~5, (2022) 055003},
  \href{http://arxiv.org/abs/2203.15499}{{\ttfamily arXiv:2203.15499
  [hep-ph]}}.

\bibitem{ATLAS:2020zms}
{\bfseries ATLAS} Collaboration, G.~Aad {\em et~al.}, ``{Search for heavy Higgs
  bosons decaying into two tau leptons with the ATLAS detector using $pp$
  collisions at $\sqrt{s}=13$ TeV},''
  \href{http://dx.doi.org/10.1103/PhysRevLett.125.051801}{{\em Phys. Rev.
  Lett.} {\bfseries 125} no.~5, (2020) 051801},
  \href{http://arxiv.org/abs/2002.12223}{{\ttfamily arXiv:2002.12223
  [hep-ex]}}.

\bibitem{CMS-PAS-EXO-19-019}
{\bfseries CMS} Collaboration, ``{Search for a narrow resonance in high-mass
  dilepton final states in proton-proton collisions using
  140$~\mathrm{fb}^{-1}$ of data at $\sqrt{s}=13~\mathrm{TeV}$},'' tech. rep.,
  CERN, Geneva, 2019.
\newblock \url{https://cds.cern.ch/record/2684757}.

\bibitem{Borsanyi:2020mff}
S.~Borsanyi {\em et~al.}, ``{Leading hadronic contribution to the muon magnetic
  moment from lattice QCD},''
  \href{http://dx.doi.org/10.1038/s41586-021-03418-1}{{\em Nature} {\bfseries
  593} no.~7857, (2021) 51--55},
  \href{http://arxiv.org/abs/2002.12347}{{\ttfamily arXiv:2002.12347
  [hep-lat]}}.

\bibitem{Ce:2022kxy}
M.~C\`e {\em et~al.}, ``{Window observable for the hadronic vacuum polarization
  contribution to the muon g-2 from lattice QCD},''
  \href{http://dx.doi.org/10.1103/PhysRevD.106.114502}{{\em Phys. Rev. D}
  {\bfseries 106} no.~11, (2022) 114502},
  \href{http://arxiv.org/abs/2206.06582}{{\ttfamily arXiv:2206.06582
  [hep-lat]}}.

\bibitem{Alexandrou:2022amy}
{\bfseries Extended Twisted Mass} Collaboration, C.~Alexandrou {\em et~al.},
  ``{Lattice calculation of the short and intermediate time-distance hadronic
  vacuum polarization contributions to the muon magnetic moment using
  twisted-mass fermions},''
  \href{http://dx.doi.org/10.1103/PhysRevD.107.074506}{{\em Phys. Rev. D}
  {\bfseries 107} no.~7, (2023) 074506},
  \href{http://arxiv.org/abs/2206.15084}{{\ttfamily arXiv:2206.15084
  [hep-lat]}}.

\bibitem{Diaz:2017lit}
B.~Diaz, M.~Schmaltz, and Y.-M. Zhong, ``{The leptoquark
  Hunter\textquoteright{}s guide: Pair production},''
  \href{http://dx.doi.org/10.1007/JHEP10(2017)097}{{\em JHEP} {\bfseries 10}
  (2017) 097}, \href{http://arxiv.org/abs/1706.05033}{{\ttfamily
  arXiv:1706.05033 [hep-ph]}}.

\bibitem{Dorsner:2018ynv}
I.~Dor\v{s}ner and A.~Greljo, ``{Leptoquark toolbox for precision collider
  studies},'' \href{http://dx.doi.org/10.1007/JHEP05(2018)126}{{\em JHEP}
  {\bfseries 05} (2018) 126}, \href{http://arxiv.org/abs/1801.07641}{{\ttfamily
  arXiv:1801.07641 [hep-ph]}}.

\bibitem{ATLAS:2020xov}
{\bfseries ATLAS} Collaboration, G.~Aad {\em et~al.}, ``{Search for pair
  production of scalar leptoquarks decaying into first- or second-generation
  leptons and top quarks in proton\textendash{}proton collisions at $\sqrt{s}$
  = 13 TeV with the ATLAS detector},''
  \href{http://dx.doi.org/10.1140/epjc/s10052-021-09009-8}{{\em Eur. Phys. J.
  C} {\bfseries 81} no.~4, (2021) 313},
  \href{http://arxiv.org/abs/2010.02098}{{\ttfamily arXiv:2010.02098
  [hep-ex]}}.

\bibitem{CMS:2018yiq}
{\bfseries CMS} Collaboration, A.~M. Sirunyan {\em et~al.}, ``{Search for dark
  matter in events with a leptoquark and missing transverse momentum in
  proton-proton collisions at 13 TeV},''
  \href{http://dx.doi.org/10.1016/j.physletb.2019.05.046}{{\em Phys. Lett. B}
  {\bfseries 795} (2019) 76--99},
  \href{http://arxiv.org/abs/1811.10151}{{\ttfamily arXiv:1811.10151
  [hep-ex]}}.

\bibitem{Eboli:1987vb}
O.~J.~P. Eboli and A.~V. Olinto, ``{Composite Leptoquarks in Hadronic
  Colliders},'' \href{http://dx.doi.org/10.1103/PhysRevD.38.3461}{{\em Phys.
  Rev. D} {\bfseries 38} (1988) 3461}.

\bibitem{Schwaller:2013baa}
P.~Schwaller and J.~Zurita, ``{Compressed electroweakino spectra at the LHC},''
  \href{http://dx.doi.org/10.1007/JHEP03(2014)060}{{\em JHEP} {\bfseries 03}
  (2014) 060}, \href{http://arxiv.org/abs/1312.7350}{{\ttfamily arXiv:1312.7350
  [hep-ph]}}.

\bibitem{Farrar:1978xj}
G.~R. Farrar and P.~Fayet, ``{Phenomenology of the Production, Decay, and
  Detection of New Hadronic States Associated with Supersymmetry},''
  \href{http://dx.doi.org/10.1016/0370-2693(78)90858-4}{{\em Phys. Lett. B}
  {\bfseries 76} (1978) 575--579}.

\bibitem{ATLAS:2019gqq}
{\bfseries ATLAS} Collaboration, M.~Aaboud {\em et~al.}, ``{Search for heavy
  charged long-lived particles in the ATLAS detector in 36.1 fb$^{-1}$ of
  proton-proton collision data at $\sqrt{s} = 13$ TeV},''
  \href{http://dx.doi.org/10.1103/PhysRevD.99.092007}{{\em Phys. Rev. D}
  {\bfseries 99} no.~9, (2019) 092007},
  \href{http://arxiv.org/abs/1902.01636}{{\ttfamily arXiv:1902.01636
  [hep-ex]}}.

\bibitem{Criado:2019mvu}
J.~C. Criado and M.~Perez-Victoria, ``{Vector-like quarks with
  non-renormalizable interactions},''
  \href{http://dx.doi.org/10.1007/JHEP01(2020)057}{{\em JHEP} {\bfseries 01}
  (2020) 057}, \href{http://arxiv.org/abs/1908.08964}{{\ttfamily
  arXiv:1908.08964 [hep-ph]}}.

\bibitem{Buttazzo:2020ibd}
D.~Buttazzo and P.~Paradisi, ``{Probing the muon $g-2$ anomaly with the Higgs
  boson at a muon collider},''
  \href{http://dx.doi.org/10.1103/PhysRevD.104.075021}{{\em Phys. Rev. D}
  {\bfseries 104} no.~7, (2021) 075021},
  \href{http://arxiv.org/abs/2012.02769}{{\ttfamily arXiv:2012.02769
  [hep-ph]}}.

\bibitem{Giacchino:2015hvk}
F.~Giacchino, A.~Ibarra, L.~Lopez~Honorez, M.~H.~G. Tytgat, and S.~Wild,
  ``{Signatures from Scalar Dark Matter with a Vector-like Quark Mediator},''
  \href{http://dx.doi.org/10.1088/1475-7516/2016/02/002}{{\em JCAP} {\bfseries
  02} (2016) 002}, \href{http://arxiv.org/abs/1511.04452}{{\ttfamily
  arXiv:1511.04452 [hep-ph]}}.

\bibitem{Belanger:2018sti}
G.~B\'elanger {\em et~al.}, ``{LHC-friendly minimal freeze-in models},''
  \href{http://dx.doi.org/10.1007/JHEP02(2019)186}{{\em JHEP} {\bfseries 02}
  (2019) 186}, \href{http://arxiv.org/abs/1811.05478}{{\ttfamily
  arXiv:1811.05478 [hep-ph]}}.

\bibitem{Davidson:2008bu}
S.~Davidson, E.~Nardi, and Y.~Nir, ``{Leptogenesis},''
  \href{http://dx.doi.org/10.1016/j.physrep.2008.06.002}{{\em Phys. Rept.}
  {\bfseries 466} (2008) 105--177},
  \href{http://arxiv.org/abs/0802.2962}{{\ttfamily arXiv:0802.2962 [hep-ph]}}.

\bibitem{gordon}
W.~Gordon, ``Zur berechnung der matrizen beim wasserstoffatom,''
  \href{http://dx.doi.org/10.1002/andp.19293940807}{{\em Annalen der Physik}
  {\bfseries 394} no.~8, (1929) 1031--1056}.

\bibitem{stobbe}
M.~Stobbe, ``Zur quantenmechanik photoelektrischer prozesse,''
  \href{http://dx.doi.org/10.1002/andp.19303990604}{{\em Annalen der Physik}
  {\bfseries 399} no.~6, (1930) 661}.

\end{thebibliography}\endgroup
\end{document}